\documentclass{aastex631}

\shorttitle{PEARLS: JWST view of ALMA sources}
\shortauthors{Cheng et al.}

\graphicspath{{./}{figures/}}

\newcommand{\etal}{{et~al.}}

\newcommand{\Msol}{M$_{\odot}$}
\newcommand{\GALFIT}{{\sc galfit}}
\newcommand{\MAGPHYS}{{\sc magphys}}
\newcommand{\no}{\nodata}
\newcommand{\js}{}
\newcommand{\HST}{{\js HST}}
\newcommand{\JWST}{{\js JWST}}

\begin{document}

\title{JWST's PEARLS: A JWST/NIRCam view of ALMA sources}

\correspondingauthor{Haojing Yan}
\email{yanha@missouri.edu}

\author[0000-0003-0202-0534]{Cheng Cheng}
\affiliation{Chinese Academy of Sciences South America Center for Astronomy, National Astronomical Observatories, CAS, Beijing 100101, China}
\affiliation{CAS Key Laboratory of Optical Astronomy, National Astronomical Observatories, Chinese Academy of Sciences, Beijing 100101, China}

\author[0000-0001-6511-8745]{Jia-Sheng Huang} 
\affiliation{Chinese Academy of Sciences South America Center for Astronomy, National Astronomical Observatories, CAS, Beijing 100101, China}
\affiliation{Center for Astrophysics \textbar\ Harvard \& Smithsonian, 60 Garden St., Cambridge, MA 02138 USA}

\author[0000-0003-3037-257X]{Ian Smail} 
\affiliation{Centre for Extragalactic Astronomy, Department of Physics, Durham
University, South Road, Durham DH1 3LE, UK}

\author[0000-0001-7592-7714]{Haojing Yan} 
\affiliation{Department of Physics and Astronomy, University of Missouri,
Columbia, MO 65211}

\author[0000-0003-3329-1337]{Seth H. Cohen} 
\affiliation{School of Earth and Space Exploration, Arizona State University,
Tempe, AZ 85287-1404, USA}

\author[0000-0003-1268-5230]{Rolf A. Jansen} 
\affiliation{School of Earth and Space Exploration, Arizona State University,
Tempe, AZ 85287-1404, USA}

\author[0000-0001-8156-6281]{Rogier A. Windhorst}
\affiliation{School of Earth and Space Exploration, Arizona State University,
Tempe, AZ 85287-1404, USA}

\author[0000-0003-3270-6844]{Zhiyuan Ma} 
\affiliation{Department of Astronomy, University of Massachusetts, Amherst, MA
01003, USA}

\author[0000-0002-6610-2048]{Anton Koekemoer} 
\affiliation{Space Telescope Science Institute, 3700 San Martin Drive, Baltimore, MD 21218, USA}

\author[0000-0001-9262-9997]{Christopher N. A. Willmer} 
\affiliation{Steward Observatory, University of Arizona, 933 N Cherry Ave,
Tucson, AZ, 85721-0009}

\author[0000-0002-9895-5758]{S. P. Willner} 
\affiliation{Center for Astrophysics \textbar\ Harvard \& Smithsonian, 60 Garden St., Cambridge, MA 02138 USA}

\author[0000-0001-9065-3926]{Jose M. Diego} 
\affiliation{Instituto de F\'isica de Cantabria (CSIC-UC). Avenida. Los Castros
s/n. 39005 Santander, Spain}

\author[0000-0003-1625-8009]{Brenda Frye} 
\affiliation{University of Arizona, Department of Astronomy/Steward
Observatory, 933 N Cherry Ave, Tucson, AZ85721}

\author[0000-0003-1949-7638]{Christopher J. Conselice} 
\affiliation{Jodrell Bank Centre for Astrophysics, Alan Turing Building,
University of Manchester, Oxford Road, Manchester M13 9PL, UK}

\author[0000-0002-8919-079X]{Leonardo Ferreira} 
\affiliation{University of Nottingham, School of Physics \& Astronomy,
Nottingham, NG7 2RD, UK}

\author[0000-0003-4030-3455]{Andreea Petric} 
\affiliation{Space Telescope Science Institute, 3700 San Martin Drive, Baltimore, MD 21218, USA}

\author[0000-0001-7095-7543]{Min Yun} 
\affiliation{Department of Astronomy, University of Massachusetts, Amherst, MA
01003, USA}

\author[0000-0003-1436-7658]{Hansung B. Gim}
\affiliation{Department of Physics, Montana State University, P. O. Box 173840, Bozeman, MT 59717, USA}

\author[0000-0001-7411-5386]{Maria del Carmen Polletta} 
\affiliation{INAF, Istituto di Astrofisica Spaziale e Fisica cosmica (IASF)
Milano, Via A. Corti 12, 20133 Milan, Italy}

\author[0000-0001-6889-8388]{Kenneth J. Duncan} 
\affiliation{Institute for Astronomy, Royal Observatory, Blackford Hill, Edinburgh, EH9 3HJ, UK}

\author[0000-0002-4884-6756]{Benne W. Holwerda} 
\affiliation{Department of Physics and Astronomy, University of Louisville,
Louisville KY 40292, USA} 

\author[0000-0001-8887-2257]{Huub J. A. R\"ottgering} 
\affiliation{Leiden Observatory, PO Box 9513, 2300 RA Leiden, The Netherlands}

\author{Rachel Honor}
\affiliation{School of Earth and Space Exploration, Arizona State University,
Tempe, AZ 85287-1404, USA}

\author[0000-0001-6145-5090]{Nimish P. Hathi} 
\affiliation{Space Telescope Science Institute, 3700 San Martin Drive, Baltimore, MD 21218, USA}

\author[0000-0001-9394-6732]{Patrick S. Kamieneski} 
\affiliation{Department of Astronomy, University of Massachusetts, Amherst, MA
01003, USA}

\author[0000-0003-4875-6272]{Nathan J. Adams} 
\affiliation{Jodrell Bank Centre for Astrophysics, Alan Turing Building, 
University of Manchester, Oxford Road, Manchester M13 9PL, UK}

\author[0000-0001-7410-7669]{Dan Coe} 
\affiliation{Space Telescope Science Institute, 3700 San Martin Drive, Baltimore, MD 21218, USA}
\affiliation{Association of Universities for Research in Astronomy (AURA) for the European Space Agency (ESA), STScI, Baltimore, MD 21218, USA}
\affiliation{Center for Astrophysical Sciences, Department of Physics and Astronomy, The Johns Hopkins University, 3400 N Charles St. Baltimore, MD 21218, USA}

\author[0000-0002-5807-4411]{Tom~Broadhurst} 
\affiliation{Department of Theoretical Physics, University of the Basque
Country UPV-EHU, 48040 Bilbao, Spain}
\affiliation{Donostia International Physics Center (DIPC), 20018 Donostia, The
Basque Country}
\affiliation{IKERBASQUE, Basque Foundation for Science, Alameda Urquijo, 36-5
48008 Bilbao, Spain}


\author[0000-0002-7265-7920]{Jake Summers} 
\affiliation{School of Earth and Space Exploration, Arizona State University,
Tempe, AZ 85287-1404, USA}

\author[0000-0001-9052-9837]{Scott Tompkins} 
\affiliation{School of Earth and Space Exploration, Arizona State University,
Tempe, AZ 85287-1404, USA}

\author[0000-0001-9491-7327]{Simon P. Driver} 
\affiliation{International Centre for Radio Astronomy Research (ICRAR) and the
International Space Centre (ISC), The University of Western Australia, M468,
35 Stirling Highway, Crawley, WA 6009, Australia}

\author[0000-0001-9440-8872]{Norman A. Grogin} 
\affiliation{Space Telescope Science Institute,
3700 San Martin Drive, Baltimore, MD 21218, USA}

\author[0000-0001-6434-7845]{Madeline A. Marshall} 
\affiliation{National Research Council of Canada, Herzberg Astronomy \&
Astrophysics Research Centre, 5071 West Saanich Road, Victoria, BC V9E 2E7,
Canada}
\affiliation{ARC Centre of Excellence for All Sky Astrophysics in 3 Dimensions
(ASTRO 3D), Australia}

\author[0000-0003-3382-5941]{Nor Pirzkal} 
\affiliation{Space Telescope Science Institute,
3700 San Martin Drive, Baltimore, MD 21218, USA}

\author[0000-0003-0429-3579]{Aaron Robotham} 
\affiliation{International Centre for Radio Astronomy Research (ICRAR) and the
International Space Centre (ISC), The University of Western Australia, M468,
35 Stirling Highway, Crawley, WA 6009, Australia}

\author[0000-0003-0894-1588]{Russell E. Ryan, Jr.} 
\affiliation{Space Telescope Science Institute,
3700 San Martin Drive, Baltimore, MD 21218, USA}

\begin{abstract}

    We report the results of James Webb Space Telescope/NIRCam
observations of 19 (sub)millimeter (submm/mm) sources detected by the 
Atacama Large Millimeter Array (ALMA)\null. The accurate ALMA positions 
allowed unambiguous identifications of their NIRCam counterparts. Taking 
gravitational lensing into account, these represent 16 distinct galaxies in 
three fields and constitute the largest sample of its kind to date. The 
counterparts' spectral energy distributions, which cover from rest-frame 
ultraviolet to near-infrared provide photometric redshifts ($1<z<4.5$) and stellar
masses ($M_*>10^{10.5}$~\Msol), which are similar to sub-millimeter galaxies (SMGs) 
studied previously. However, our sample is fainter in submm/mm than the
classic SMG samples are, and our sources exhibit a wider range of properties.
They have dust-embedded star-formation rates as low as 10~\Msol~yr$^{-1}$,
and the sources populate both the star-forming main sequence and the 
quiescent categories. The deep NIRCam data allow us to study the 
rest-frame near-IR morphologies. Excluding two multiply imaged systems and
one quasar, the majority of the remaining sources are disk-like and show 
either little or no disturbance. This suggests that secular growth is a 
potential route for the assembly of high-mass disk galaxies.
While a few objects have large disks, the majority have small disks (median 
half-mass radius of 1.6~kpc). At this time, it is unclear whether this is due
to the prevalence of small disks at these redshifts or some unknown selection
effects of deep ALMA observations. A larger sample of ALMA sources with 
NIRCam observations will be able to address this question.

\end{abstract}


\section{Introduction} \label{sec:intro}

     The submillimeter/millimeter (submm/mm) window offers a view of the 
high-redshift galaxy population complementary to that seen in visible light 
\citep[e.g.,][]{ 1997ApJ...490L...5S, 2014PhR...541...45C}. Sources detected 
in this window are commonly referred to as submillimeter galaxies (SMGs). 
Followup studies have shown that SMGs are mainly high-redshift ($z\sim 2$--3),
dusty, and star-forming galaxies, and most of the brightest SMGs are 
gravitationally magnified by foreground massive galaxies or clusters
\citep{1996MNRAS.283.1340B, 2002MNRAS.329..445P, 2002PhR...369..111B, 
2010Sci...330..800N, 2017MNRAS.465.3558N}. Over the years, finding SMGs has 
become one of the most efficient methods to select 
dusty star-forming galaxies (DSFGs) at high redshifts.

    Despite many successes, SMG studies have been limited by the poor angular 
resolution of single-dish telescopes through which SMGs have been selected. 
The low resolution makes the source positions uncertain. More recently, 
interferometry arrays such as the Atacama Large Millimeter/submillimeter Array
(ALMA) have been used to locate SMGs and allow unambiguous identification of their 
counterparts \citep[e.g.,][]{2013ApJ...768...91H}. This has allowed studies
of SMGs using exquisite {\HST} images
\citep[e.g.,][]{2015ApJ...799..194C, 2016ApJ...833..103H,
2017MNRAS.466..861D, 2019ApJ...876..130H, 2019MNRAS.487.4648S, 
2020MNRAS.499.5241C, 2022arXiv221004437C}. Due to the high redshifts where SMGs reside, however,
{\HST} only sees them in the rest-frame UV-to-visible wavelengths, where SMGs
have severe dust extinction. This makes it difficult to obtain a comprehensive 
picture of their underlying stellar populations. 
Spitzer/IRAC images of SMGs probe them in the rest-frame near-infrared 
but have only $\sim$2\arcsec\ angular resolution, which is
insufficient to study the host morphologies. The low resolution also often
leads to blended IRAC images, which can make it difficult to derive accurate
spectral energy distributions (SEDs).

    {\JWST} promises to overcome prior limitations in infrared studies of SMGs 
\citep{2022ApJ...936L..19C, 2022arXiv220805296C, 2022arXiv220801816Z}. The 
NIRCam instrument offers $\sim$0.7 to 5~\micron\ images with angular resolution
comparable to that of {\HST}. Provided that accurate positions are known (e.g., 
from ALMA), NIRCam images can reveal the host stellar distributions and allow 
accurate SED measurements to enable detailed diagnostics of the underlying
stellar populations. 

   This paper presents the {\JWST}/NIRCam view of 19 submm/mm sources (16 distinct
galaxies) from the ALMA archival data in three well-studied fields, which form
the largest sample of its kind to date. The {\JWST}/NIRCam data come from the 
Prime Extragalactic Areas for Reionization and Lensing Science program (PEARLS;
\citealt{2022arXiv220904119W}) and the archival data from the Public Release
IMaging for Extragalactic Research 
\citep[PRIMER;][PI: James Dunlop,]{2021jwst.prop.1837D}. Accurate positions
are available from the ALMA 92~GHz, 260~GHz, or 340~GHz maps, making
counterpart identification possible.  Section~2 of this paper describes the 
data, and Section~3 presents the results. A summary is in Section~4.  
Throughout this paper, magnitudes are in the AB system, and we adopt a flat 
$\Lambda$CDM cosmology with $H_0 = 70$~km~s$^{-1}$~Mpc$^{-1}$, $\Omega_M=0.3$, 
and $\Omega_\Lambda=0.7$.

\setcounter{figure}{0}

\begin{figure}
    \centering
    \includegraphics[width=0.97\textwidth]{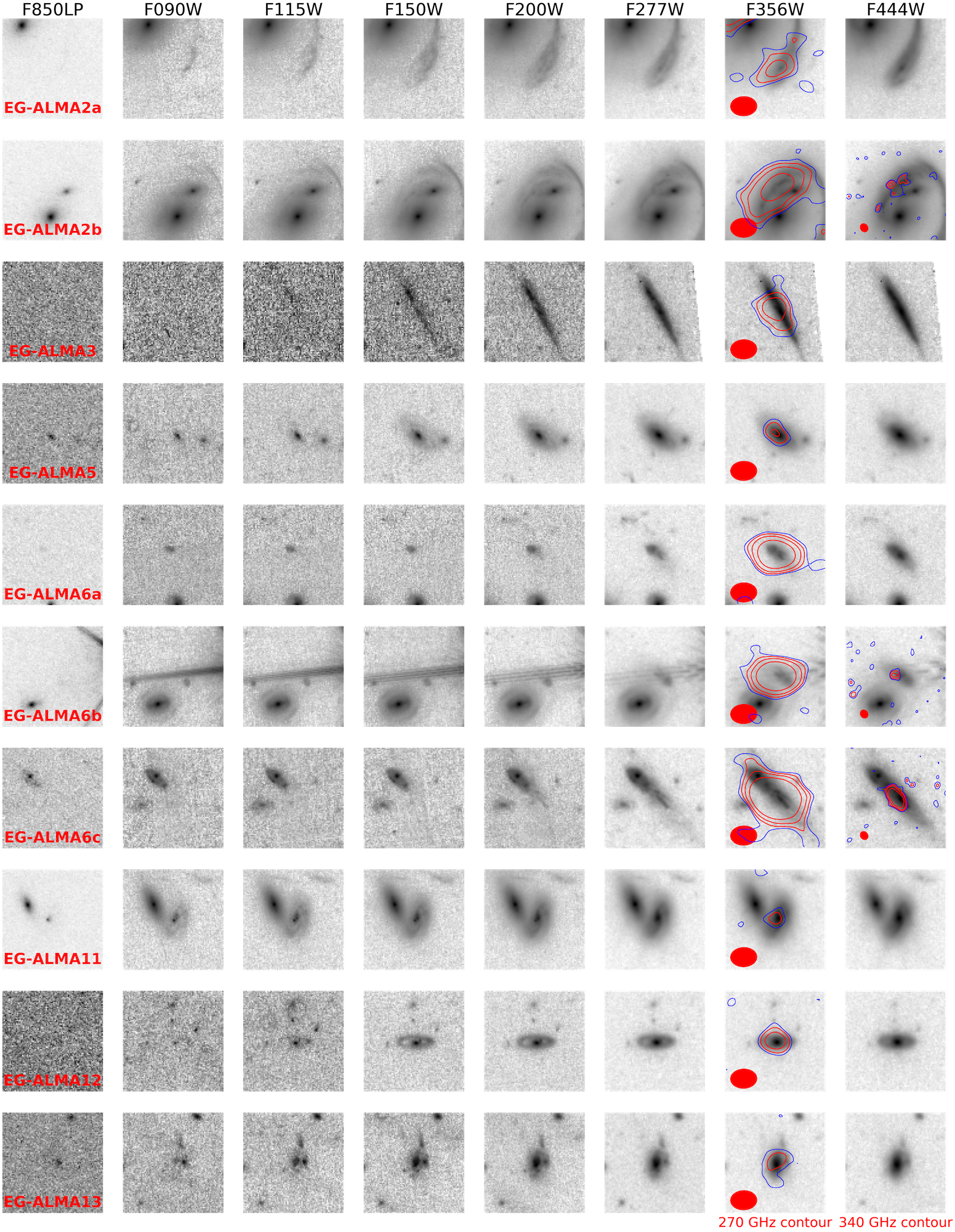}
    \caption{Images of the ALMA sources in El Gordo, one source in each row. 
     Each stamp is 5\arcsec$\times$5\arcsec\ in size.
     Both the archival HST ACS images (F850LP or F814W) and the new JWST 
     NIRCam images (F090W to F444W)
     are shown with bandpass filters noted at top. The ALMA Band~6 (1.1~mm) 
     contours (2$\sigma$ in blue; 3, 5, and 10$\sigma$ in red) are superposed 
     on the F356W images, and the Band~7 (870~$\mu$m) contours when available
     are superposed on the F444W images (using the same contour color-coding 
     as for Band~6). The ALMA beams are shown as the red-filled ellipses. }
    \label{EGstamps}
\end{figure}

\setcounter{figure}{0}

\begin{figure}
    \centering
    \includegraphics[width=0.97\textwidth]{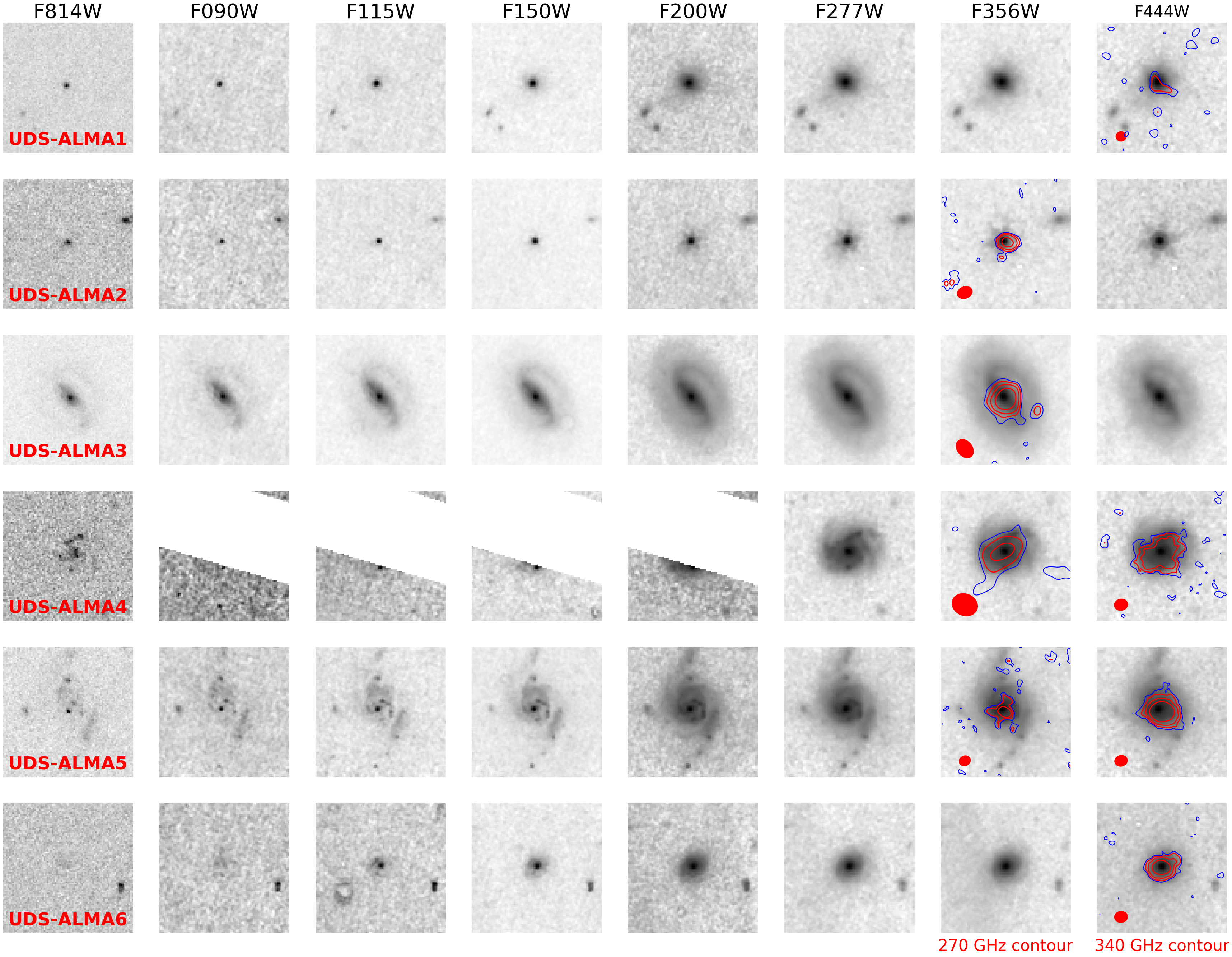}
    \caption{(Continued) Images of the ALMA sources in UDS\null. The images 
    are 5\arcsec$\times$5\arcsec\ in size.}
    \label{UDSstamps}
\end{figure}

\setcounter{figure}{0}

\begin{figure}
    \centering
    \includegraphics[width=0.97\textwidth]{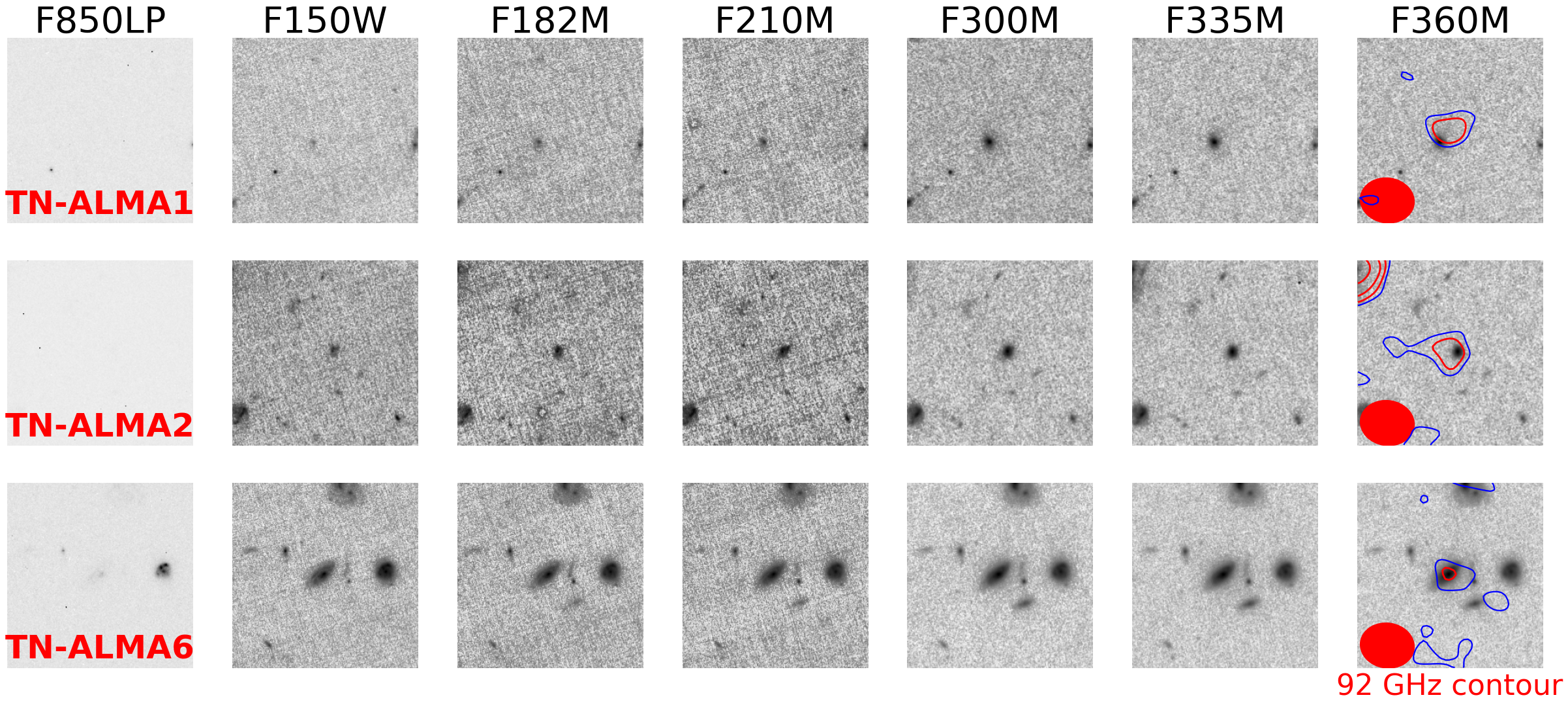}
    \caption{(Continued) Images of the ALMA sources in TNJ1338. The images
    are 10\arcsec$\times$10\arcsec\ in size.}
    \label{TNstamps}
\end{figure}

\begin{figure}
    \centering
    \includegraphics[width=0.355\textwidth]{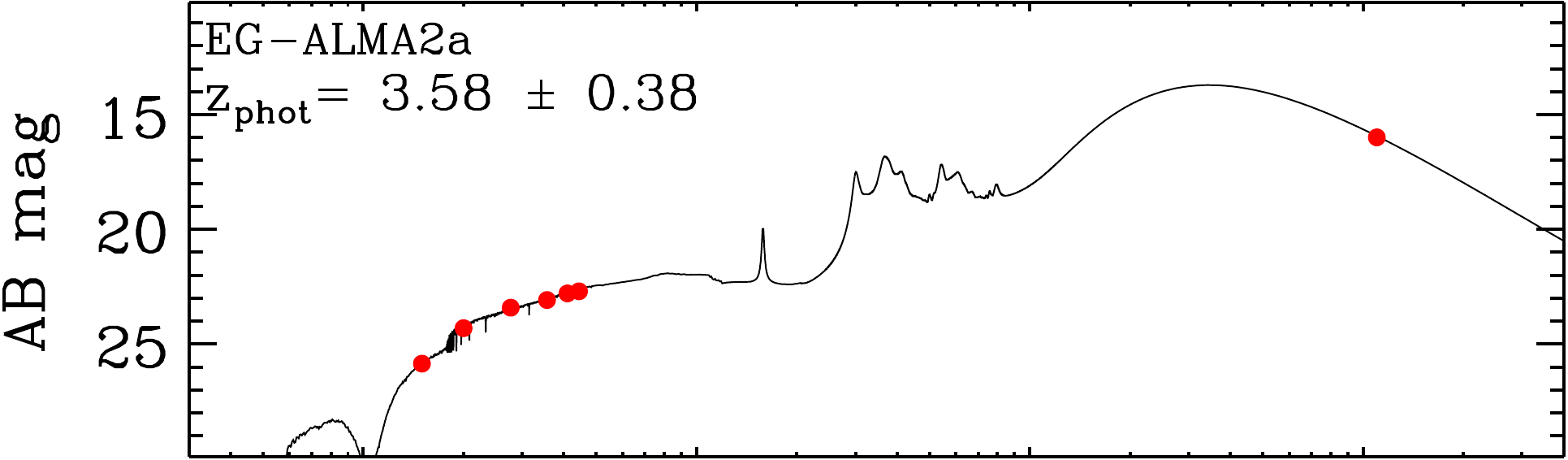}
    \includegraphics[width=0.315\textwidth]{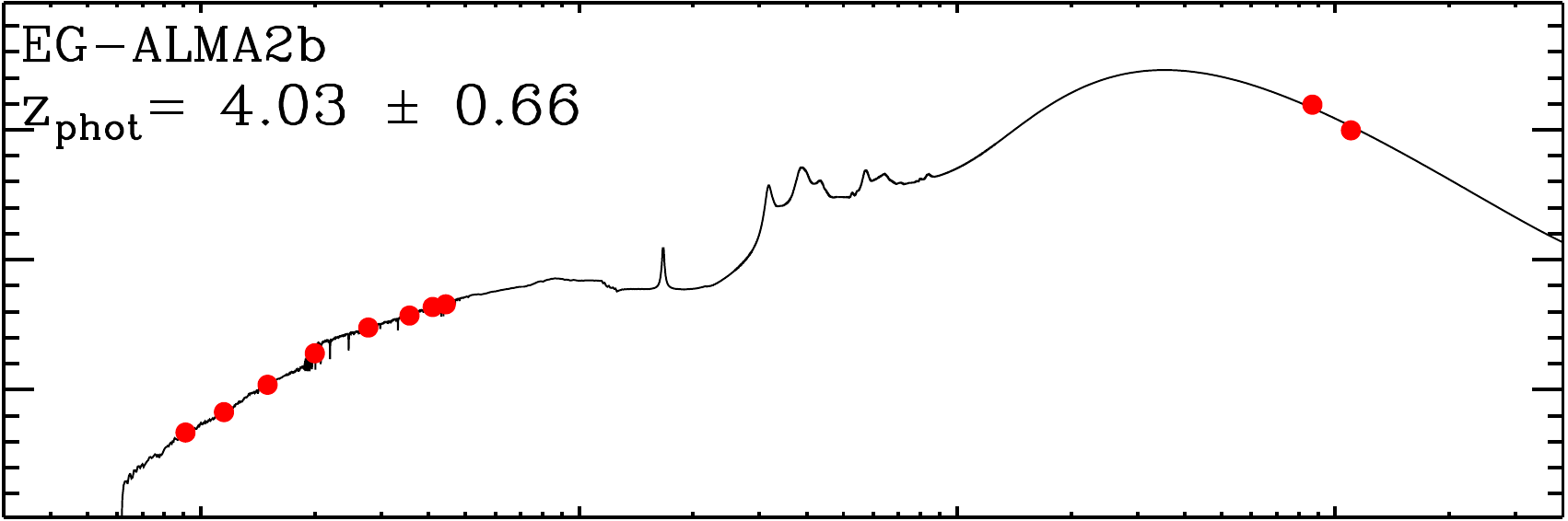}
    \includegraphics[width=0.315\textwidth]{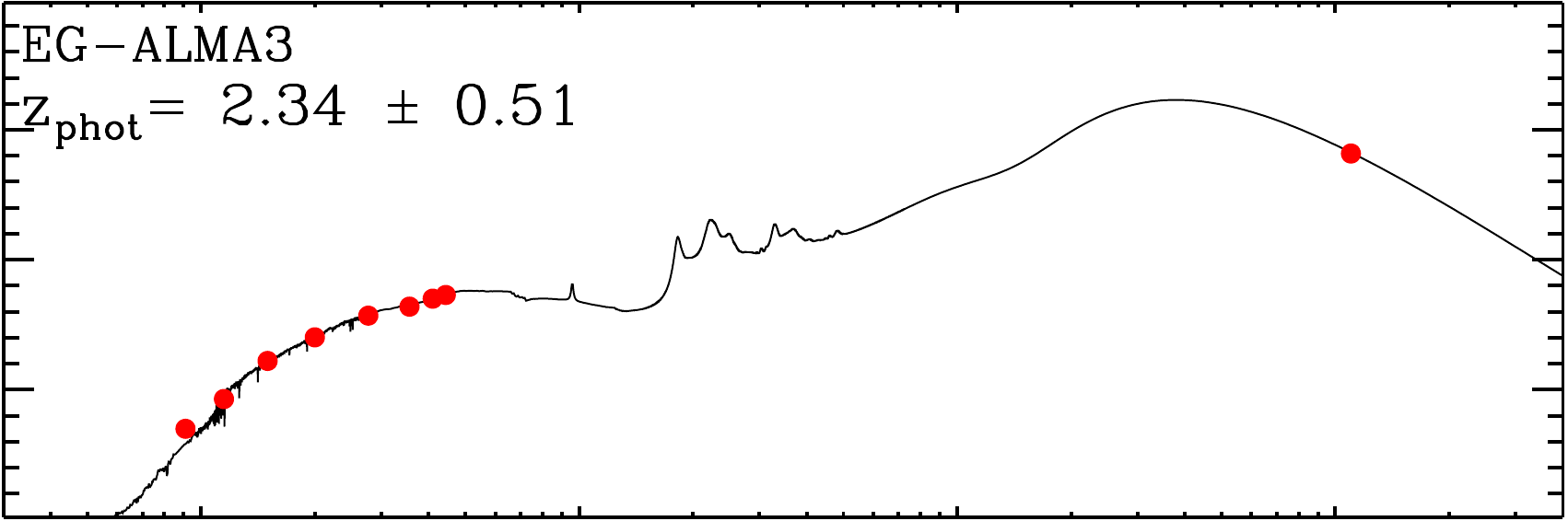}
    \includegraphics[width=0.355\textwidth]{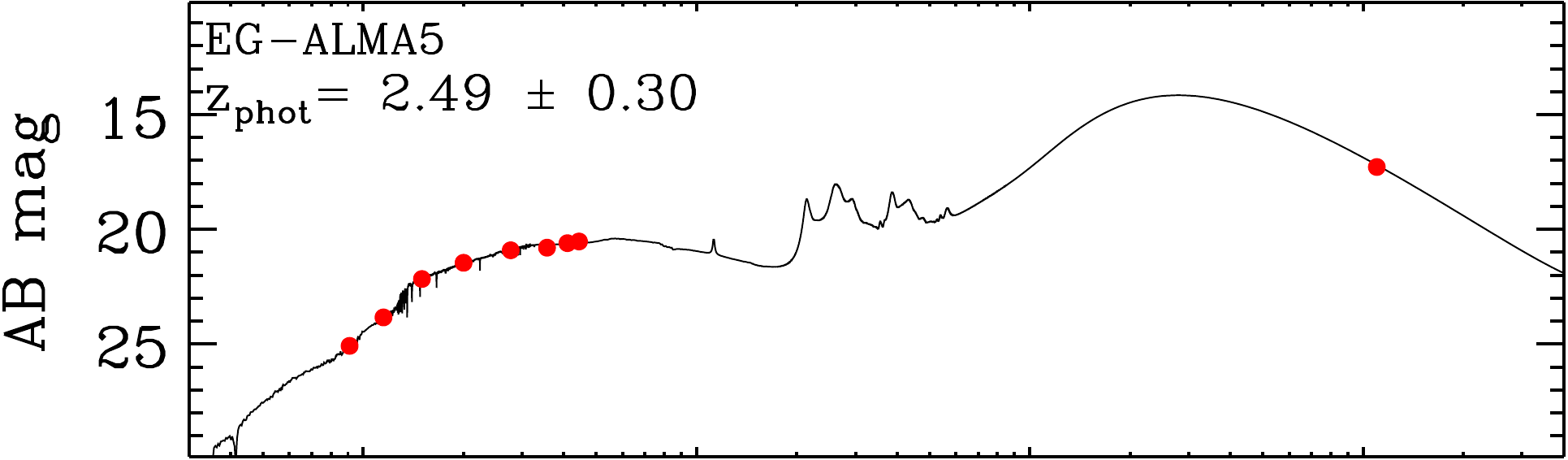}
    \includegraphics[width=0.315\textwidth]{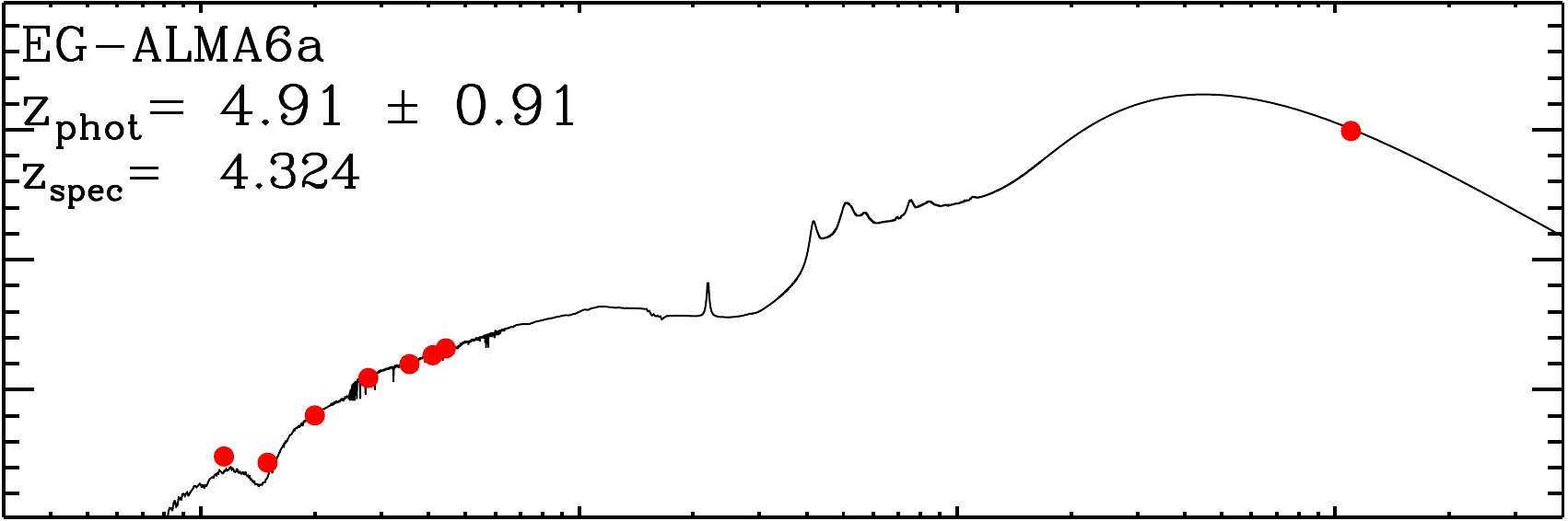}
    \includegraphics[width=0.315\textwidth]{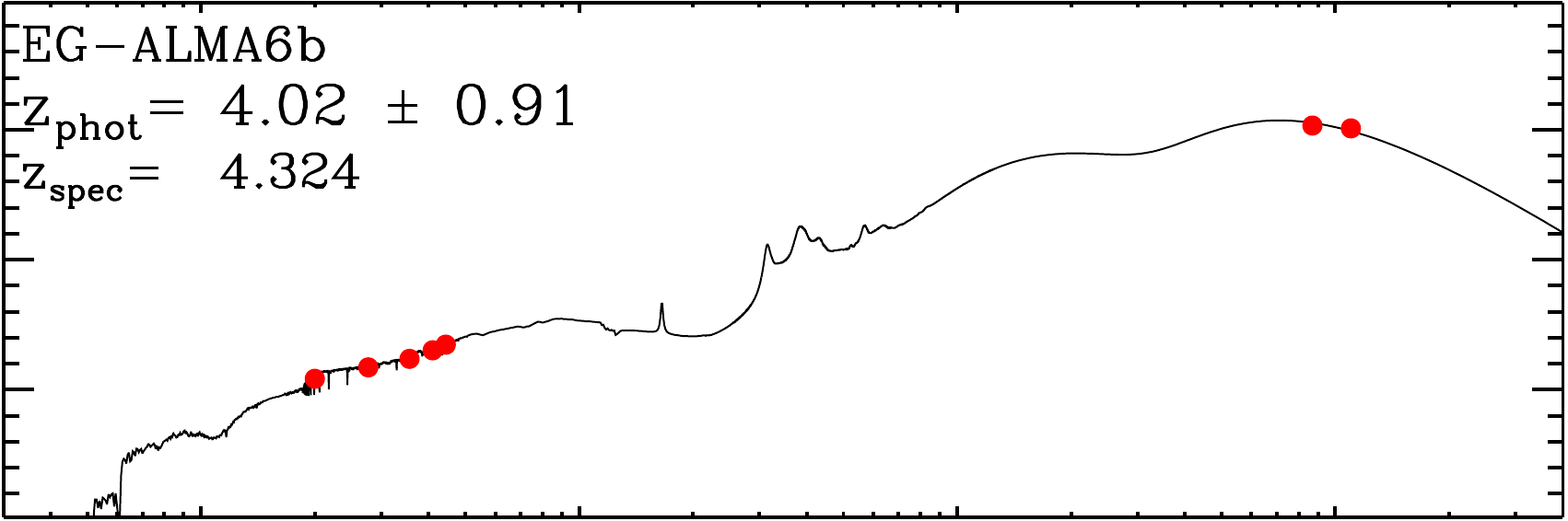}
    \includegraphics[width=0.355\textwidth]{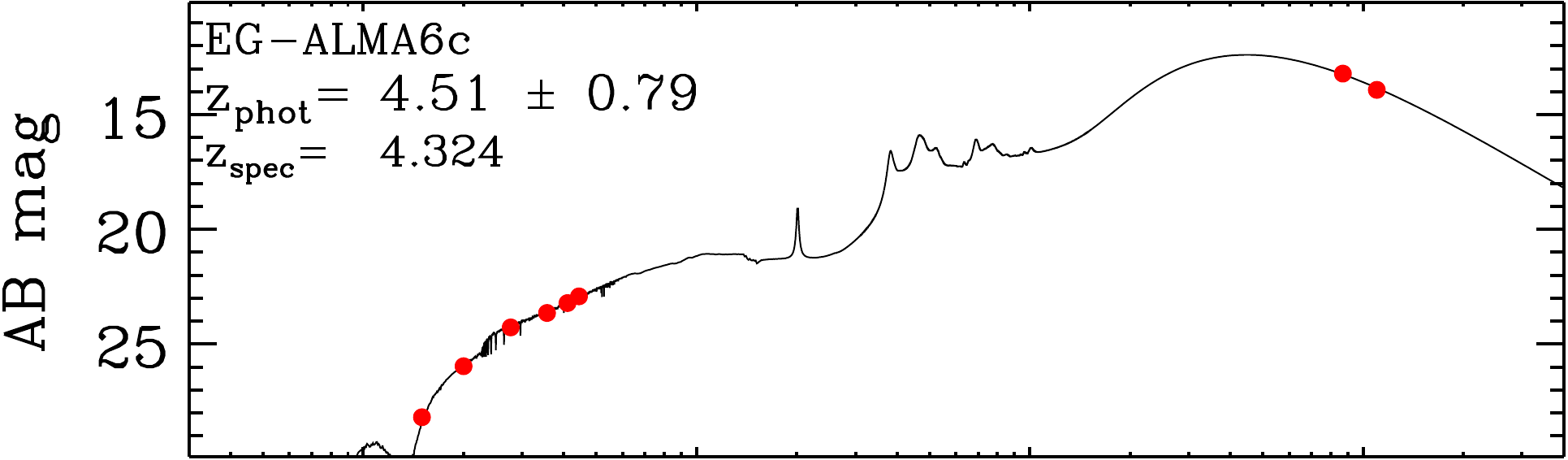}
    \includegraphics[width=0.315\textwidth]{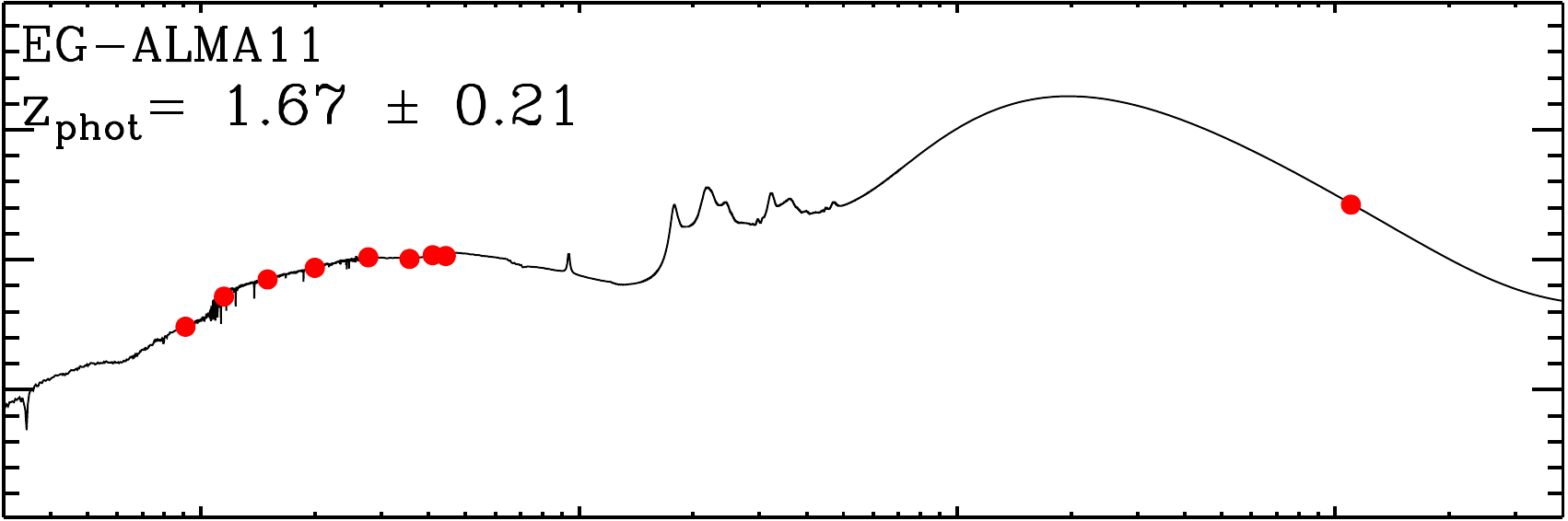}
    \includegraphics[width=0.315\textwidth]{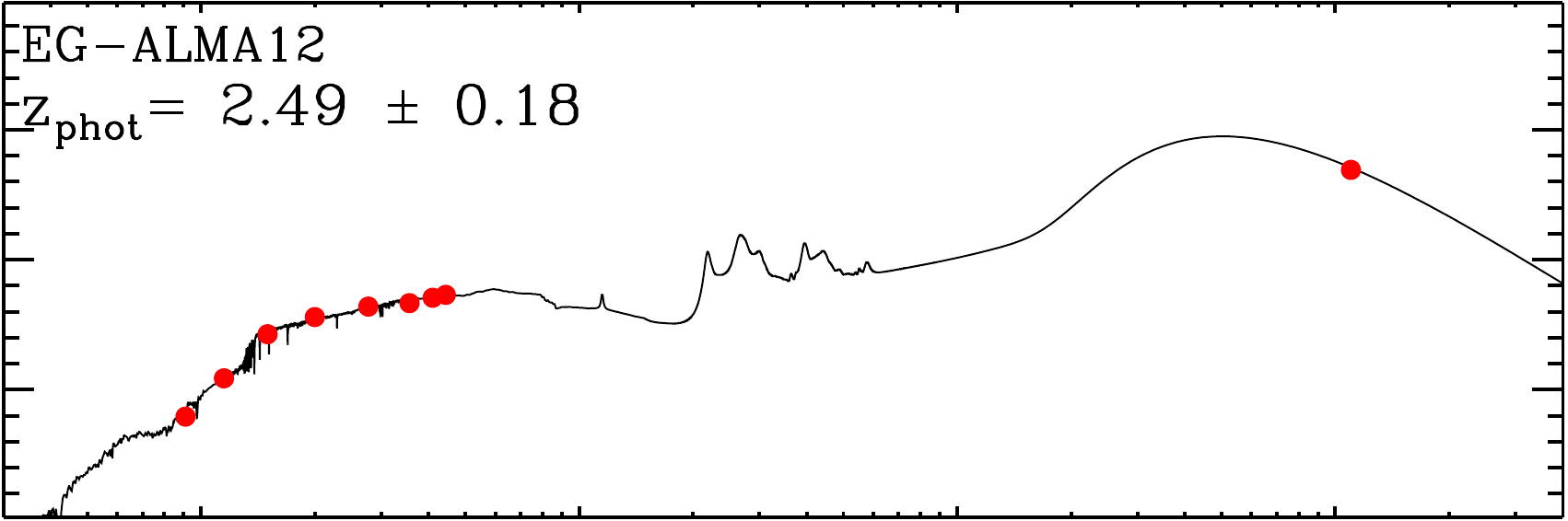}
    \includegraphics[width=0.355\textwidth]{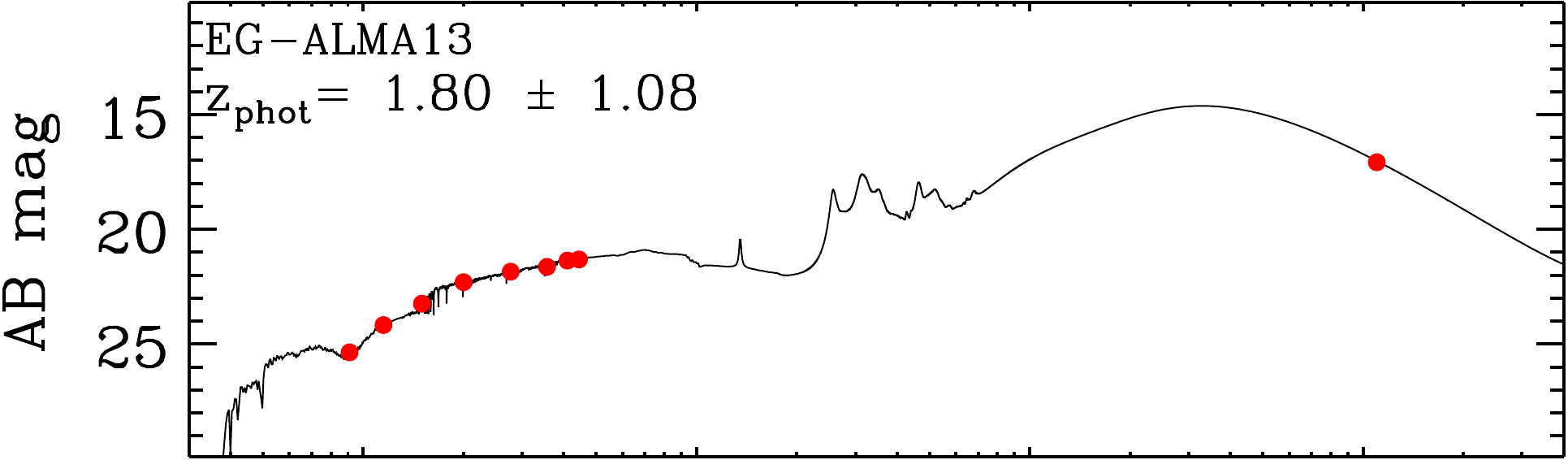}
    \includegraphics[width=0.315\textwidth]{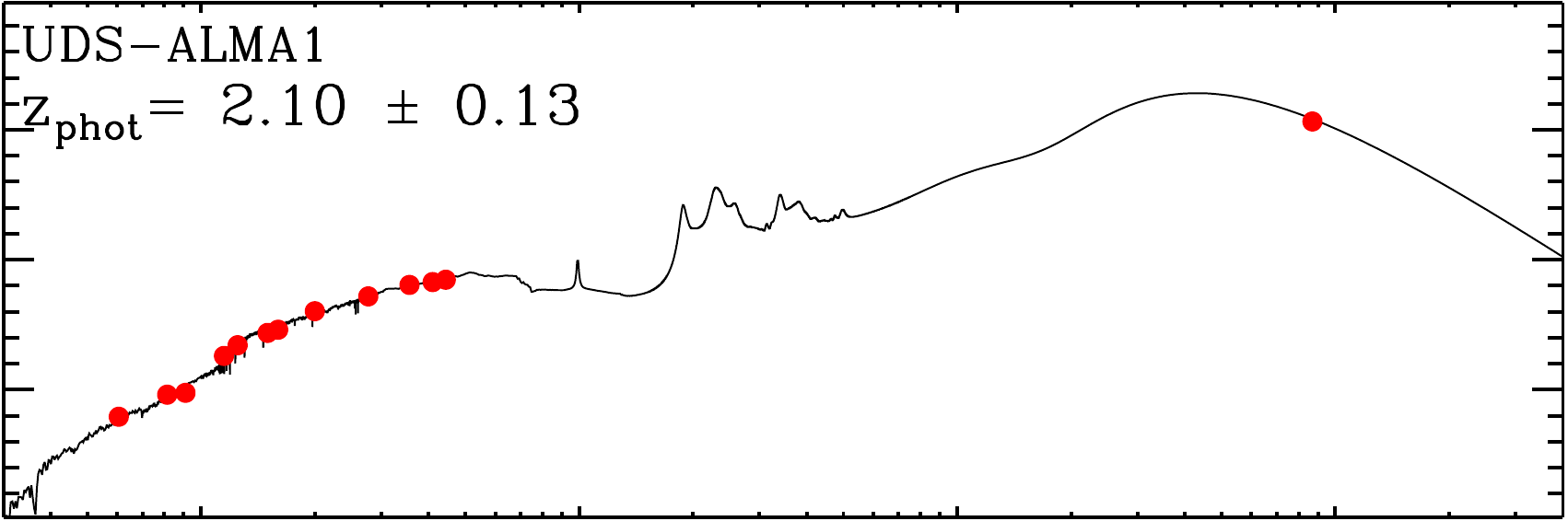}
    \includegraphics[width=0.315\textwidth]{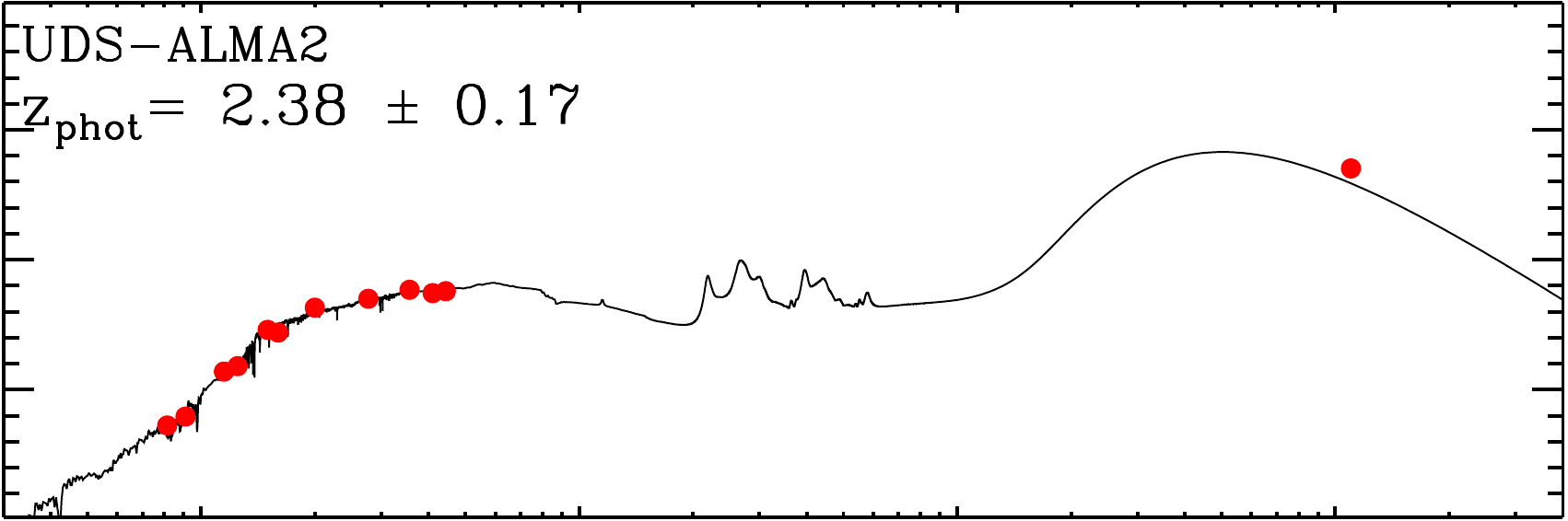}
    \includegraphics[width=0.355\textwidth]{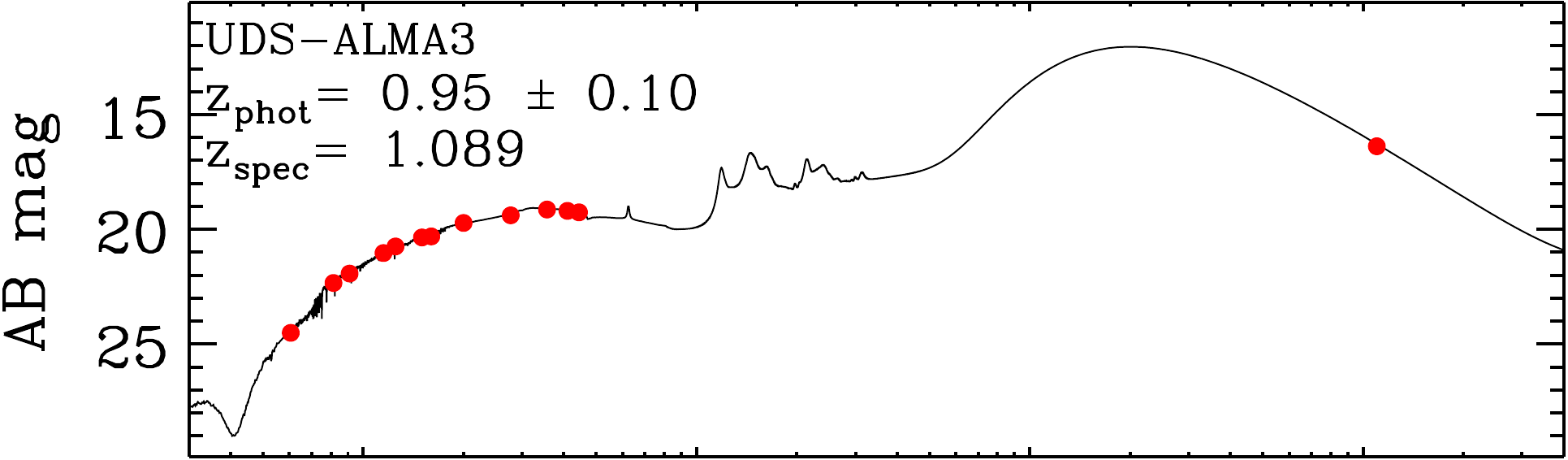}
    \includegraphics[width=0.315\textwidth]{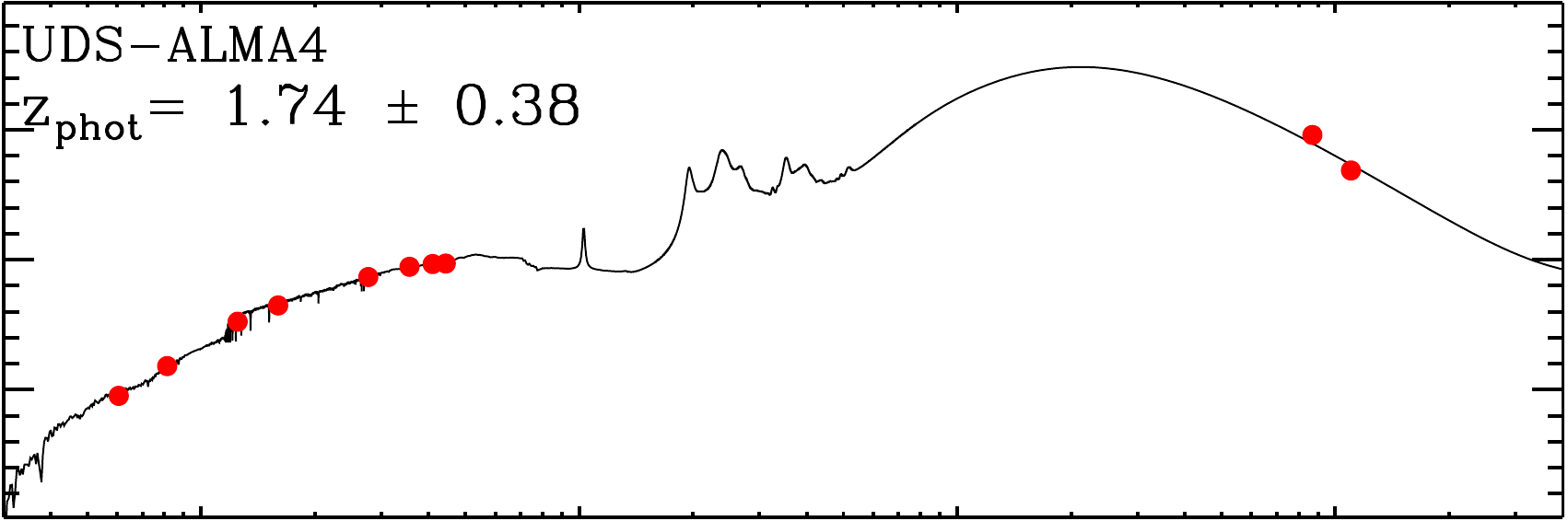}
    \includegraphics[width=0.315\textwidth]{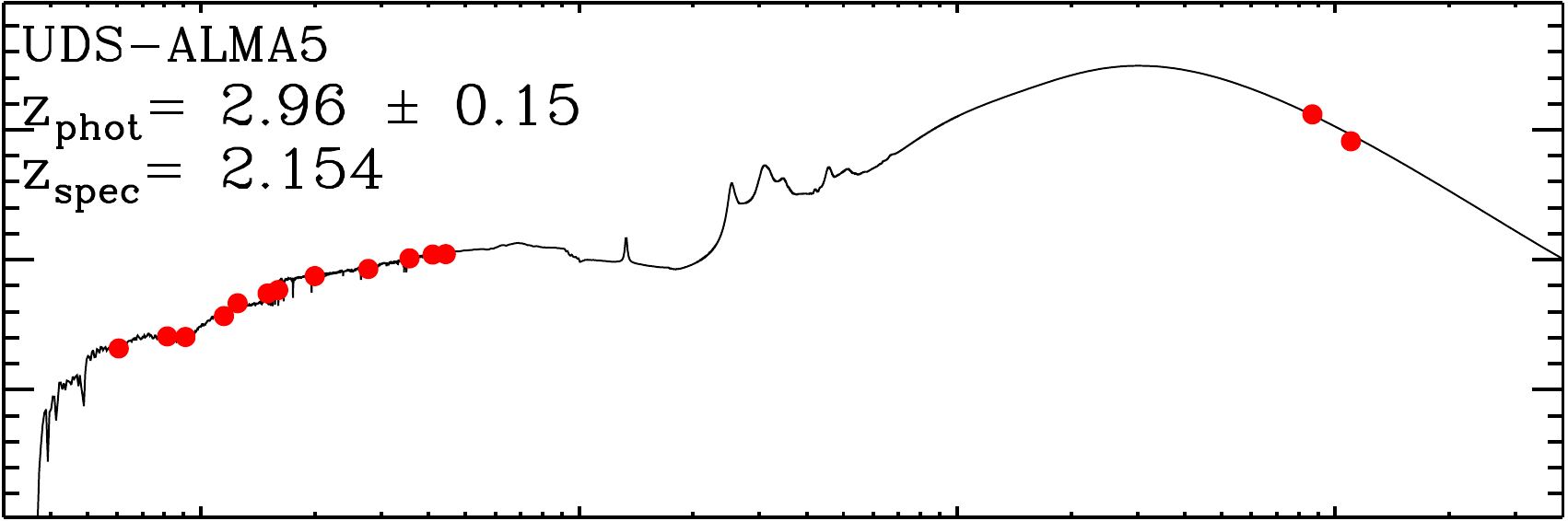}
    \includegraphics[width=0.355\textwidth]{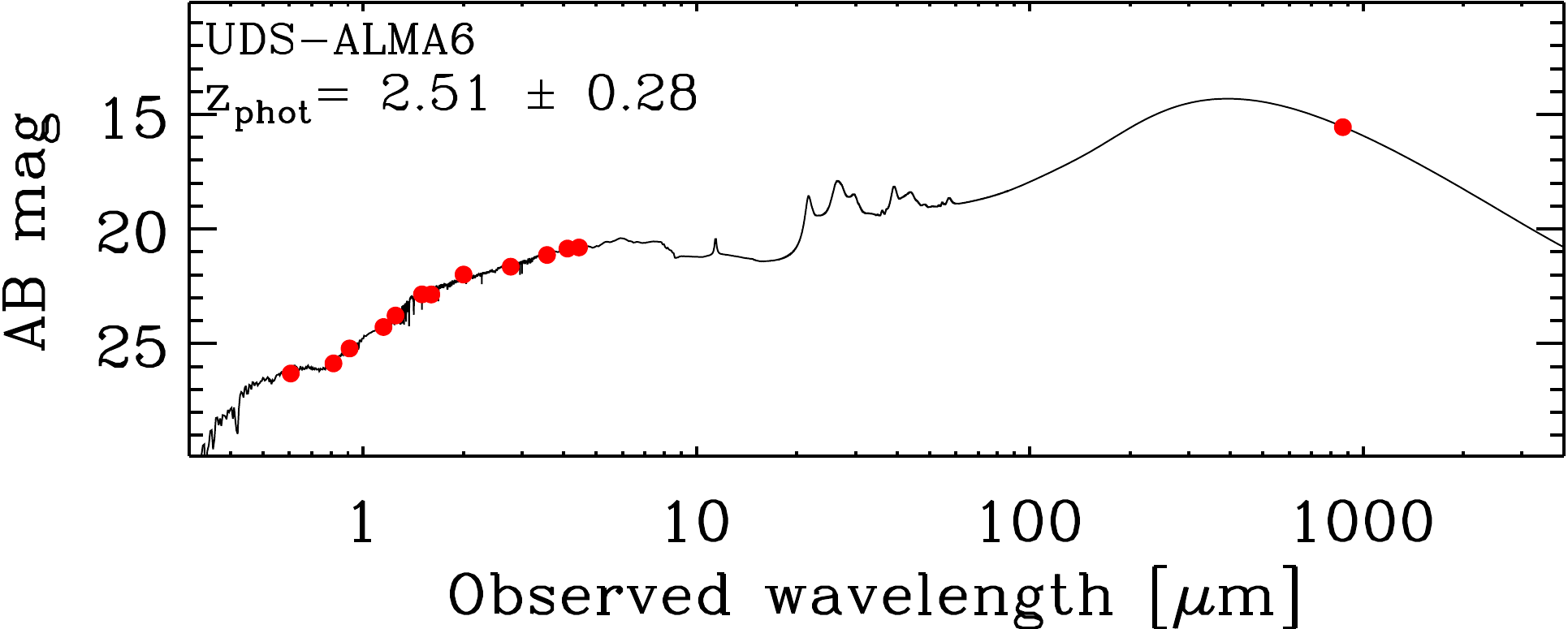}
    \includegraphics[width=0.315\textwidth]{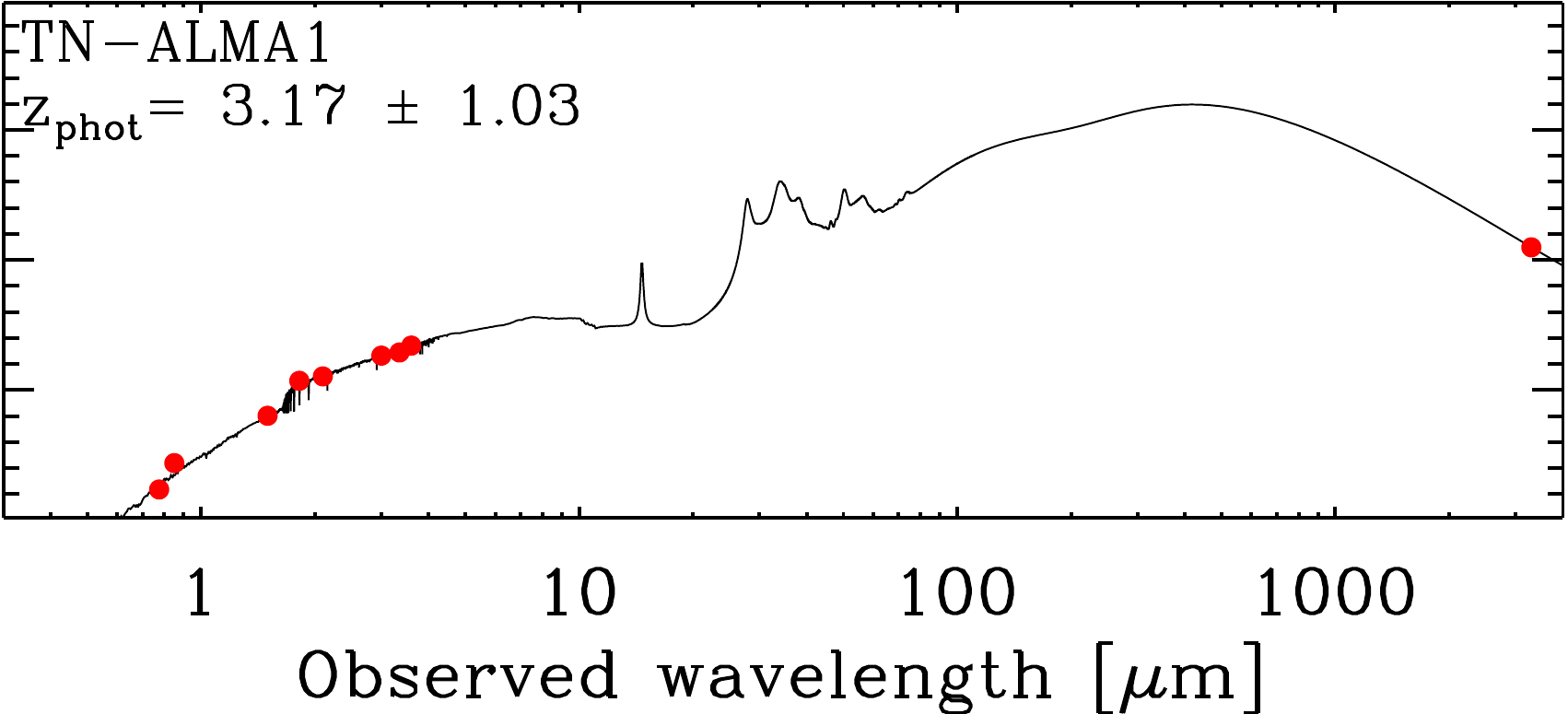}
    \includegraphics[width=0.315\textwidth]{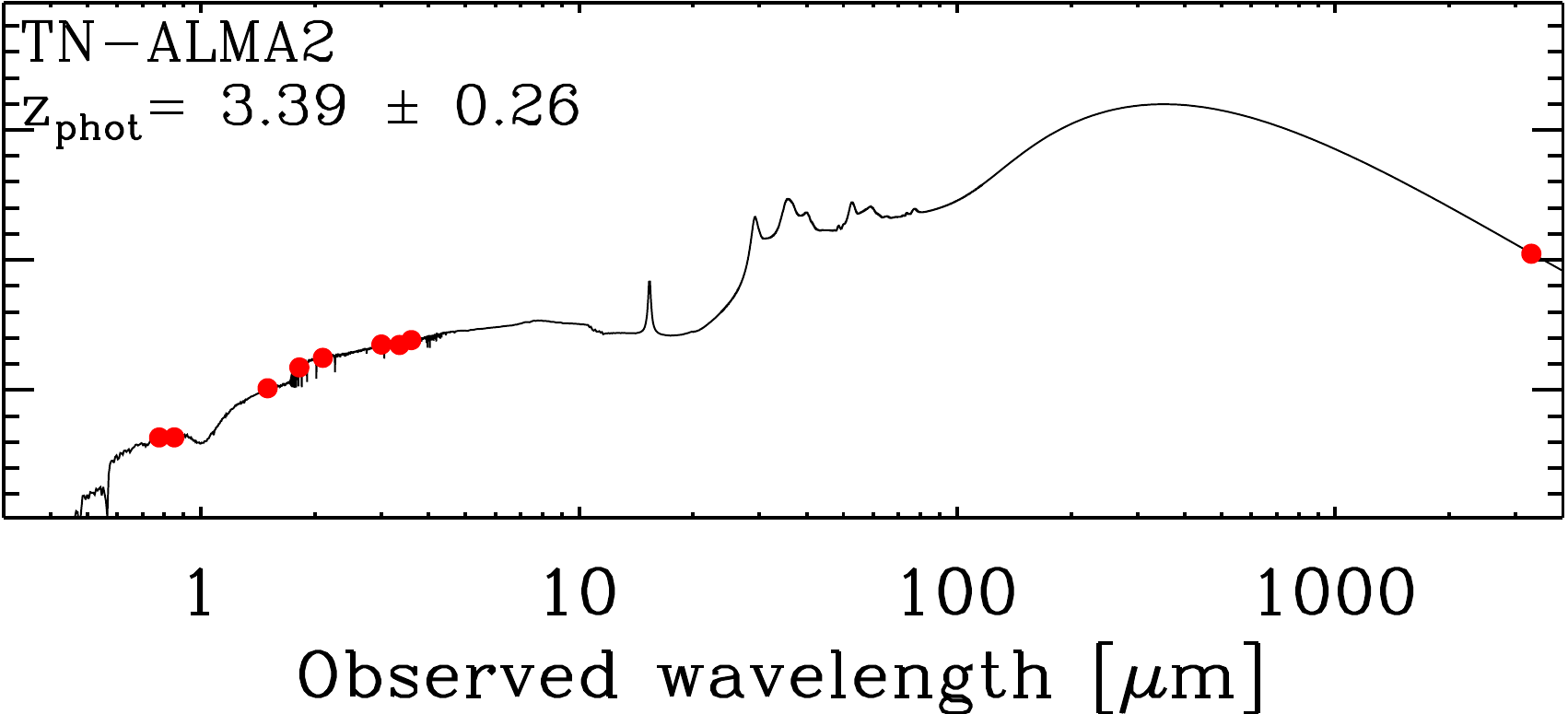}
    \includegraphics[width=0.355\textwidth]{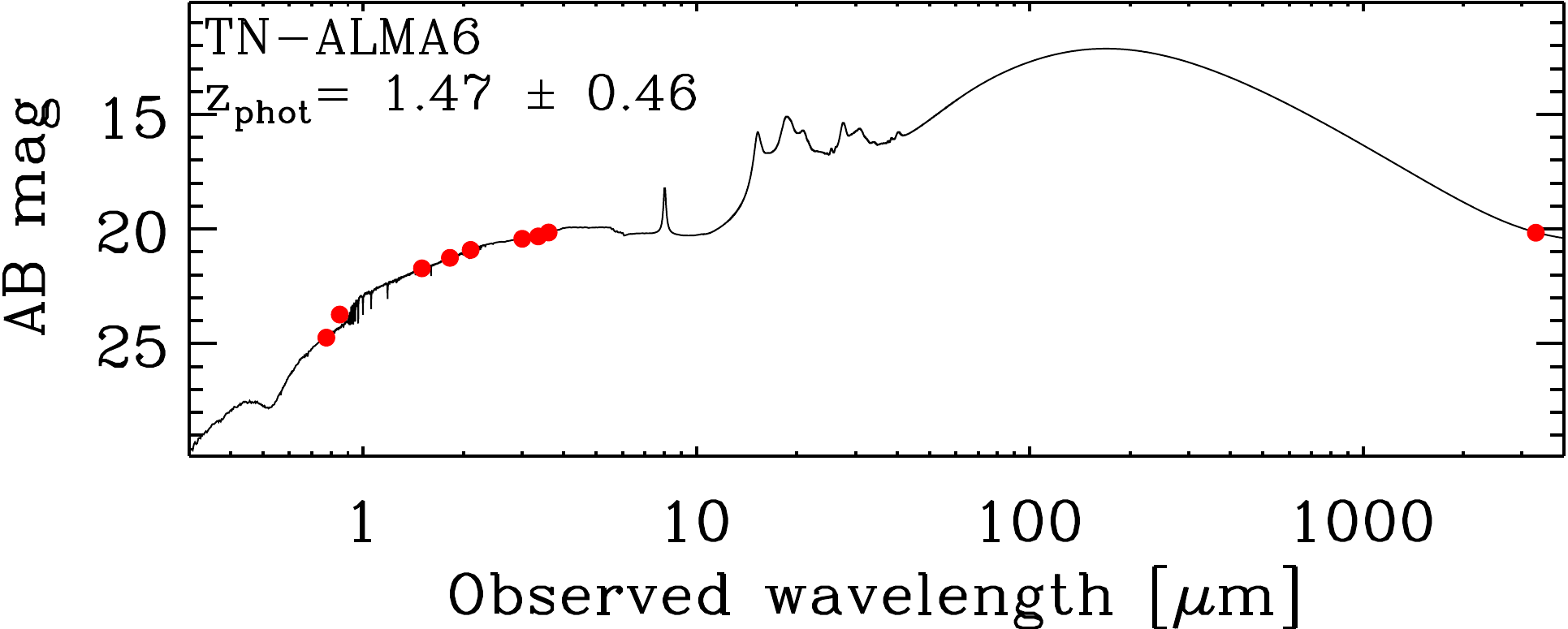}
\caption{
SED fitting results using {\sc MAGPHYS-photo-z}. The legend in each panel gives the $z_{\rm spec}$
value when available. The red dots show the observed data, and the thick black  
lines show the best-fit templates. The multiply imaged systems EG-ALMA 2a/2b 
and EG-ALMA 6a/6b/6c suffer from unreliable photometry because of the lensing 
distortion, and therefore
the fits for individual images for the individual images have large error bars. UDS-ALMA~2 is a 
point source and very likely a quasar, and therefore the stellar templates used by
{\sc MAGPHYS-photo-z} might not be applicable. It is still shown here for completeness.
We note that the photometric redshifts and uncertainties agree well with the spectroscopic redshifts when these are available.
}
\label{fig:magphysfits}
\end{figure}

\begin{figure}
    \centering
    \includegraphics[width = 0.25\textwidth]{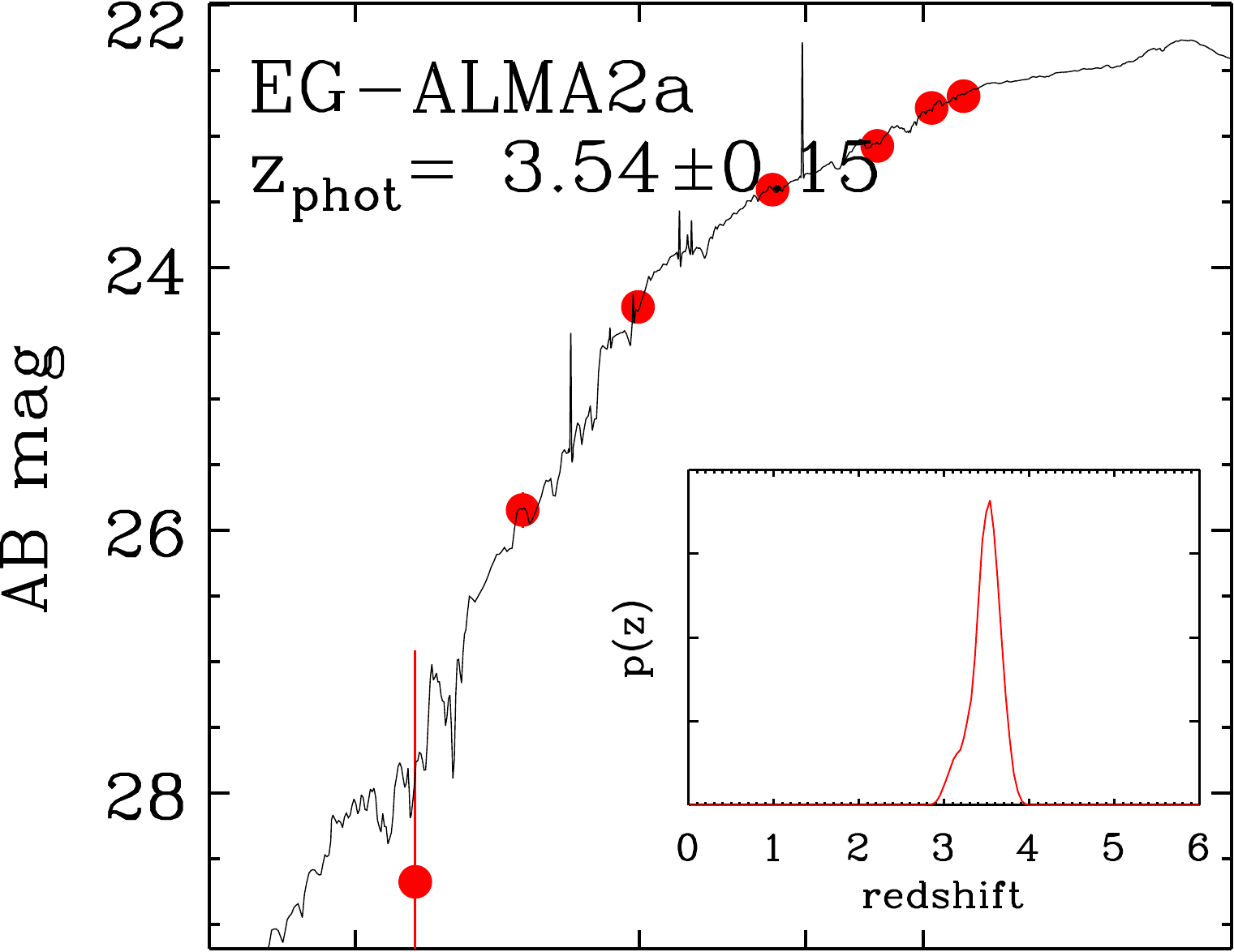}
    \includegraphics[width = 0.2275\textwidth]{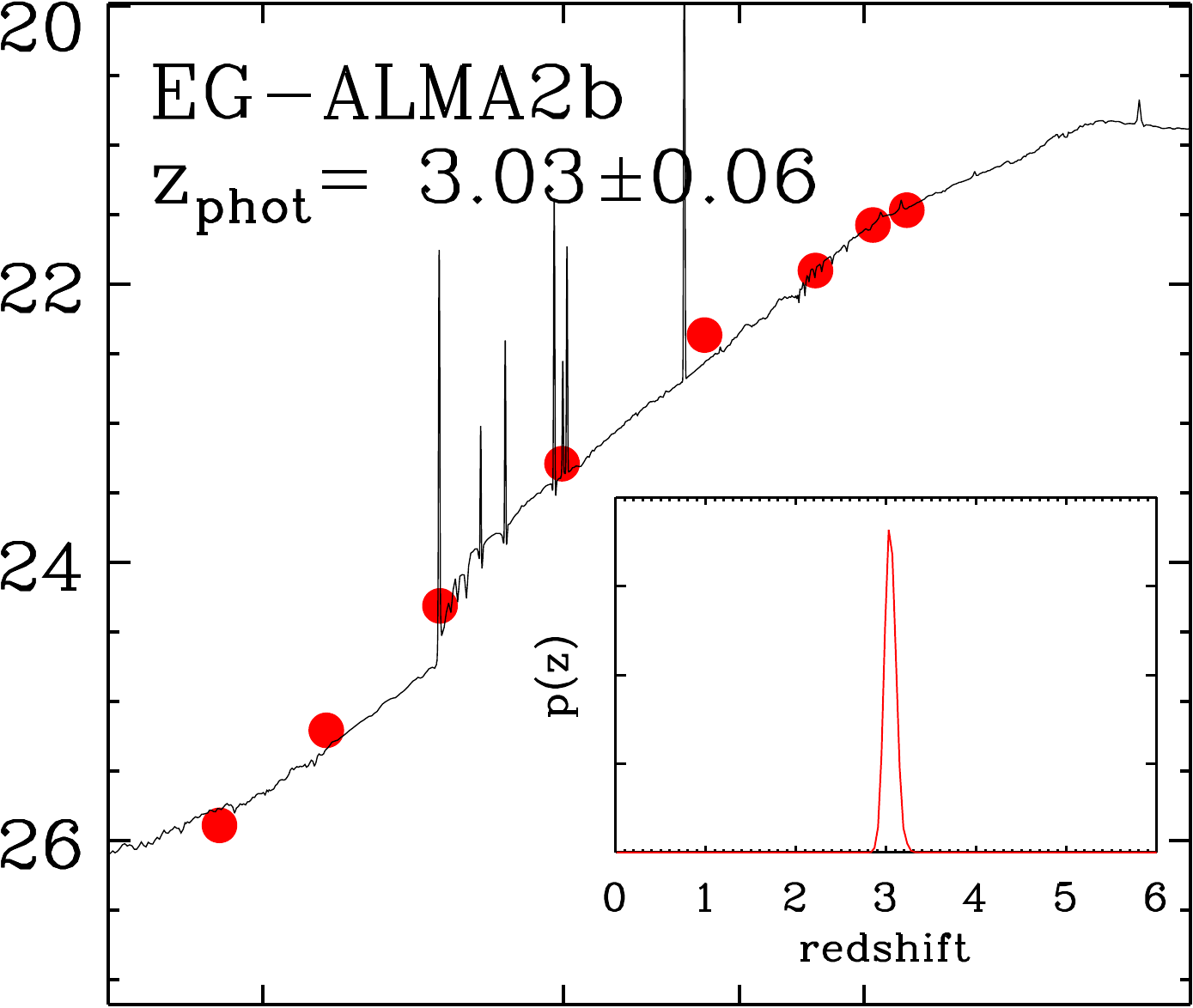}
    \includegraphics[width = 0.2275\textwidth]{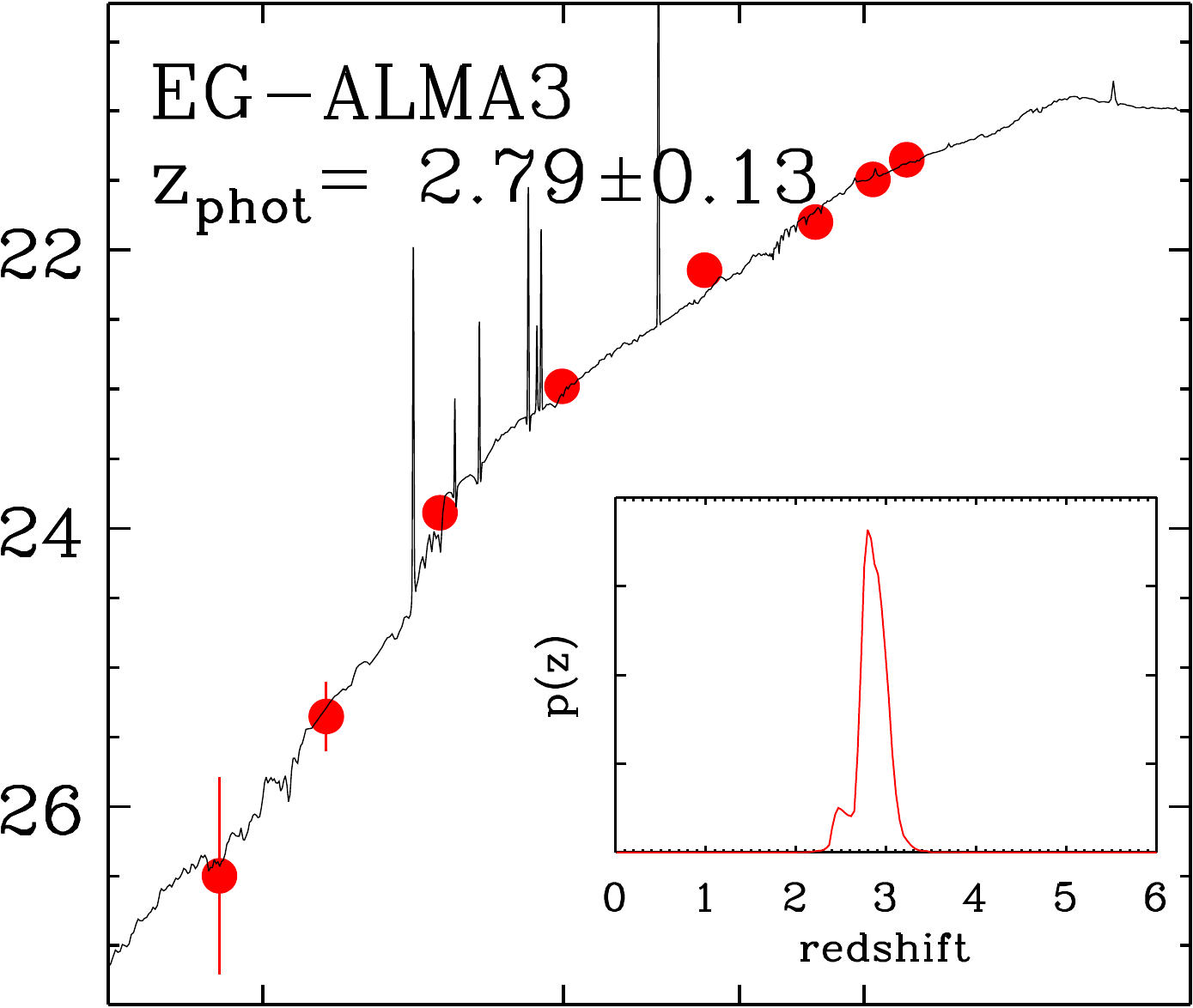}
    \includegraphics[width = 0.2275\textwidth]{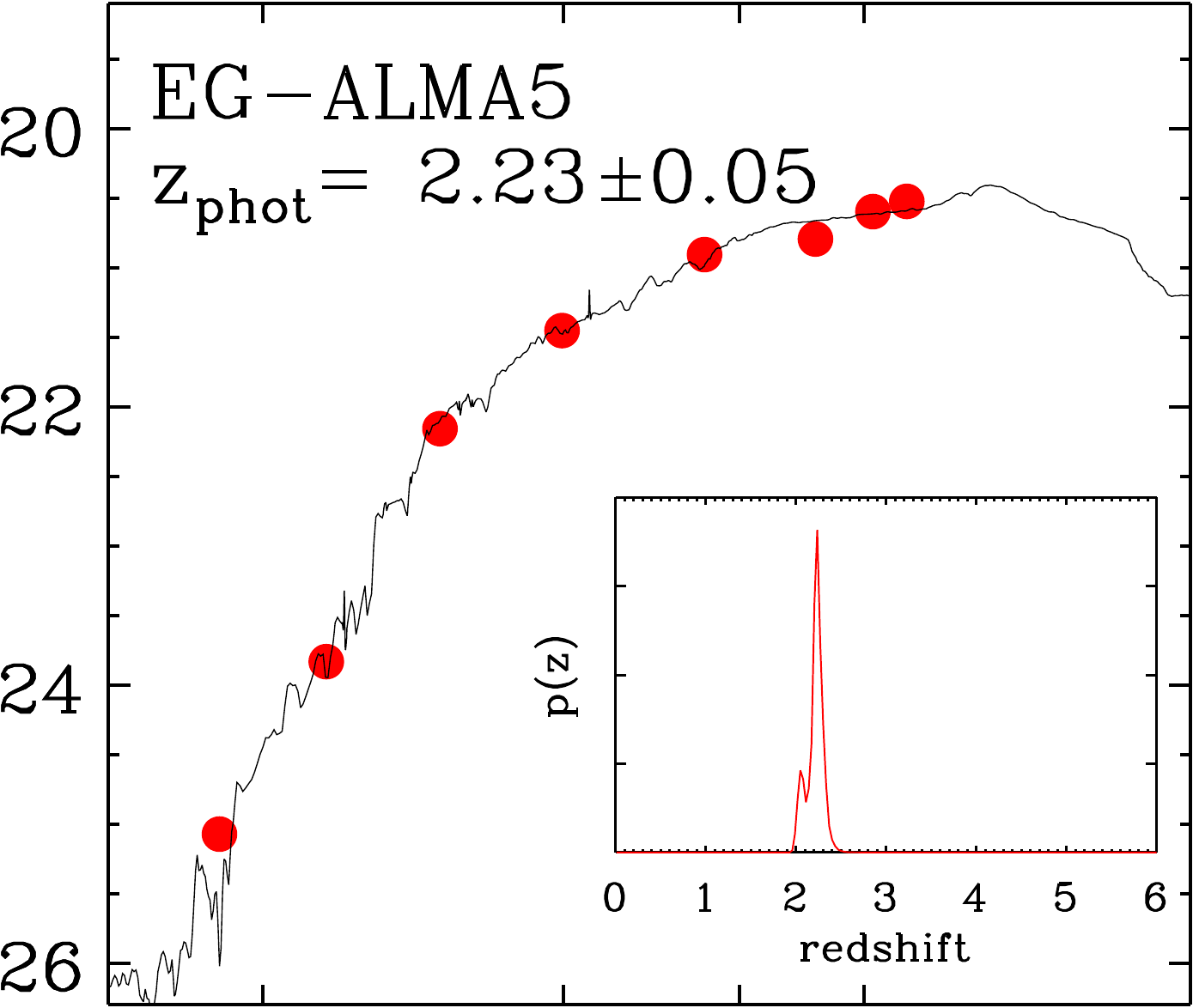}
    \includegraphics[width = 0.25\textwidth]{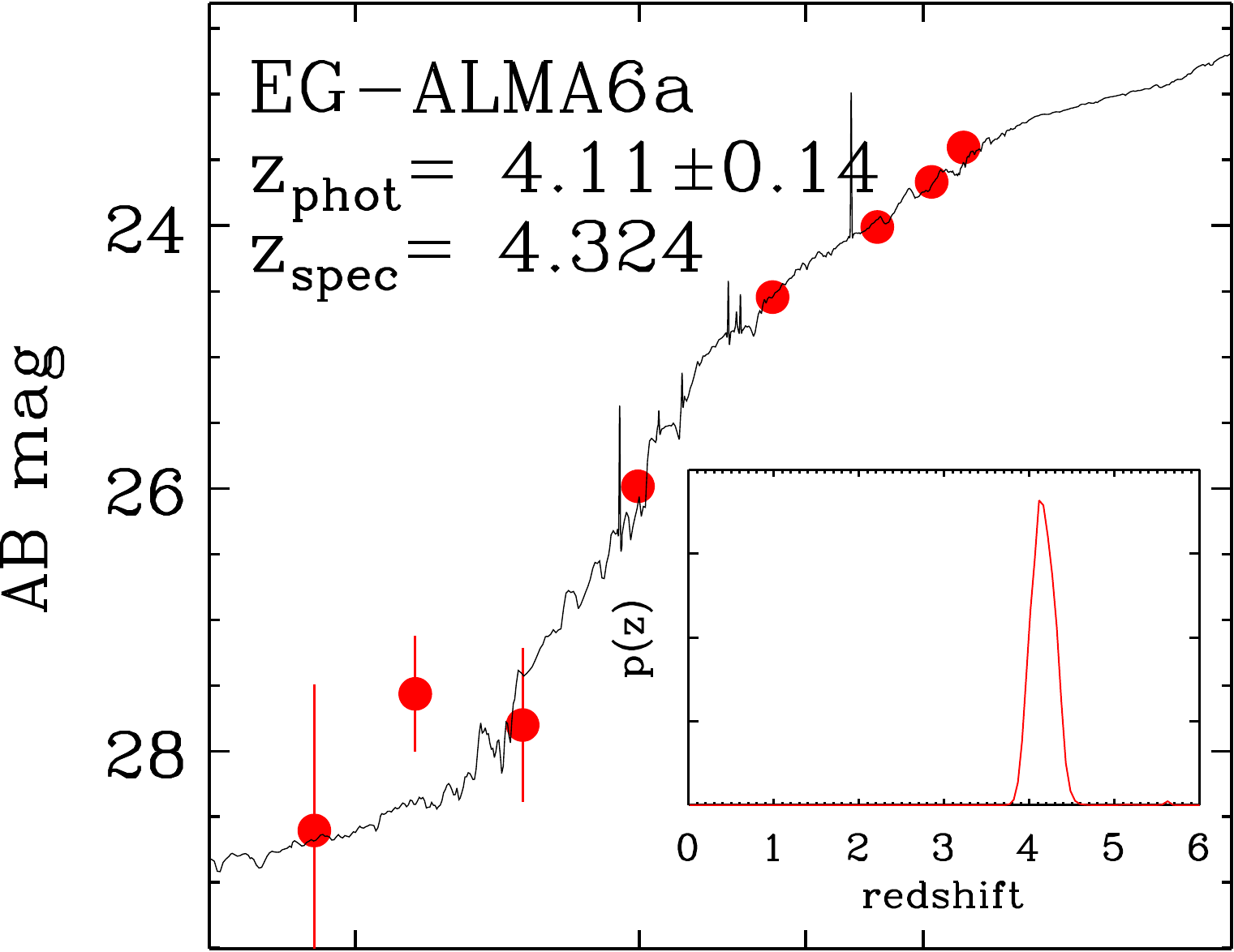}
    \includegraphics[width = 0.2275\textwidth]{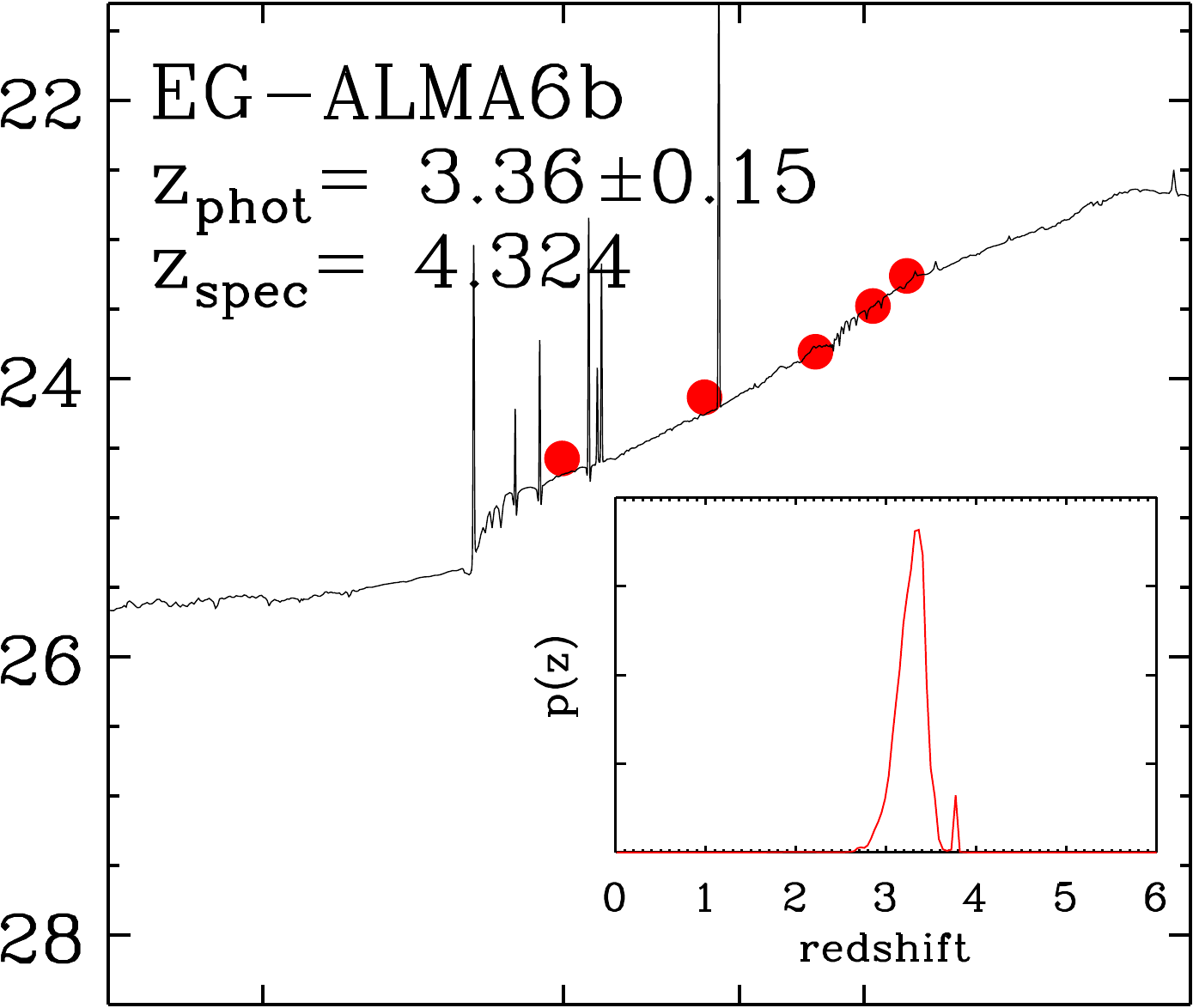}
    \includegraphics[width = 0.2275\textwidth]{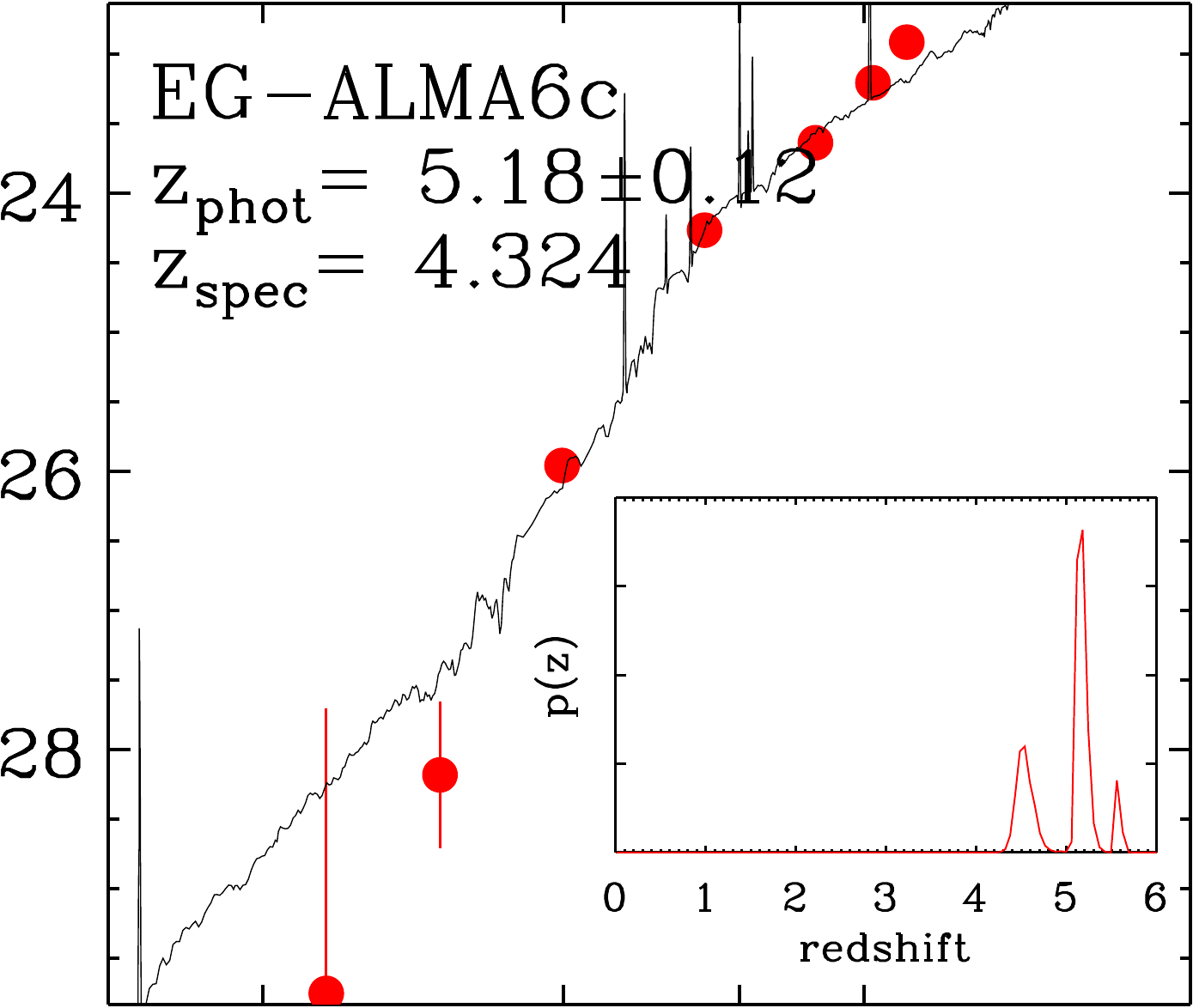}
    \includegraphics[width = 0.2275\textwidth]{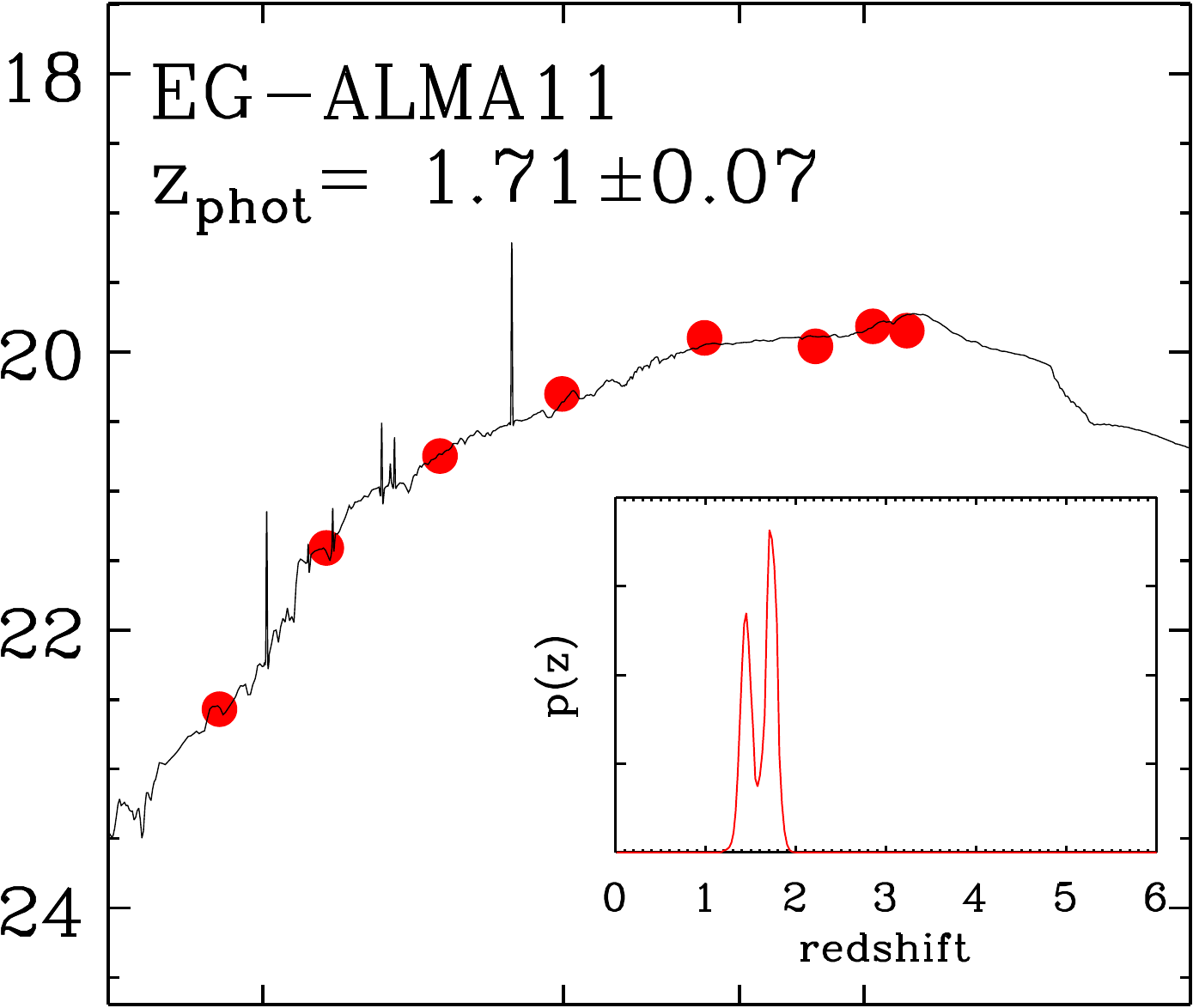}
    \includegraphics[width = 0.25\textwidth]{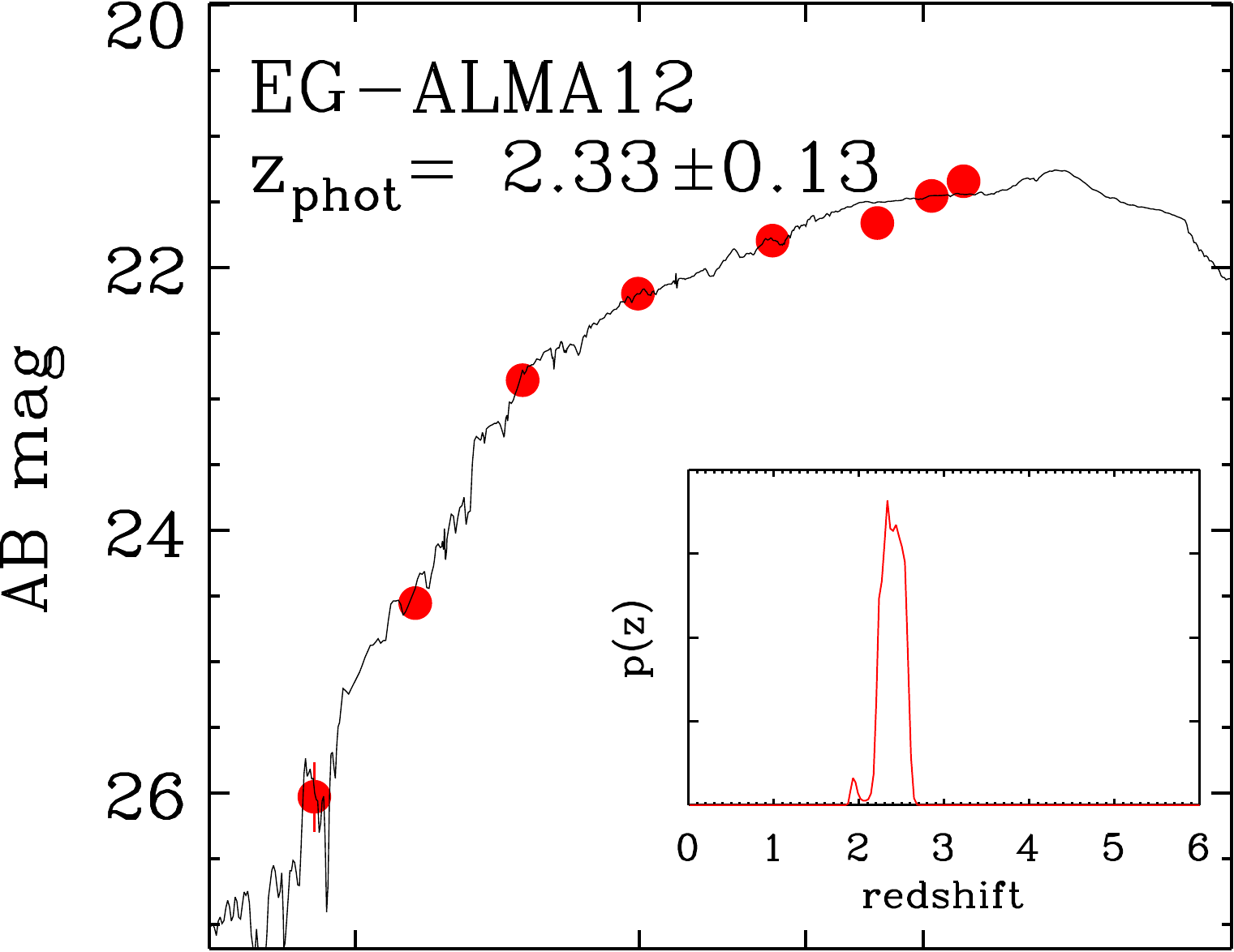}
    \includegraphics[width = 0.2275\textwidth]{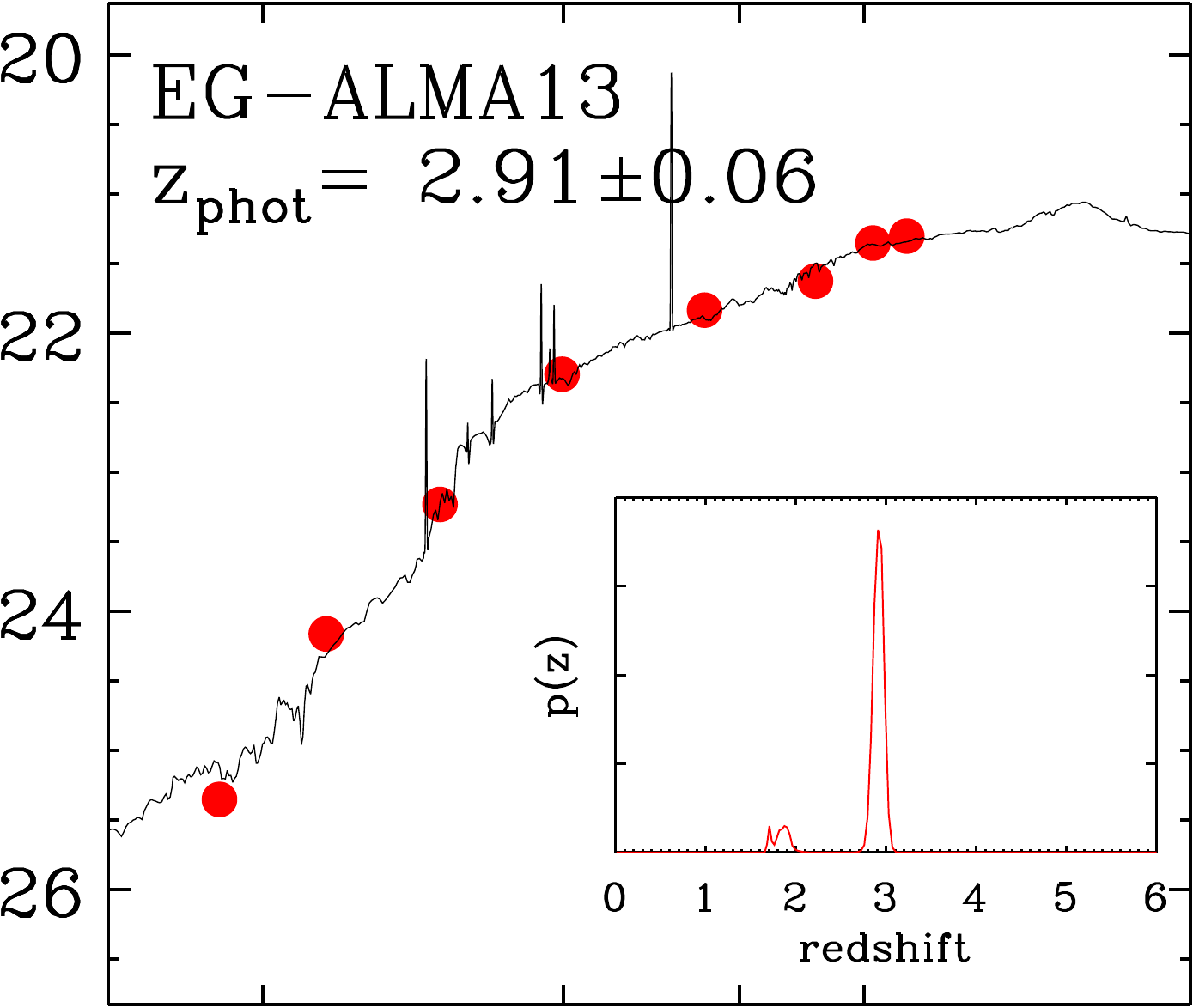}
    \includegraphics[width = 0.2275\textwidth]{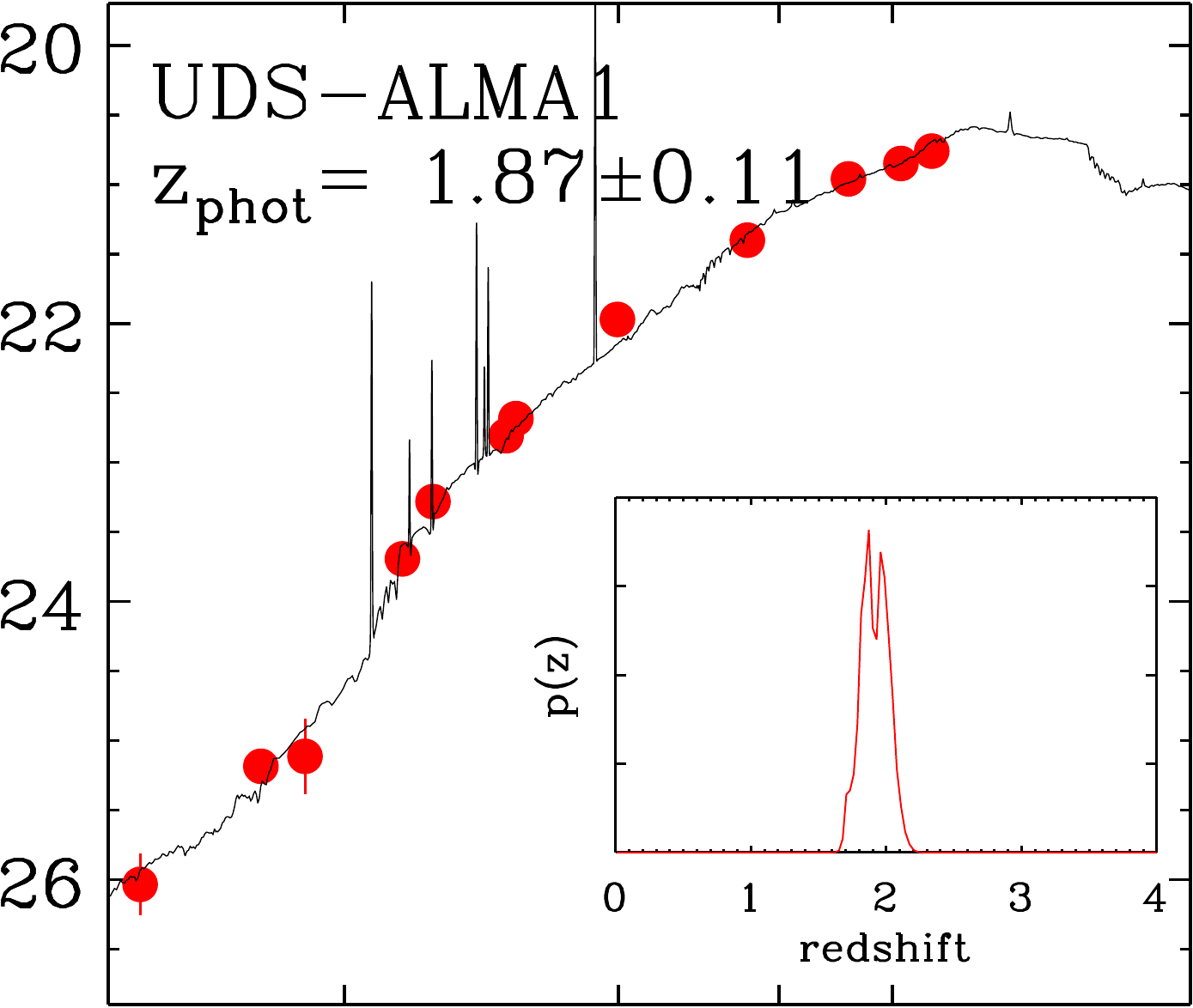}
    \includegraphics[width = 0.2275\textwidth]{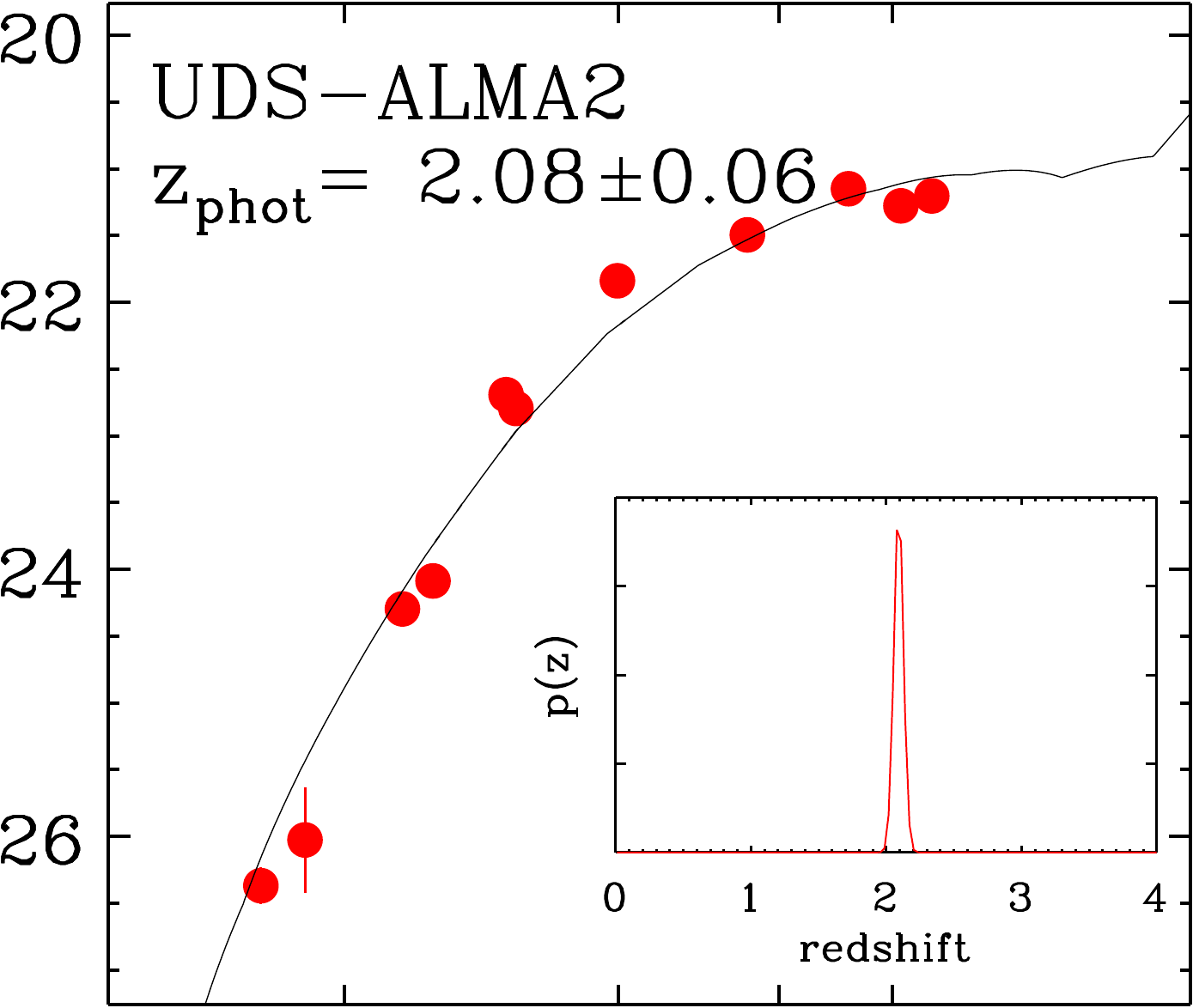}
    \includegraphics[width = 0.25\textwidth]{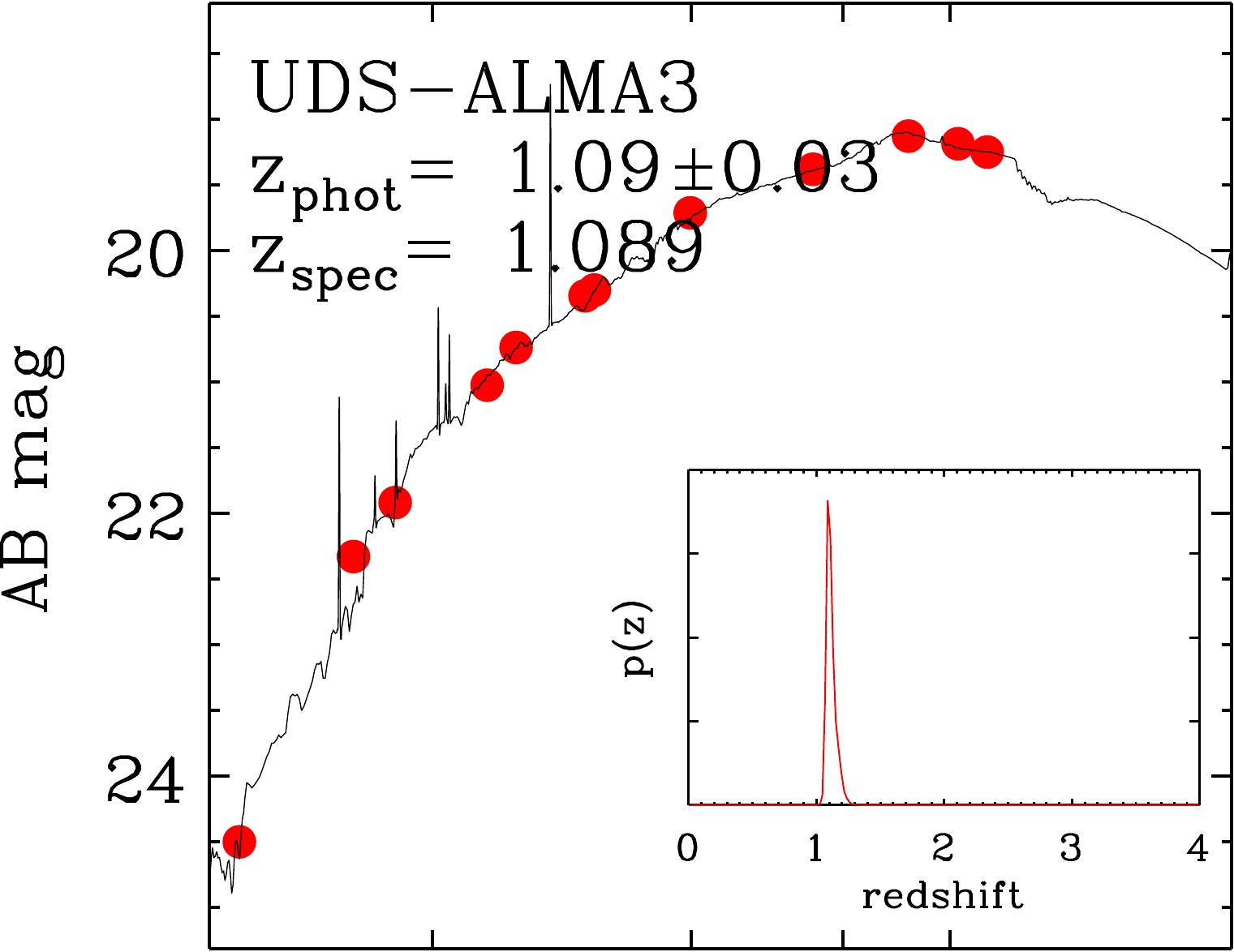}
    \includegraphics[width = 0.2275\textwidth]{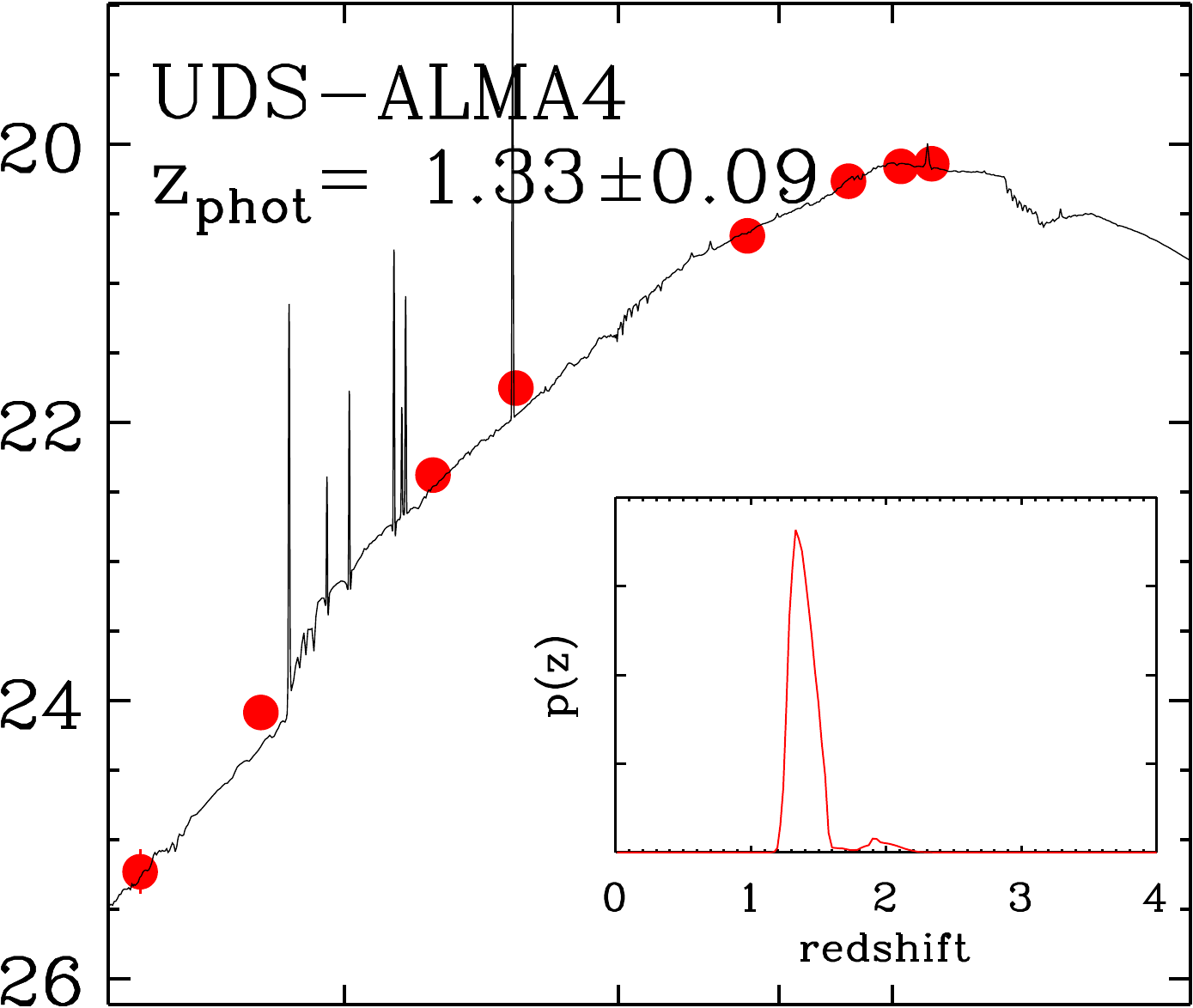}
    \includegraphics[width = 0.2275\textwidth]{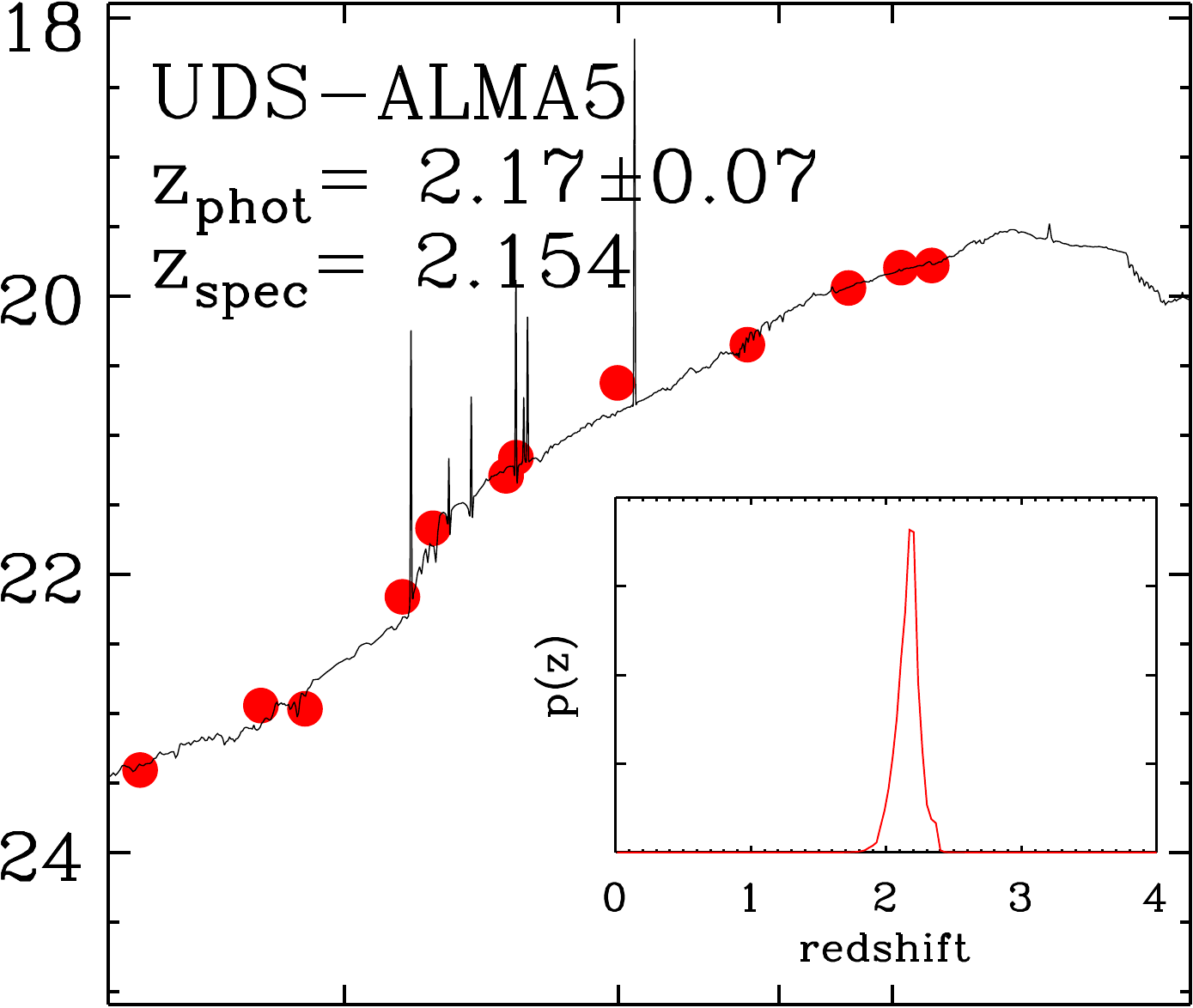}
    \includegraphics[width = 0.2275\textwidth]{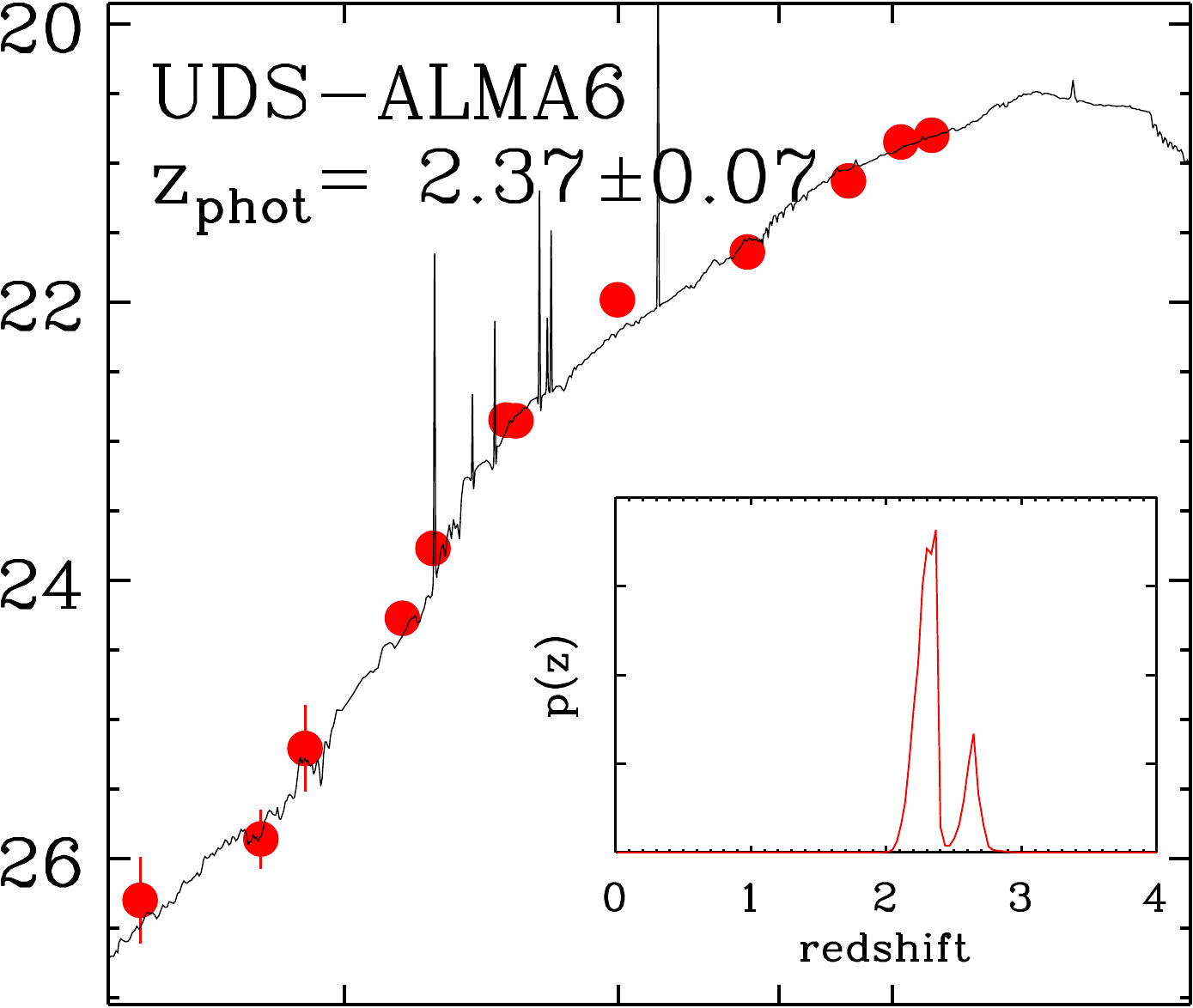}
    \includegraphics[width = 0.25\textwidth]{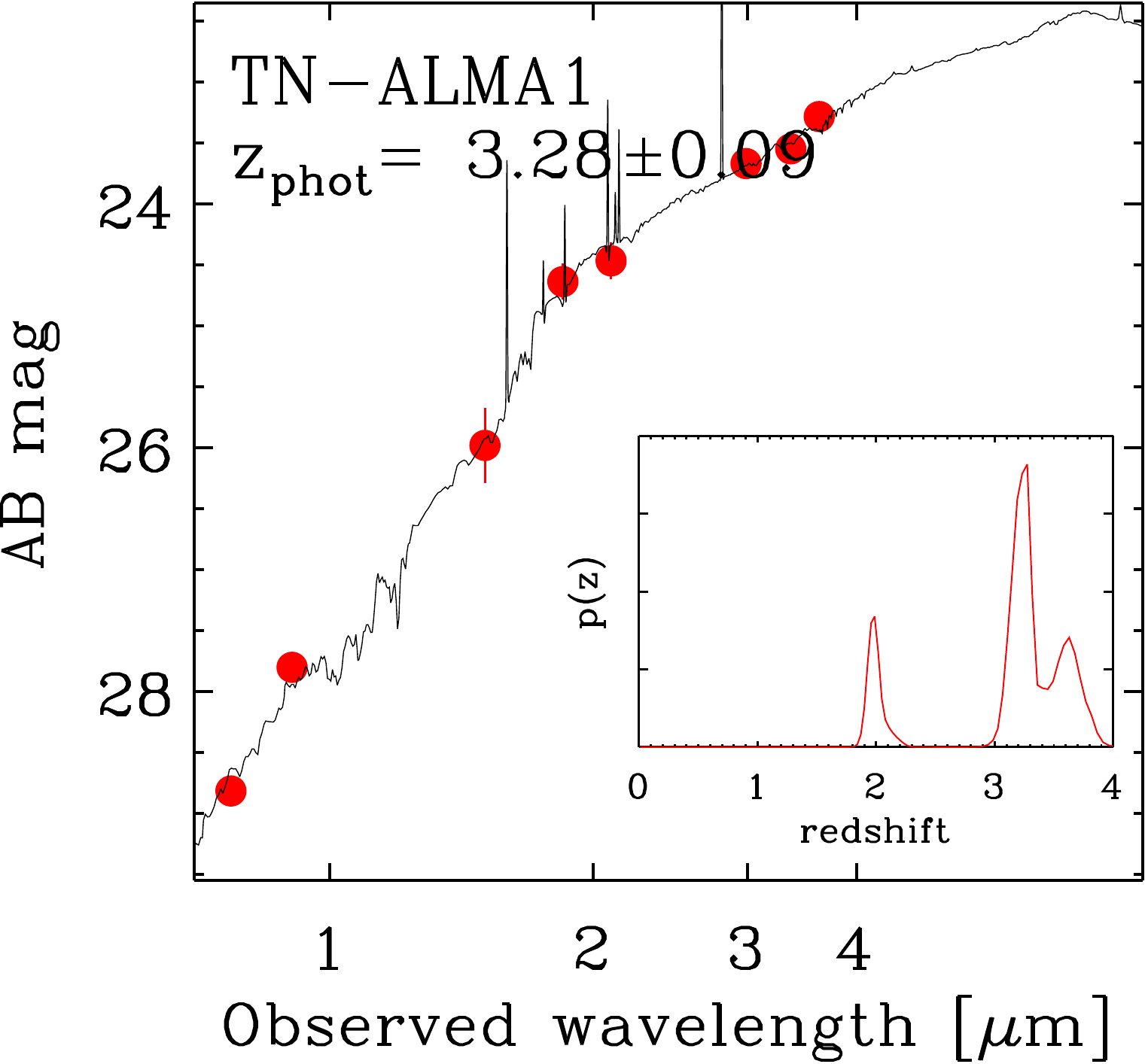}
    \includegraphics[width = 0.2275\textwidth]{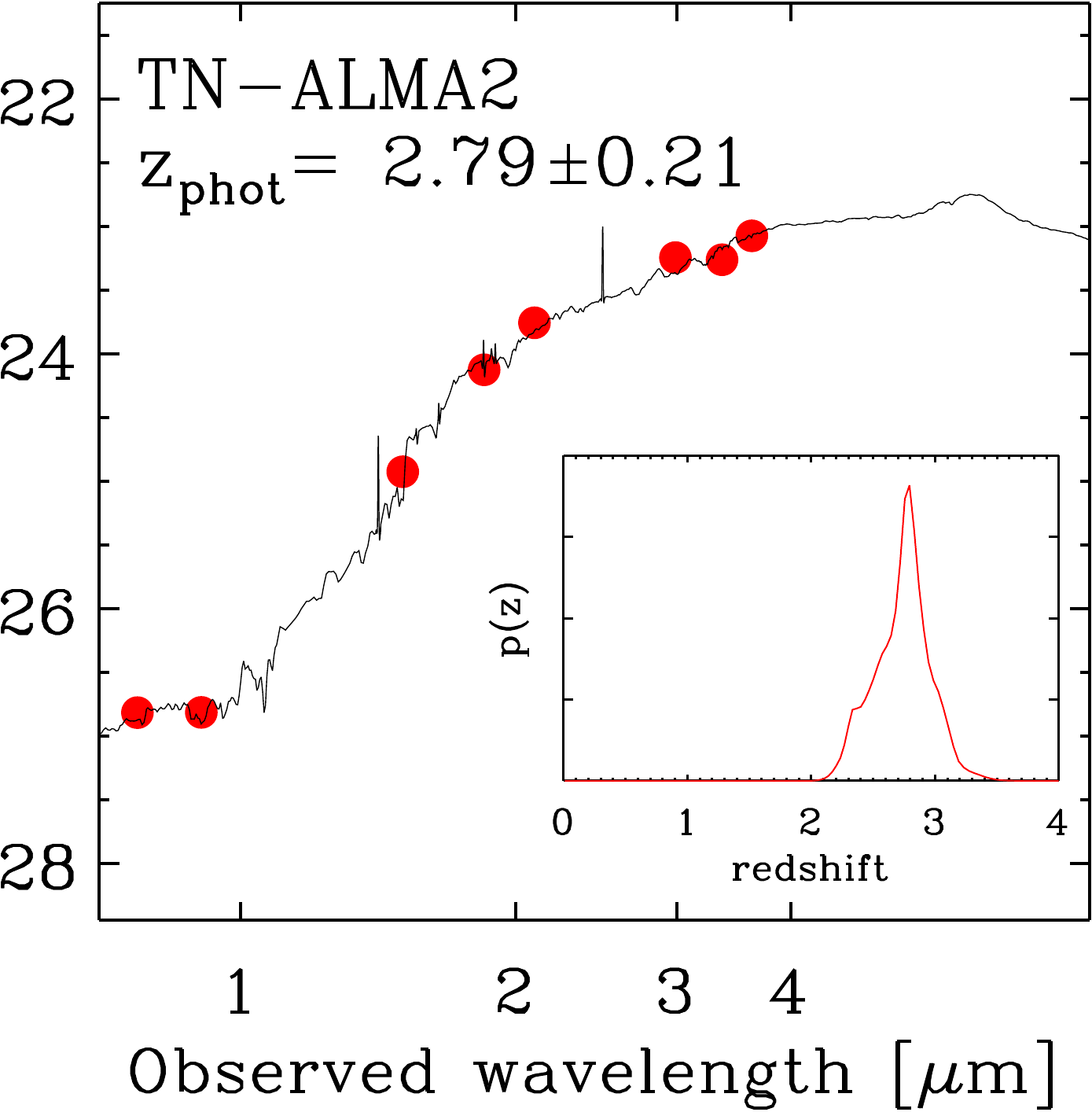}
    \includegraphics[width = 0.2275\textwidth]{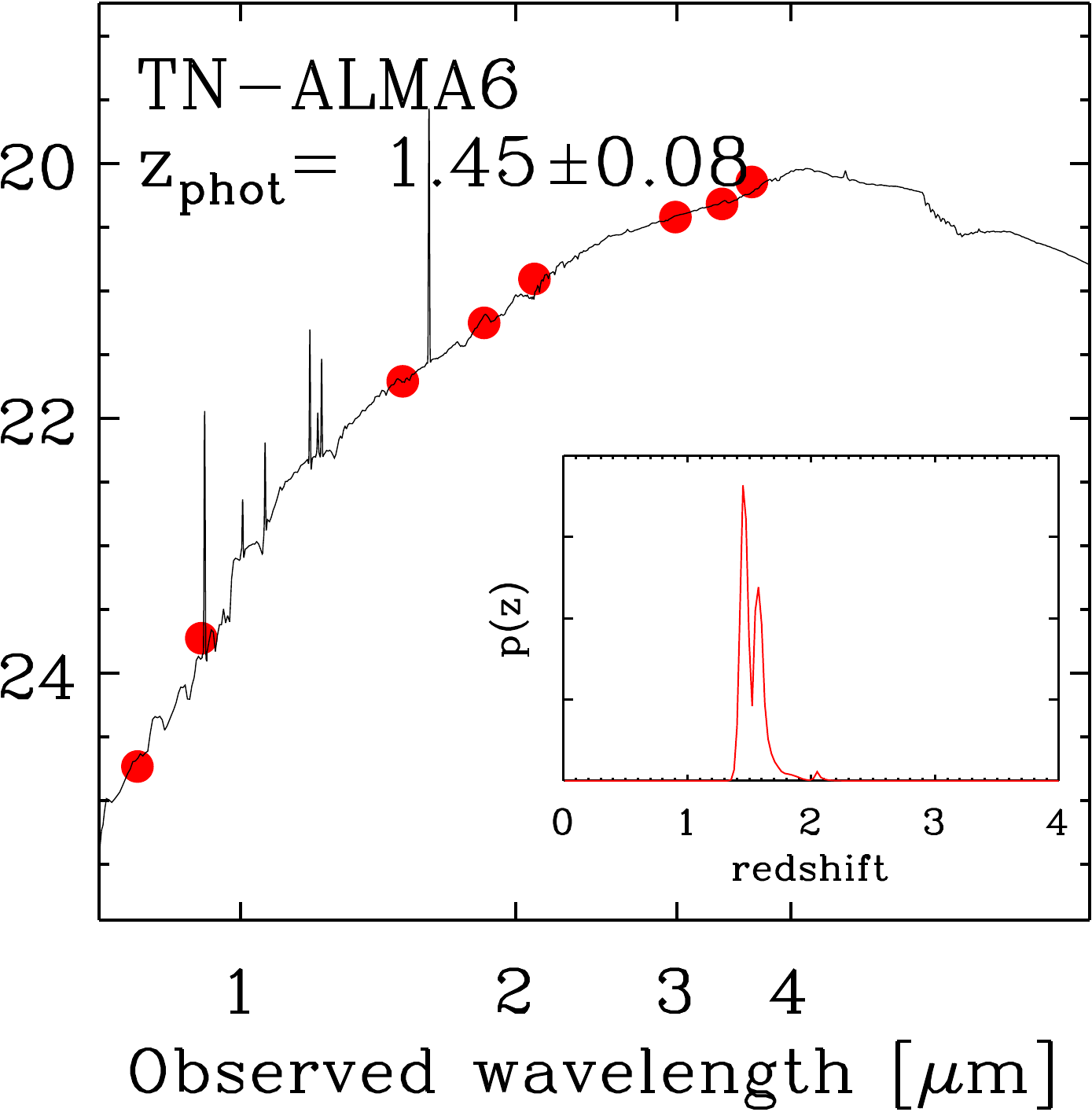}
    \includegraphics[width = 0.2275\textwidth]{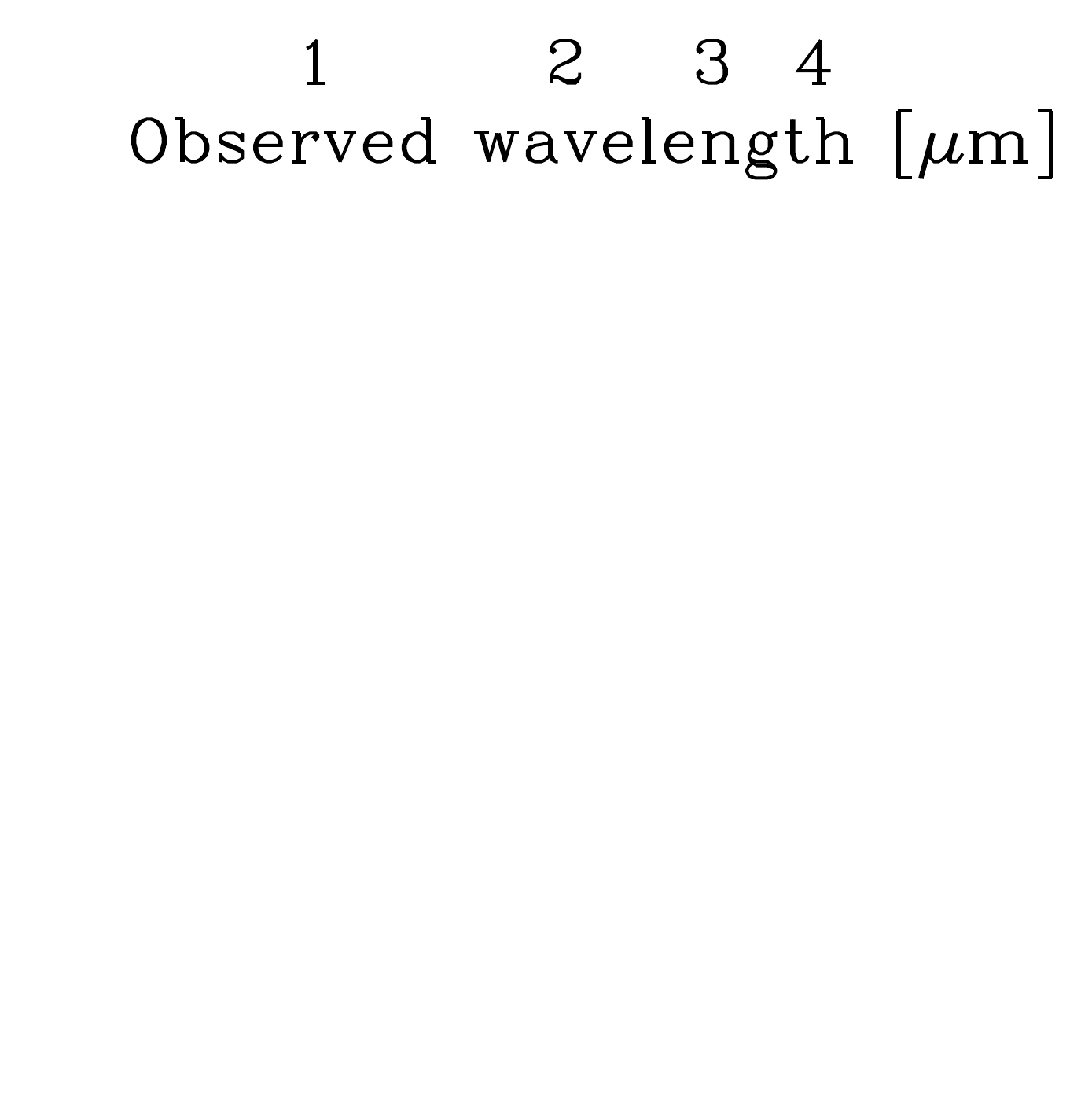}
    \caption{
    Results of {\sc eazy} fitting to derive $z_{\rm phot}$ independently. 
    Points show observed data, and solid lines show the best-fit templates.  
    Best-fit redshift is indicated in each panel. Insets show the redshift 
    probability distributions. For UDS-ALMA~3, $z_{\rm spec} = 1.089$ and for
    UDS-ALMA~5, $z_{\rm spec} = 2.154$, which are both in excellent agreement
    with the photometric redshift estimates from {\sc eazy} listed in their 
    figure panels. UDS-ALMA~2 is fitted using the AGN templates, and the
    best-fit template is the obscured AGN Abell~2690\#75 from 
    \citet{2008ApJ...672...94F}. The vertical scales of all panels cover a
    7-magnitude range, though with differing offsets, to allow comparison of
    the SEDs.  
    }
     \label{fig:eazyphotz}
\end{figure}

\section{Data and Sample} 

\subsection{Field Description}

   Our sample is collected from three fields that have both ALMA archival data
and new {JWST}/NIRCam data. 
   ACT-CL J0102$-$4915 (nicknamed ``El Gordo'') is a high-mass
($M_{\rm tot}\sim 10^{15}$~\Msol), colliding galaxy cluster at $z=0.87$
\citep{2010ApJ...723.1523M, 2011ApJ...737...61M, 2012ApJ...748....7M, 
2014ApJ...786...49L, 2014ApJ...785...20J}.
It was selected as a PEARLS field because it is a powerful cosmic lens.
Its lensing properties have been modeled based on lensed background objects
detected by {\HST} \citep{2013ApJ...770L..15Z, 2020ApJ...904..106D} and more 
recently by the PEARLS team \citep{Frye2022,Diego2022} using \JWST\ data.

   TN J1338+1942 (hereafter ``TNJ1338'') is a radio galaxy at $z=4.11$ 
\citep{1999A&A...352L..51D} and is the dominant member of a proto-cluster
\citep{2002ApJ...569L..11V, 2004Natur.427...47M}. It was selected as a PEARLS
field primarily for the study of galaxy properties in a proto-cluster 
environment. 

   The UKIDSS Ultra-Deep Survey field (hereafter ``UDS''---\citealt{Stach2018}) is one of the 
most-studied extragalactic survey fields and is where PRIMER will carry out
the majority of its {\JWST} observations. 

\begin{deluxetable*}{lcccccccccc}
\tabletypesize{\scriptsize}
\tablewidth{0pt}
\tablecaption{ALMA source catalog.}
\label{t:targets}
\tablehead{
Name    & RA     & Dec             & $S_{\rm 870\mu m}$ & $S_{\rm 1.1mm}$   & $S_{\rm 3.3mm}$      & Redshift                 & $\log(M_*)$       & log(SFR) & $R_e({\rm NIR})$  & $\mu$\\
        & J2000  & J2000 & mJy              & mJy              & mJy             &           & log($M_\odot$) & log($M_\odot \rm yr^{-1}$)  & kpc   & }
\startdata
EG-ALMA~2a  &  01:02:49.2 & -49:15:08.7  & \no               &  0.32$\pm$0.03  &   \no             &  3.58$\pm$0.38\rlap{\tablenotemark{a}} & \no & \no  &   \no            &  4.6  \\     
EG-ALMA~2b  &  01:02:49.4 & -49:15:05.3  & 1.62$\pm$0.12     &  0.65$\pm$0.07  &   \no             &  3.58$\pm$0.38\rlap{\tablenotemark{a}} & \no & \no  &   \no            &  5.5  \\     
EG-ALMA~3   &  01:02:49.3 & -49:14:38.1  & \no               &  0.75$\pm$0.08  &   \no             &  2.34$\pm$0.51                         & 10.94$\pm$ 0.11 & 1.94$\pm$ 0.49  &  5.35$\pm$0.27   &  2.1  \\  
EG-ALMA~5   &  01:02:50.5 & -49:15:41.7  & \no               &  0.25$\pm$0.03  &   \no             &  2.49$\pm$0.30                         & 11.13$\pm$ 0.14 & 1.09$\pm$ 0.44  &  1.53$\pm$0.30   &  1.8  \\   
EG-ALMA~6a  &  01:02:51.1 & -49:15:38.8  & \no               &  0.90$\pm$0.09  &   \no             &  4.324\rlap{\tablenotemark{b}}    & \no & \no  &  \no   &  3.9  \\     
EG-ALMA~6b  &  01:02:54.9 & -49:15:14.7  & 1.37$\pm$0.15     &  1.24$\pm$0.12  &   \no             &  4.324\rlap{\tablenotemark{b}}    & \no             & \no             &   \no            &  3.1  \\     
EG-ALMA~6c  &  01:02:55.7 & -49:15:09.0  & 1.89$\pm$0.19     &  0.99$\pm$0.10  &   \no             &  4.324\rlap{\tablenotemark{b}}    & 10.14$\pm$ 0.10 & 1.97$\pm$ 0.14  &  1.80$\pm$0.10   &  10.0 \\  
EG-ALMA~11  &  01:02:57.7 & -49:15:20.0  & \no               &  0.11$\pm$0.01  &   \no             &  1.67$\pm$0.21                         & 10.96$\pm$ 0.11 & 1.08$\pm$ 0.46  &  2.27$\pm$0.27   &  2.3  \\  
EG-ALMA~12  &  01:02:58.1 & -49:14:56.2  & \no               &  0.63$\pm$0.07  &   \no             &  2.49$\pm$0.18                         & 10.91$\pm$ 0.15 & 0.98$\pm$ 0.58  &  1.63$\pm$0.34   &  1.4  \\   
EG-ALMA~13  &  01:03:00.2 & -49:16:03.4  & \no               &  0.17$\pm$0.02  &   \no             &  1.80$\pm$1.08                         & 10.48$\pm$ 0.12 & 1.32$\pm$ 0.62  &  1.08$\pm$0.21   &  3.2  \\   
UDS-ALMA~1  &  02:17:19.6 & -05:09:41.4  & 4.91$\pm$1.54     &  \no            &   \no             &  2.10$\pm$0.13                         & 11.15$\pm$ 0.08 & 2.30$\pm$ 0.38  &  1.18$\pm$0.42   &  1    \\   
UDS-ALMA~2  &  02:17:21.0 & -05:08:37.2  & \no               &  0.93$\pm$0.16  &   \no             &  2.08$\pm$0.06\rlap{\tablenotemark{c}} & \no             & \no             &   $<$ 0.80       &  1    \\   
UDS-ALMA~3  &  02:17:22.3 & -05:10:38.6  & \no               &  1.02$\pm$0.13  &   \no             &  1.089\rlap{\tablenotemark{d}}     & 11.55$\pm$ 0.14 & 1.75$\pm$ 0.64  &  2.54$\pm$0.39   &  1    \\   
UDS-ALMA~4  &  02:17:26.1 & -05:10:58.3  & 3.04$\pm$0.38     &  0.86$\pm$0.09  &   \no             &  1.74$\pm$0.38                         & 11.18$\pm$ 0.19 & 2.42$\pm$ 0.37  &  4.64$\pm$0.42   &  1    \\   
UDS-ALMA~5  &  02:17:27.2 & -05:11:57.8  & 6.31$\pm$0.65     &  2.43$\pm$0.57  &   \no             &  2.154\rlap{\tablenotemark{d}}     & 11.32$\pm$ 0.08 & 2.33$\pm$ 0.22  &  3.78$\pm$0.41   &  1    \\   
UDS-ALMA~6  &  02:17:43.9 & -05:07:51.3  & 2.19$\pm$0.26     &  \no            &   \no             &  2.51$\pm$0.28                         & 11.31$\pm$ 0.07 & 2.28$\pm$ 0.36  &  1.44$\pm$0.40   &  1    \\   
TN-ALMA~1   &  13:38:24.9 & -19:42:15.8  & \no               &  \no            &   0.057$\pm$0.006 &  3.17$\pm$1.03                         & 10.77$\pm$ 0.40 & 2.36$\pm$ 0.33  &  1.12$\pm$0.42   &  1    \\   
TN-ALMA~2   &  13:38:25.7 & -19:42:34.6  & \no               &  \no            &   0.046$\pm$0.005 &  3.39$\pm$0.26                         & 10.58$\pm$ 0.09 & 2.17$\pm$ 0.31  &  1.20$\pm$0.39   &  1    \\   
TN-ALMA~6   &  13:38:26.9 & -19:42:30.9  & \no               &  \no            &   0.031$\pm$0.003 &  1.47$\pm$0.46                         & 11.35$\pm$ 0.10 & 2.05$\pm$ 0.46  &  2.80$\pm$0.41   &  1    \\   
\enddata
\tablecomments{The flux densities, masses, SFRs, and sizes for the El Gordo
sources have been corrected for the gravitational lensing magnification $\mu$.
For most objects, their $z_{\rm phot}$ from {\sc magphys-photo-z} are listed
in the ``Redshift'' column. The exceptions are (1) the objects that
have $z_{\rm spec}$, for which their $z_{\rm spec}$ are listed; and (2) the
point-like source UDS-ALMA~2, for which we adopt $z_{\rm phot}$ from {\sc eazy} as
it was derived using AGN templates and is more plausible.
$R_e$ is for the disk component (i.e., with a point-source core subtracted for UDS-ALMA 1/3/4/5). 
Because $R_e$ is measured in the rest-frame near-IR, it is equivalent
to the half-mass radius. 
}
\tablenotetext{a}{Lensed double images of a single source 
\citep[see][]{Diego2022}.}
\tablenotetext{b}{Lensed triple images of a single source. Its $z_{\rm spec}$ 
is based on CO J=4--3 emission line (see Appendix B) and has uncertainty 0.001.}
\tablenotetext{c}{$z_{\rm phot}$ derived by {\sc eazy} using AGN templates.}
\tablenotetext{d}{$z_{\rm spec}$ from \citet{2019ApJ...879...54L, 2019MNRAS.482.3135B}. The typical $z_{\rm spec}$ uncertainty is 0.001}.
\end{deluxetable*}

\subsection{ALMA archive data and sample construction}

   The El Gordo field has been observed by the ALMA programs 2013.1.01358.S 
(PI: A. Baker) and 2018.1.00035.L (PI: K. Kohno) in Band~6 
(270~GHz = 1.1~mm) and 2013.1.01051.S (PI: P. Aguirre) in Band~7 
(340~GHz = 870~$\mu$m). The ALMA data in the TNJ1338 field were obtained by 
program 2015.1.00530.S (PI: C. De Breuck) in Band~3 (92~GHz = 
3.3~mm). The UDS has a large number of ALMA programs. In the current PRIMER
UDS field, the ALMA Band~6 and/or Band~7 data were collected by
{\sc astroquery} in each PRIMER observation region. These archival data were 
reduced by the Chinese South American Center ALMA data processing program 
(Cheng et~al., in prep.). Briefly, the data were calibrated
by the default ALMA data reduction script {\sc ScriptForPI.py}. 
The task {\sc tclean} in {\sc casa 6.2.1} \citep{2007ASPC..376..127M} was used 
to clean the data to build the continuum images. For this set of data, the 
clean parameters were set to {\sc {weighting=``briggs''}} and {\sc robust=2.0}, and we cleaned the images to 3~$\sigma$.

    The final 1.1~mm map in El Gordo has beam size 
1\farcs28$\times$0\farcs93 at $\rm PA = 88$\arcdeg\ and reaches
$\sim$0.065~mJy~beam$^{-1}$ (root-mean-square; rms). This map has an area of
$\sim$4~arcmin$^2$ overlapped with the NIRCam coverage (see below) and was used
for the source detection in this field. 
The final 870~$\mu$m map, which
covers only a portion of the 1.1~mm map, reaches
0.25~mJy~beam$^{-1}$ with beam size  0\farcs33$\times$0\farcs40 at 
$\rm PA = -35$\arcdeg.

    In TNJ1338, the final 3.3~mm map has beam size of 
2\farcs11$\times$1\farcs74, $\rm PA = -82$\arcdeg\ and reaches 
$\sim$0.009~mJy~beam$^{-1}$ rms. It covers $\sim$2~arcmin$^2$, centered around
the radio galaxy.

    In the UDS field, the final images have beam sizes of about 0\farcs7 at 
1.1~mm and 0\farcs4 at 870~$\mu$m. The total coverage is $\sim$1~arcmin$^2$. 
Both images reach rms $\sim$0.05~mJy. The source detections were done
in both bands, and the results were merged.

    Source extraction was done by running SExtractor
\citep{1996A&AS..117..393B} in dual-image mode. The detection was done on the 
ALMA maps not corrected for the primary-beam attenuation, as these have a 
uniform noise distribution. The photometry was done on the primary-beam-corrected
images exported from {\sc casa}. The detection map was convolved using a Gaussian
kernel, and we set {\sc detect\_thresh = 4}. {\sc flux\_auto} was adopted for the
flux density measurements. To account for the calibration uncertainty, we added
in quadrature 10\% of the flux density \citep{2014Msngr.155...19F} to the reported 
{\sc fluxerr\_auto} to obtain the final uncertainty estimate. 
All these detections have signal-to-noise ratios (S/N) higher than 3, and 
all but two
have $S/N >5$ (see Table~\ref{t:targets}).

\subsection{{\JWST} NIRCam Data}

   The PEARLS NIRCam data in the TNJ1338 field and the El Gordo field were
taken on 2022 July~1 and~29, respectively, and these observations are 
described by \citet{2022arXiv220904119W}. Briefly, the observations of 
El Gordo were done in F090W, F115W, F150W, F200W, F277W, F356W, F410M, and 
F444W, and the total integration times were 2491, 2491, 1890, 2104, 2104, 1890,
2491, and 2491 seconds, respectively. The observations in TNJ1338 were done in 
F150W, F182M, F210M, F300M, F335M, and F360M with a uniform total integration
time of 1031 seconds in each band. In both fields, the areas where the ALMA 
data reside are covered by one NIRCam module (module~B).

   As of this writing, only a small fraction of the PRIMER observations in UDS were
executed, all on 2022 July~29. The NIRCam data were taken in three pointings 
as the parallels to the primary MIRI observations. One pointing is isolated, 
while the other two overlap and fill the wide gaps between the two NIRCam 
modules. The passbands used were the same as for El Gordo, but integration 
times were 837 seconds in each band.

  All data were reduced using the {\JWST} data reduction pipeline
version 1.6.1dev3+gad99335d in the context of \texttt{jwst\_0944.pmap}, starting
from the Stage~1 ``uncal'' products. A few changes and augmentations were made 
to the pipeline to improve the reduction quality \citep{2022arXiv220904092Y}.
In the El Gordo field, the final stacked images were created at a pixel scale 
0\farcs06 and were aligned to the existing {\HST} images of the same scale 
produced by the Reionization Lensing Cluster Survey 
\citep[RELICS;][]{2019ApJ...884...85C}. 
The final stacks in the TNJ1338 field were also aligned to the existing {\HST} 
images, which are available from the
High Level Science Products provided by the Mikulski Archive for Space 
Telescopes. These {\HST} images have  pixel scale of 0\farcs04, and the NIRCam 
images were created at the same scale. In the UDS field, the final
stacks were created at 0\farcs06~pixel$^{-1}$ and were aligned to the 
existing {\HST} images of the same scale from the Cosmic Assembly Near-infrared 
Deep Extragalactic Legacy Survey
\citep[CANDELS;][]{2011ApJS..197...35G, 2011ApJS..197...36K}.

   In total, we identified 10, 4, and 6 NIRCam counterparts of our ALMA sources
in the El Gordo, TNJ1338, and UDS fields, respectively. One of the TNJ1338 
sources is the central radio galaxy, which is studied by \citet{Duncan2022} 
and not discussed further here. Table~\ref{t:targets} lists the remaining 19 
sources. Figure~\ref{EGstamps}
shows the HST ACS and NIRCam stamp images for each. The sources in the El Gordo
field are affected by gravitational lensing to various degrees.
In particular, EG-ALMA~6a/6b/6c are the triple images of a single 
galaxy that is believed to be associated with the $z=4.32$ galaxy group 
\citep{2021ApJ...908..146C}. EG-ALMA~2a/2b are two images of another 
single galaxy lensed by two cluster-member galaxies as well as the 
cluster as a whole. Therefore, the 10 sources in this field correspond to seven
unique galaxies; and so our final sample consists of 16 galaxies.

\section{Analysis and Results}

\subsection{SED construction and fitting}

  For the targets in the El Gordo field,  we did not use the HST data because 
the JWST NIRCam data alone sufficiently sample the rest-frame visible to 
near-IR range. In the UDS field, UDS-ALMA~4 has no NIRCam data in the short 
wavelength bands because it falls in the gap in between the modules. 
Therefore, we used the HST data in F606W, F814W, F125W, and F160W\null. For
homogeneity, these HST images were also used for the other UDS targets. In 
TNJ1338, the bluest NIRCam band is F150W, and therefore we added HST F775W 
and F850LP data to extend the wavelength sampling to visible wavelengths.

    We used SExtractor in dual-image mode for photometry. For the sources
in El~Gordo and UDS, the detection band was set to F444W; in TNJ1338, it was
set to F360M. In most cases, we adopted {\sc mag\_iso} to optimize the
signal-to-noise ratio (S/N) for the best color measurement. The exceptions were
the two multiply imaged systems EG-ALMA~2a/2b and EG-ALMA~6a/6b/6c.
The former is blended with two cluster-member galaxies, which contribute the 
major part of the lensing effect that creates this system. The latter has a 
close neighbor that might be associated with the source but is not the 
counterpart. In both cases, their {\sc mag\_iso} apertures are severely
contaminated by the light from these neighbors. To minimize the contamination,
we had to derive colors using a circular aperture of 0\farcs6 in diameter 
centered on the position of the brightest pixel in the F444W image. 

   Figure~\ref{fig:magphysfits} shows the full SEDs, combining rest-frame visible 
to near-IR photometry with the ALMA photometry. We used {\sc magphys+photo-z} 
\citep{Cun08, 2019ApJ...882...61B} to fit these SEDs to obtain physical 
properties of the hosts, most importantly the photometric redshift
($z_{\rm phot}$), the star formation rate (SFR), and the stellar mass ($M_*$).
The best-fit results are shown in Figure~\ref{fig:magphysfits}, and the
derived values are presented in Table~\ref{t:targets}. The SFR and $M_*$ values
for the El Gordo sources have been corrected for the 
magnification factors ($\mu$) at the corresponding redshifts.
Two sources, UDS-ALMA~3 and~5,  have spectroscopic redshifts ($z_{\rm 
spec}$), which are quoted 
in Table~\ref{t:targets}; their $M_*$ and SFR  listed in the table were 
derived at their $z_{\rm spec}$, while Figure~\ref{fig:magphysfits}
still shows their SED-fitting results when treating redshift as a free
parameter.

{\sc magphys+photo-z} is designed to fit panchromatic SEDs based on the
``energy-balance'' premise, which argues that the UV-to-near-IR energy 
absorbed by dust in a system should roughly equal the far-IR-to-mm light re-emitted
by dust. This approach provides a good constraint on the amount of dust
extinction, and propagates the uncertainty of $z_{\rm phot}$ to the uncertainties
in other derived parameters.
However, the energy-balance premise might not hold if the dust component is 
not well mixed with stars in the host galaxy. In addition, we have only one or 
two ALMA bands constraining the dust emission. 
To check the robustness of the $z_{\rm phot}$ derived by 
{\sc magphys+photo-z}, we also fitted the visible-to-near-IR SEDs using {\sc 
eazy} \citep{Bra08} with {\sc eazy\_v1.1\_lines.spectra.param} templates to 
derive $z_{\rm phot}$ independently. The best-fit results are summarized in
Figure \ref{fig:eazyphotz}. For most sources, these two sets of $z_{\rm phot}$ 
are consistent with each other (see Appendix A), which gives us
confidence in the results obtained by {\sc magphys+photo-z}.

    EG-ALMA~2a/2b and EG-ALMA~6a/6b/6c are two multiply
imaged systems. (The latter has $z_{\rm spec}$ from the ALMA spectroscopy; see 
Appendix~B.) It is difficult to obtain reliable photometry for them 
because the images are highly distorted. This is reflected in the disagreement in $z_{\rm phot}$
values for the individual images and for the same image as measured by the two different methods. 
We have therefore excluded these objects from the later discussion. We
also exclude UDS-ALMA~2, which is likely a quasar, for most purposes.

\begin{figure}
    \centering
    \includegraphics[width=0.288\textwidth]{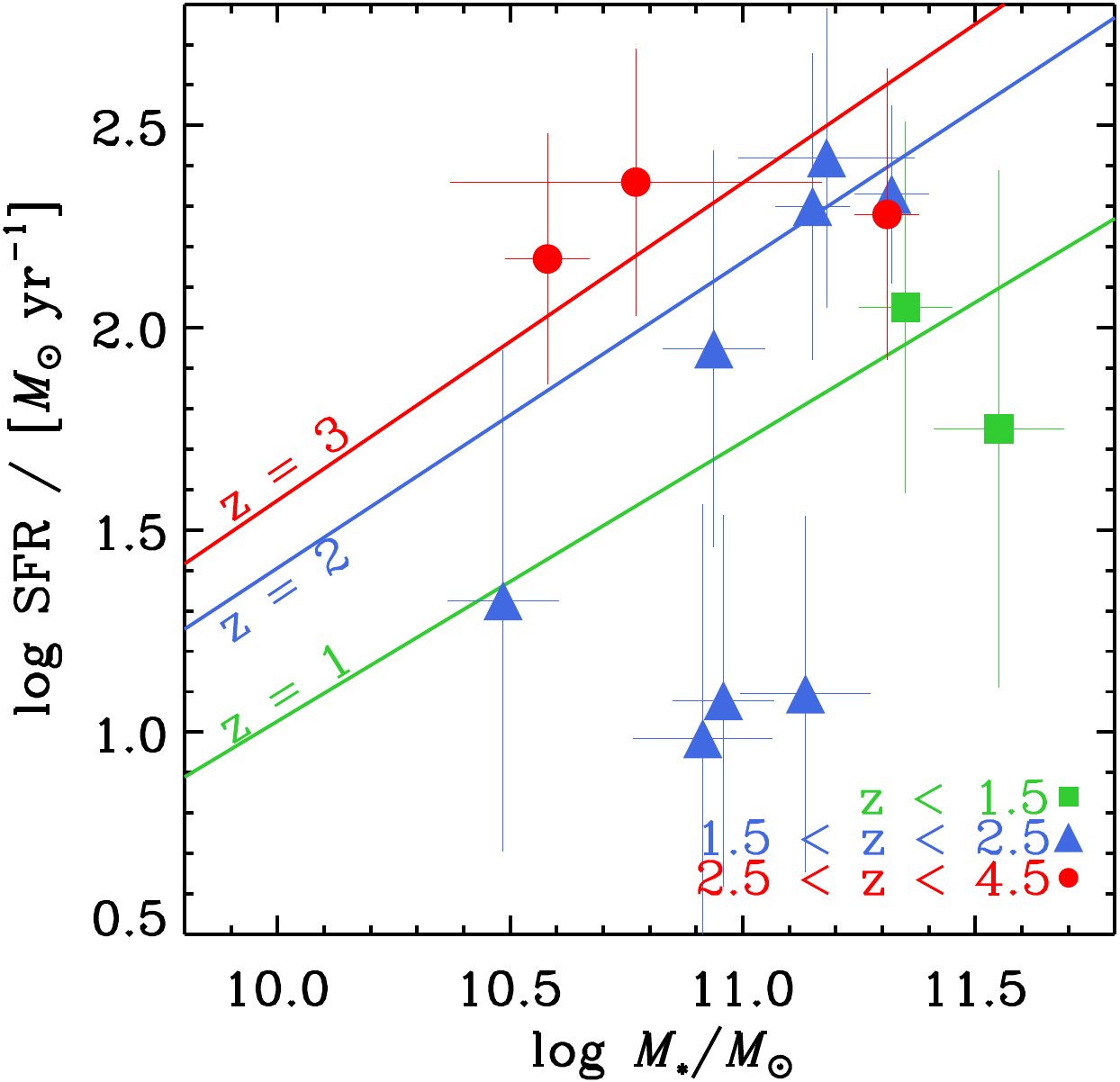}
    \includegraphics[width=0.3\textwidth]{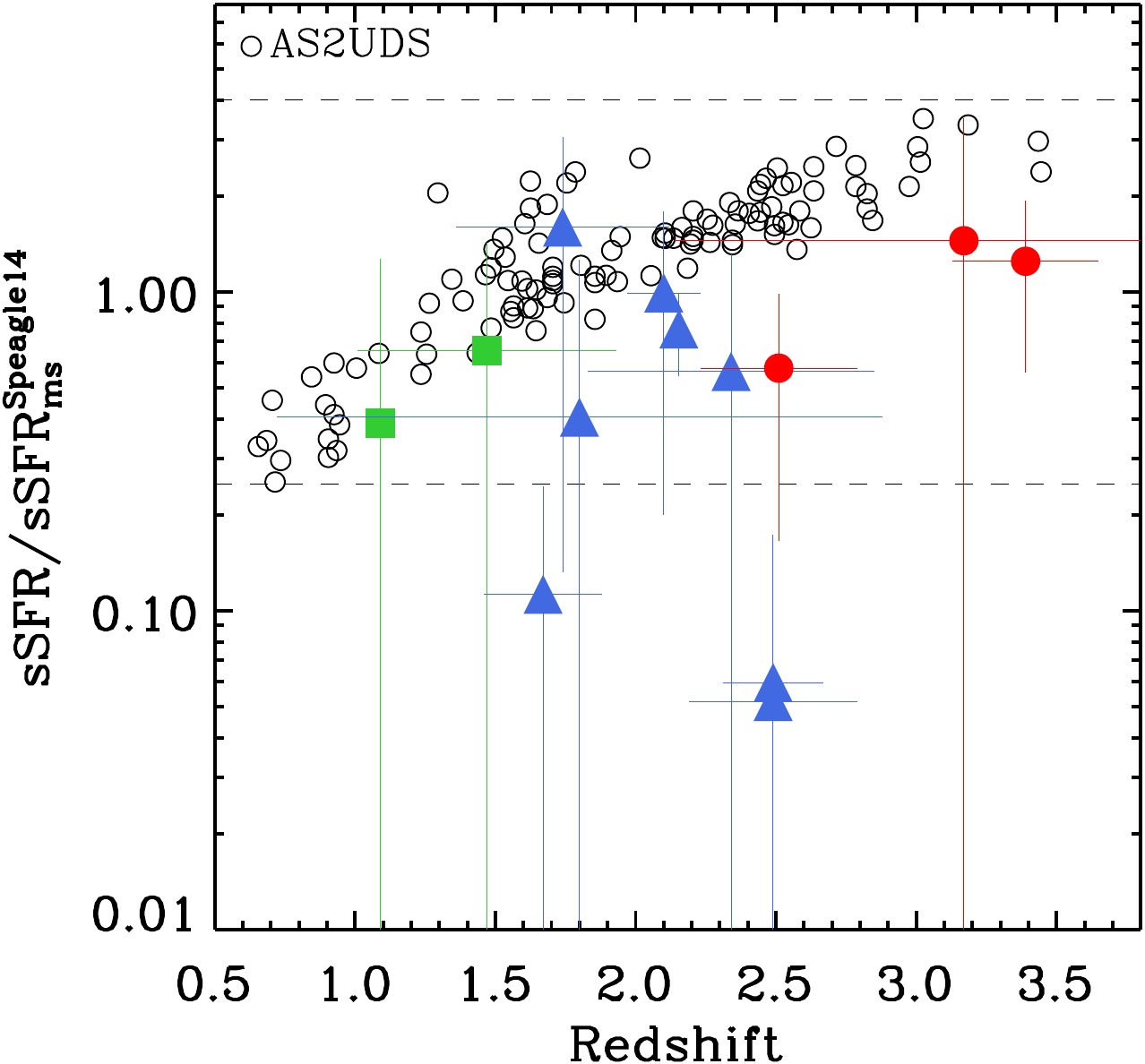}
    \includegraphics[width=0.335\textwidth]{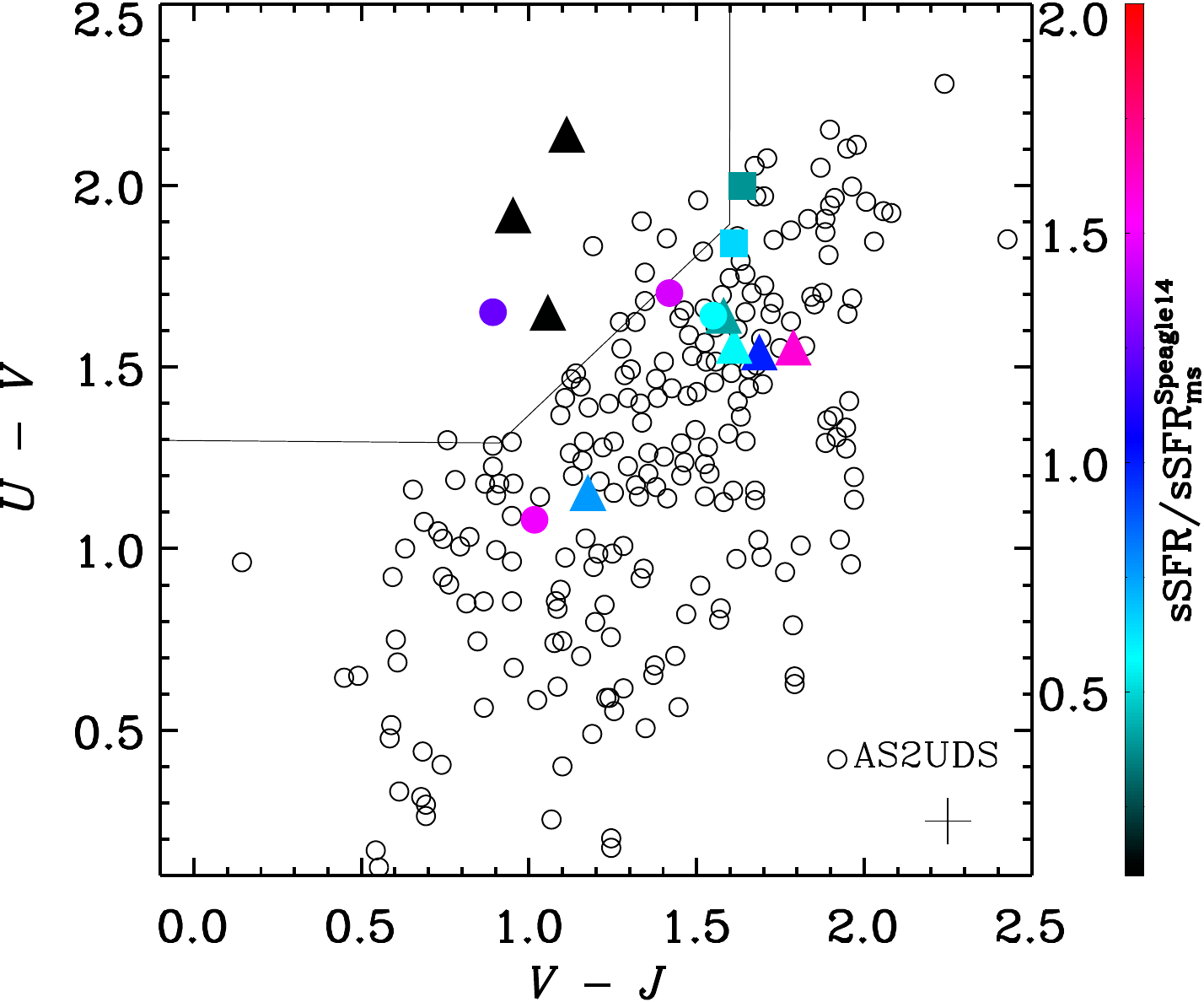}
    \caption{Stellar population properties of the JWST counterparts of the ALMA 
    sources.  The two multiply imaged systems EG-ALMA~2a/2b, EG-
    ALMA~6a/6b/6c and the point source UDS-ALMA~2 are excluded.
    {\bf Left:} star-formation rate versus stellar mass. All values have been corrected for magnification. The points with error 
    bars represent the 13 galaxies, and the lines show the star-formation main 
    sequence \citep{2014ApJS..214...15S}. The points and lines are 
    color-coded by redshift as labeled and indicated in the legend.  
    {\bf Middle:} Ratios of the specific star-formation rates of our sources
    to those of the main sequence galaxies as a function of redshift. The 
    colored points with error bars are the 13 ALMA sources, and the open 
    circles show the SMGs from the AS2UDS program \citep{2020MNRAS.494.3828D}
    for comparison. 
    The dashed horizontal lines (ratios of 1/4 and 4, respectively) indicate 
    the nominal range within which galaxies are considered to be on the main 
    sequence. 
    {\bf Right:} \textit{UVJ} color--color diagram. The points represent the ALMA-source host galaxies, color-coded by their sSFR/sSFR$_{\rm ms}$ values as
    indicated by the color bar to the right. The rest-frame \textit{UVJ} colors were
    computed by {\sc eazy} from the best-fit templates at the adopted redshifts.
    The cross in the lower right corner indicates our estimate of the typical 
    uncertainties. The black open circles show the \textit{UVJ} colors of the AS2UDS SMGs \citep{2019MNRAS.487.4648S}
    for comparison.   
    The quiescent-galaxy region marked by solid lines in the upper left is
    based on \cite{Patel2012}. The one blue data point in this region
    is TN-ALMA~2 ($z_{\rm phot}=2.79$). Its reddest passband, F360M, does not reach
    the rest-frame $J$-band, and therefore {\sc eazy}'s estimate of its $V-J$ 
    color has large uncertainty. 
    }
    \label{ssfr}
\end{figure}

\subsection{Host galaxy star formation: from star-forming to quiescent} 

   Figure~\ref{ssfr} (left) compares these ALMA source hosts to the 
``star-formation main sequence'' from \citet{2014ApJS..214...15S}. Excluding
EG-ALMA~2a/2b, EG-ALMA~6a/6b/6c and UDS-ALMA~2, all other sources have 
$M_*\geq 10^{10.5}$~\Msol,  in line with the general SMG population. 
However, three objects have $\rm SFR  \le16$~\Msol~yr$^{-1}$, which
puts them well below the main sequence.
This is more clearly seen in Figure~\ref{ssfr} (middle). 
The mean ratio of sSFR to main sequence is 0.24; the aforementioned three sources 
are below 0.10, which 
places them in the realm of quiescent galaxies. In other words, their host 
galaxies have already built up the bulk of their stellar masses, and their 
ongoing star formation, while detectable from the submm/mm emission, is not 
significantly increasing the host's stellar mass.
The usual \textit{UVJ} diagnostic (Figure~\ref{ssfr} right) shows that the 
three quiescent hosts are
indeed in the conventional quiescent-galaxy region \citep{2005ApJ...624L..81L, 2009ApJ...691.1879W}, verifying the low sSFR\null.

\begin{figure}
\centering
\includegraphics[width = 0.49\textwidth]{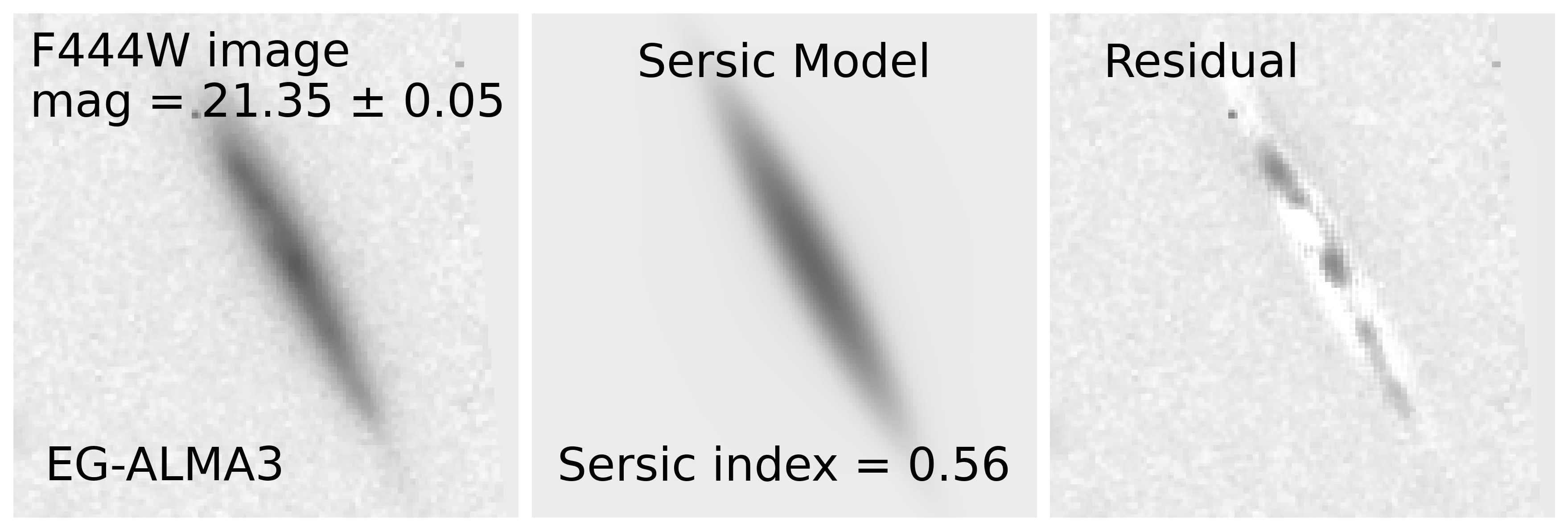}
\includegraphics[width = 0.49\textwidth]{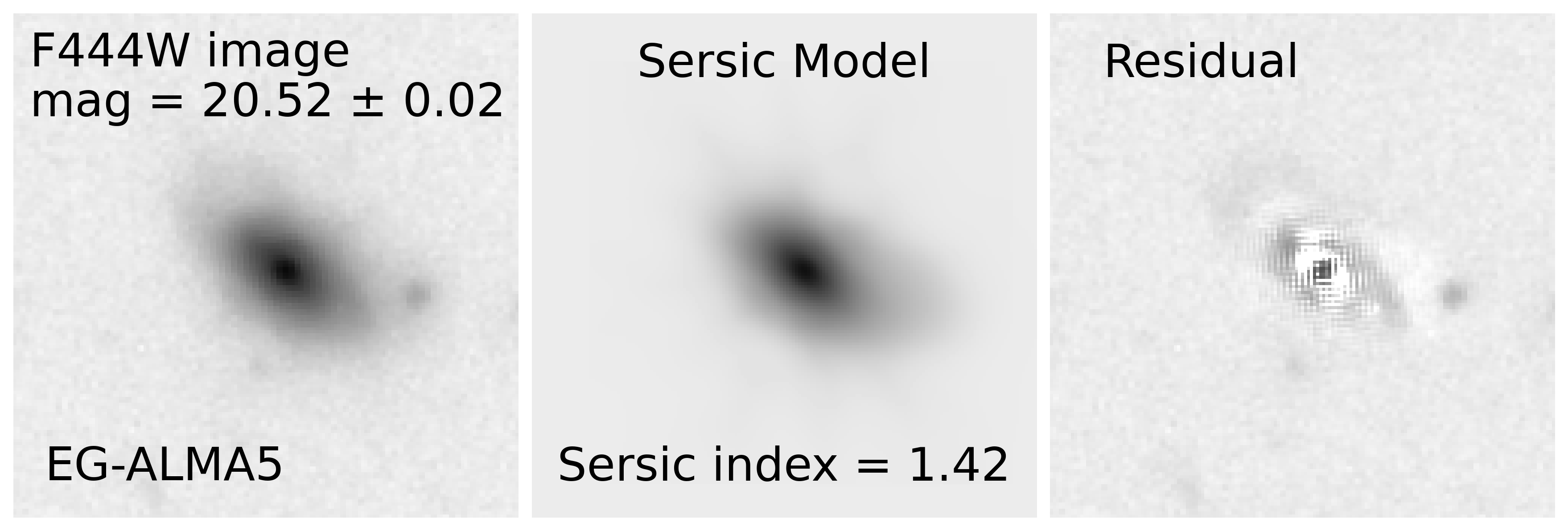}\\
\includegraphics[width = 0.49\textwidth]{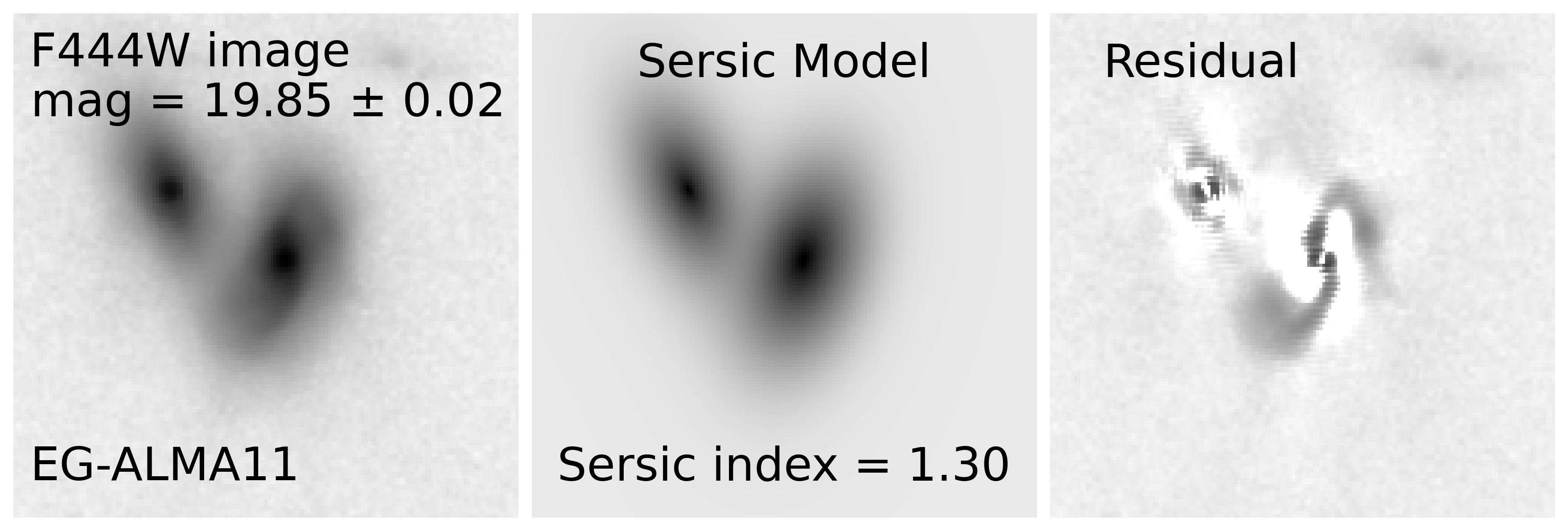}
\includegraphics[width = 0.49\textwidth]{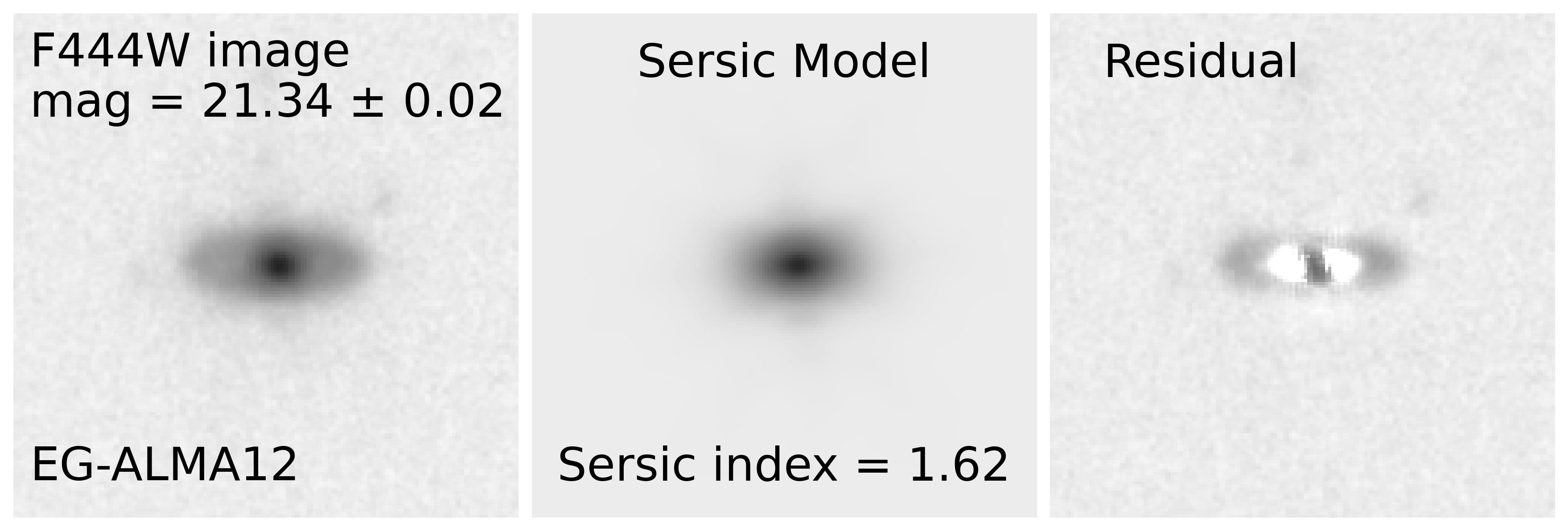}\\
\includegraphics[width = 0.49\textwidth]{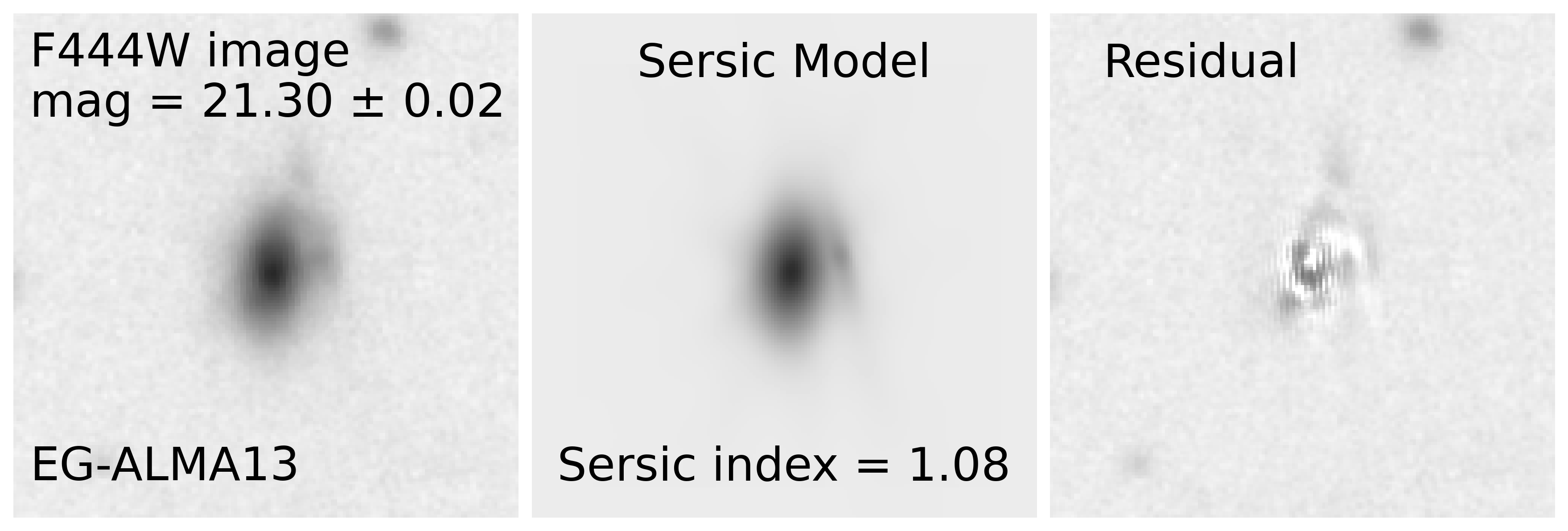}
\includegraphics[width = 0.49\textwidth]{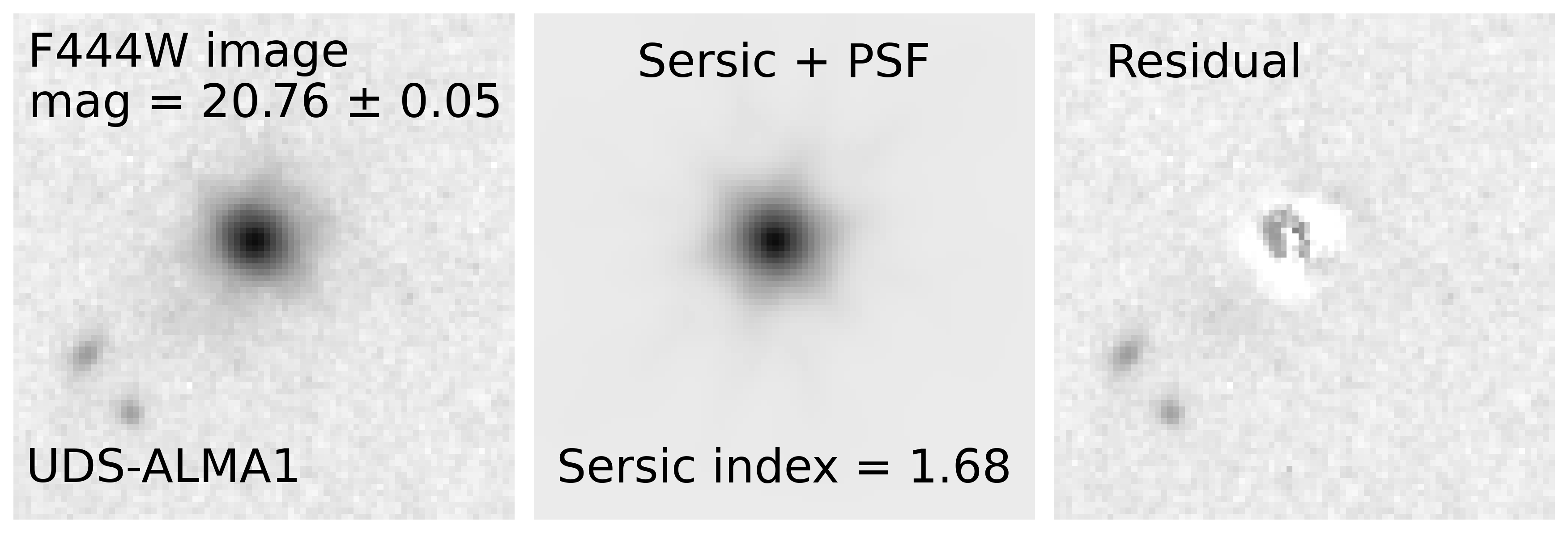}\\
\includegraphics[width = 0.49\textwidth]{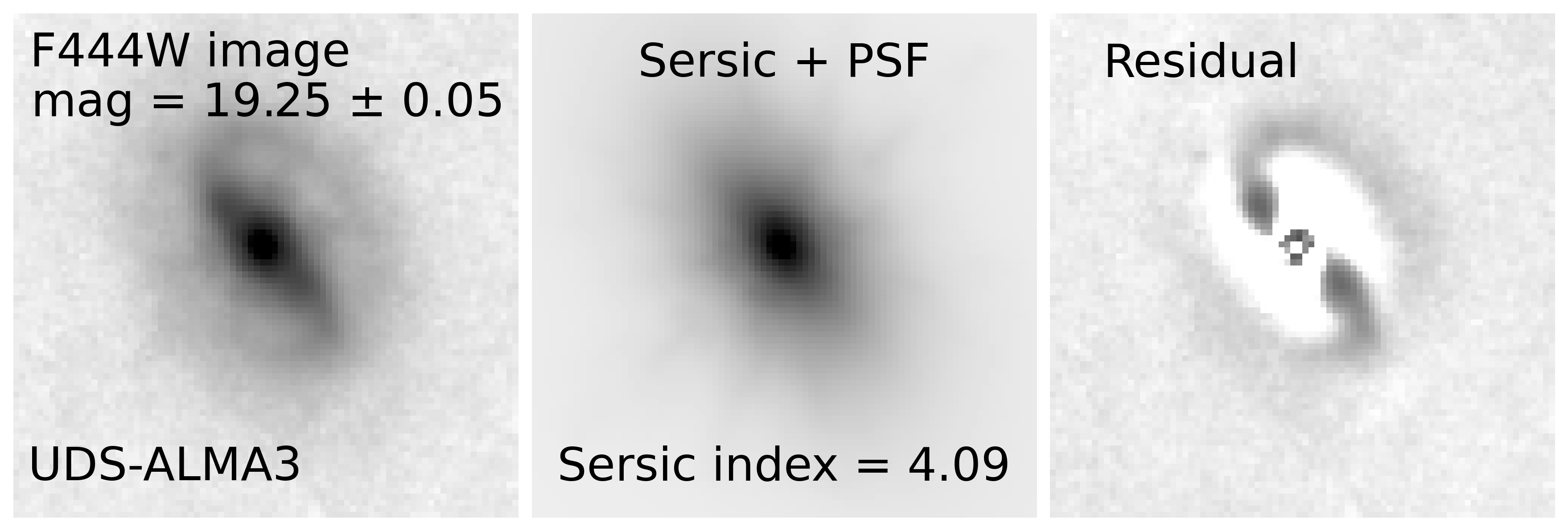}
\includegraphics[width = 0.49\textwidth]{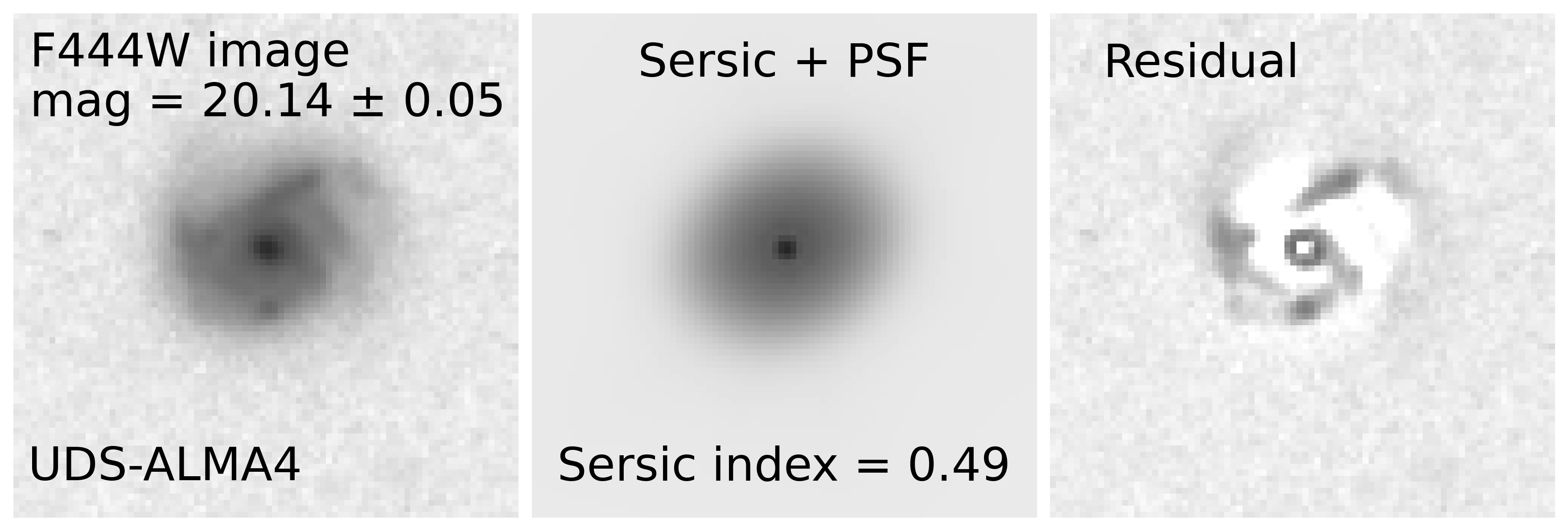}\\
\includegraphics[width = 0.49\textwidth]{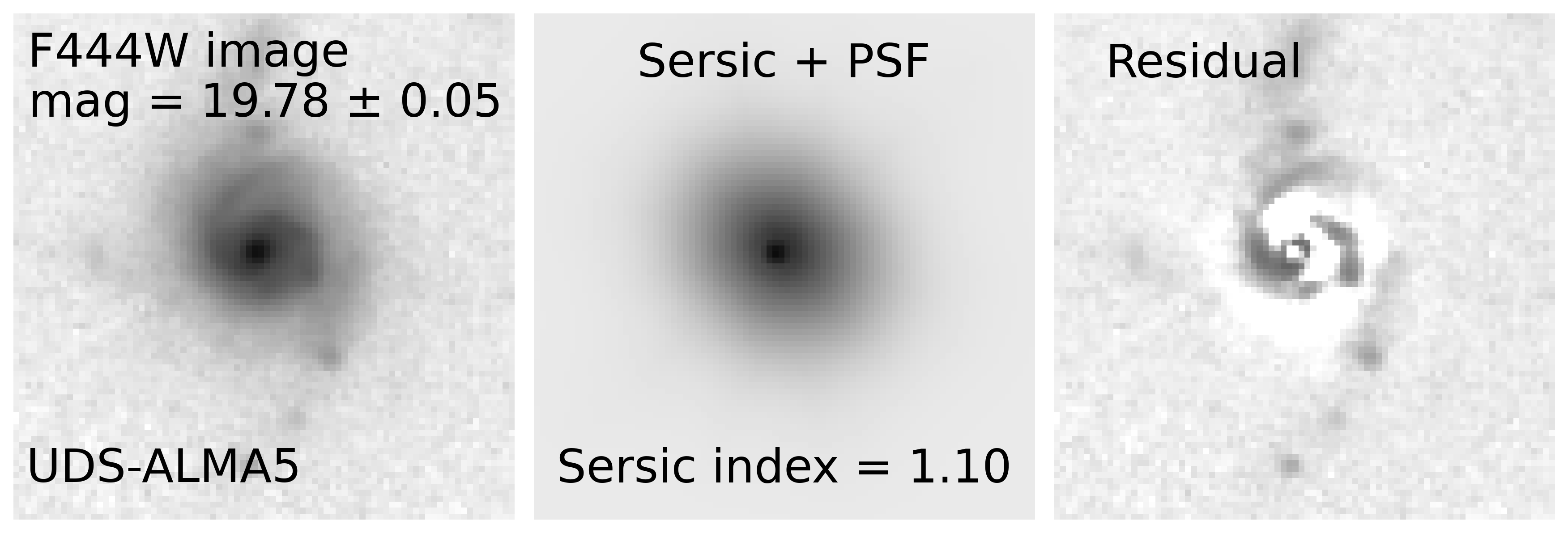}
\includegraphics[width = 0.49\textwidth]{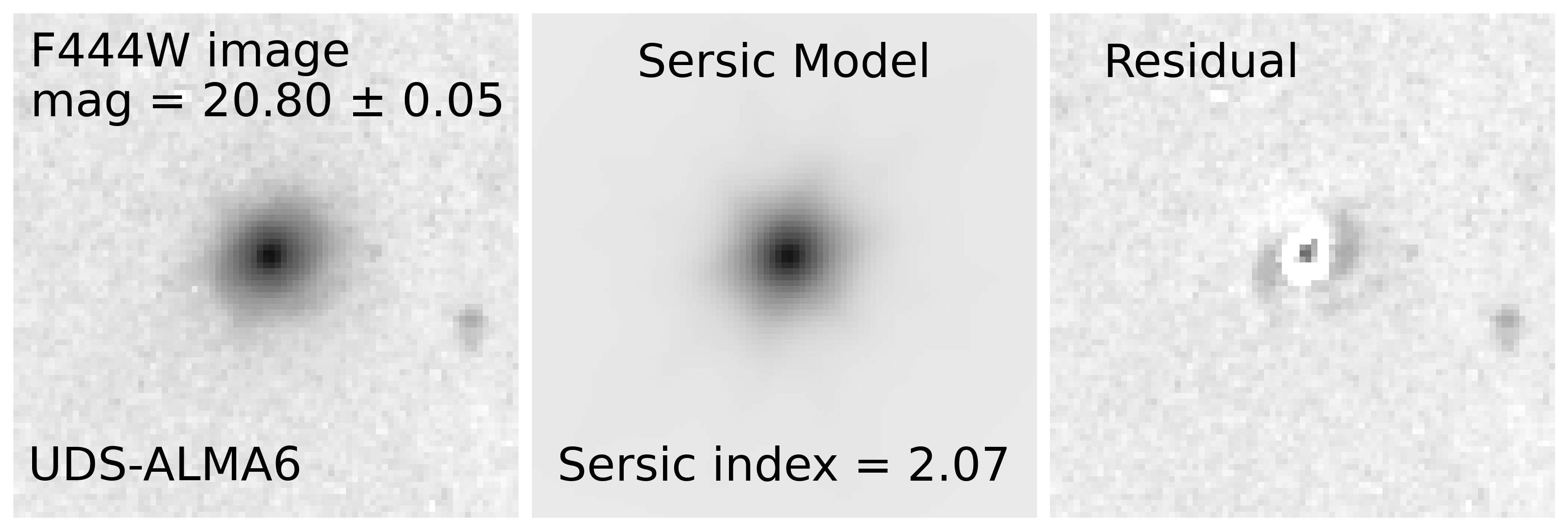}\\
\includegraphics[width = 0.49\textwidth]{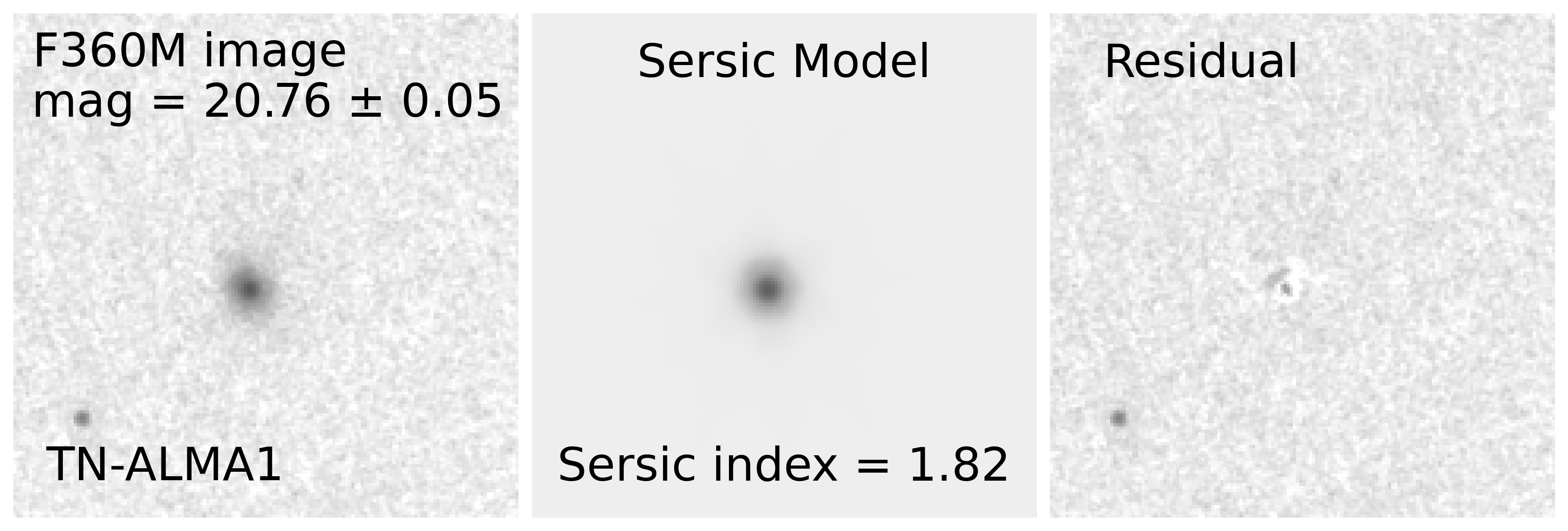}
\includegraphics[width = 0.49\textwidth]{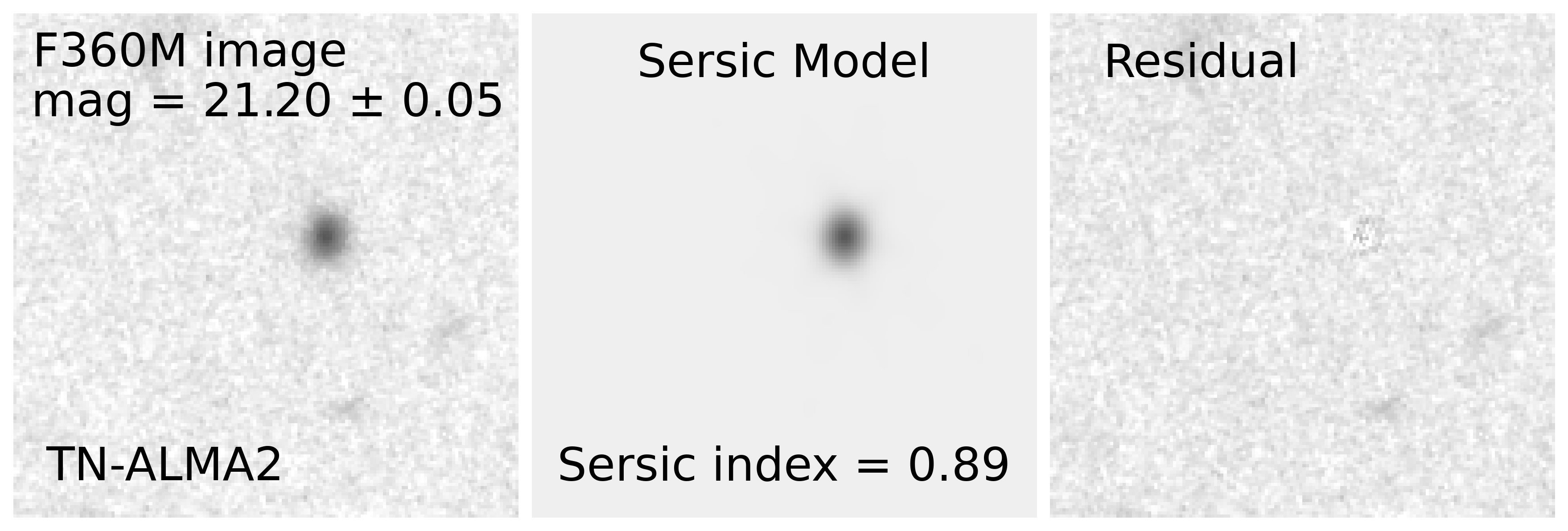}\\
\includegraphics[width = 0.49\textwidth]{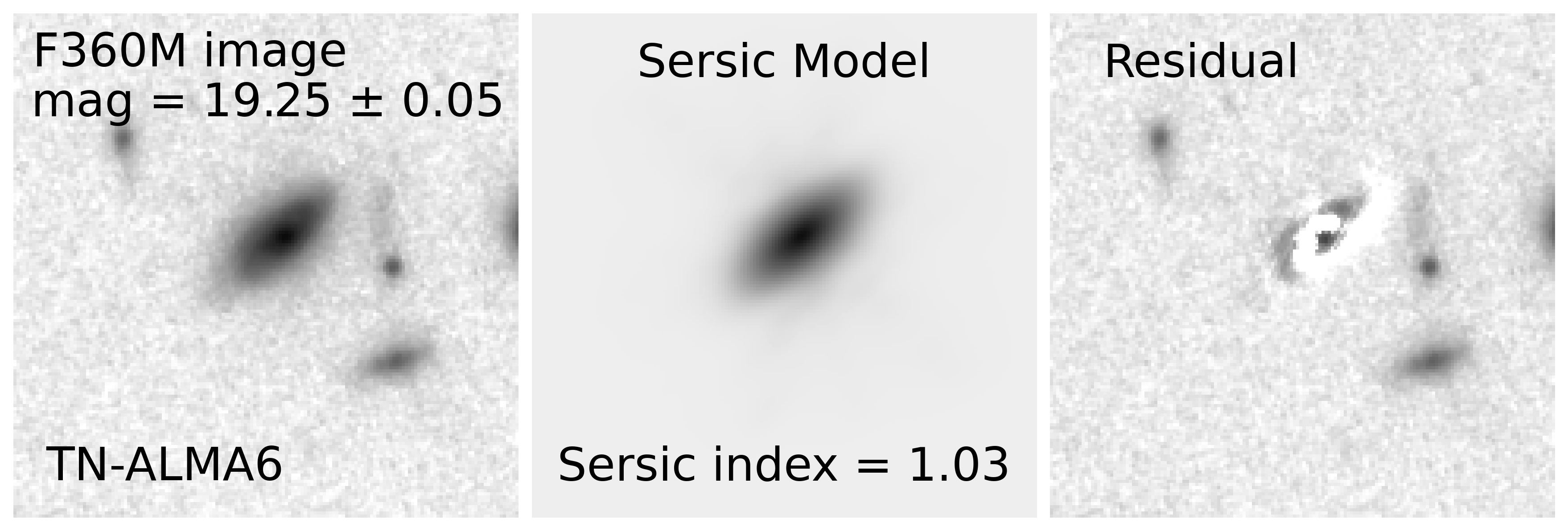}
\caption{\GALFIT\ fitting results of our sources. 
The two multiply imaged systems (EG-ALMA 2a/2b and EG-ALMA 6a/6b/6c) and the
quasar (UDS-ALMA~2) are not included. 
From left to right, three panels
show the original F444W (or F360M) negative image (with source name and filter noted), the fitted S\'ersic model (with the
index noted), and the residual (the original image with model subtracted),
respectively.  The images are centered on the ALMA positions and are 5\arcsec\ ($\sim$40~kpc) on a side. 
The depth of the TNJ1338 images is shallower than the rest, 
so we stacked the F300M, F335M, F360M images to synthesize deeper broadband images with 3$\times$ longer exposure.
Fitting models to the stacked images (not shown) gave sizes consistent with those from the F360M images.
}
    \label{galfit}
\end{figure}

\subsection{Morphology of host galaxies}

  The NIRCam images (Figure~\ref{EGstamps}) allow morphologies to be
determined for 13 of the ALMA source 
hosts.\footnote{The two multiply imaged systems in El 
Gordo are highly distorted, and UDS-ALMA~2 is a point source and potentially hosts
a quasar. Morphologies cannot be determined for these.} 
By our visual classification, all 13 show non-disturbed (or weakly disturbed) disks 
even though some of them (e.g., UDS-ALMA~4) might have features indicative of a 
recent minor 
merger.\footnote{EG-ALMA~11 has a close neighbor comparable in size, but the 
neighbor's spectroscopic redshift $z_{\rm spec} = 0.87$ shows that it is a 
foreground galaxy.}

\begin{figure}
    \centering
    \includegraphics[width=0.8\textwidth]{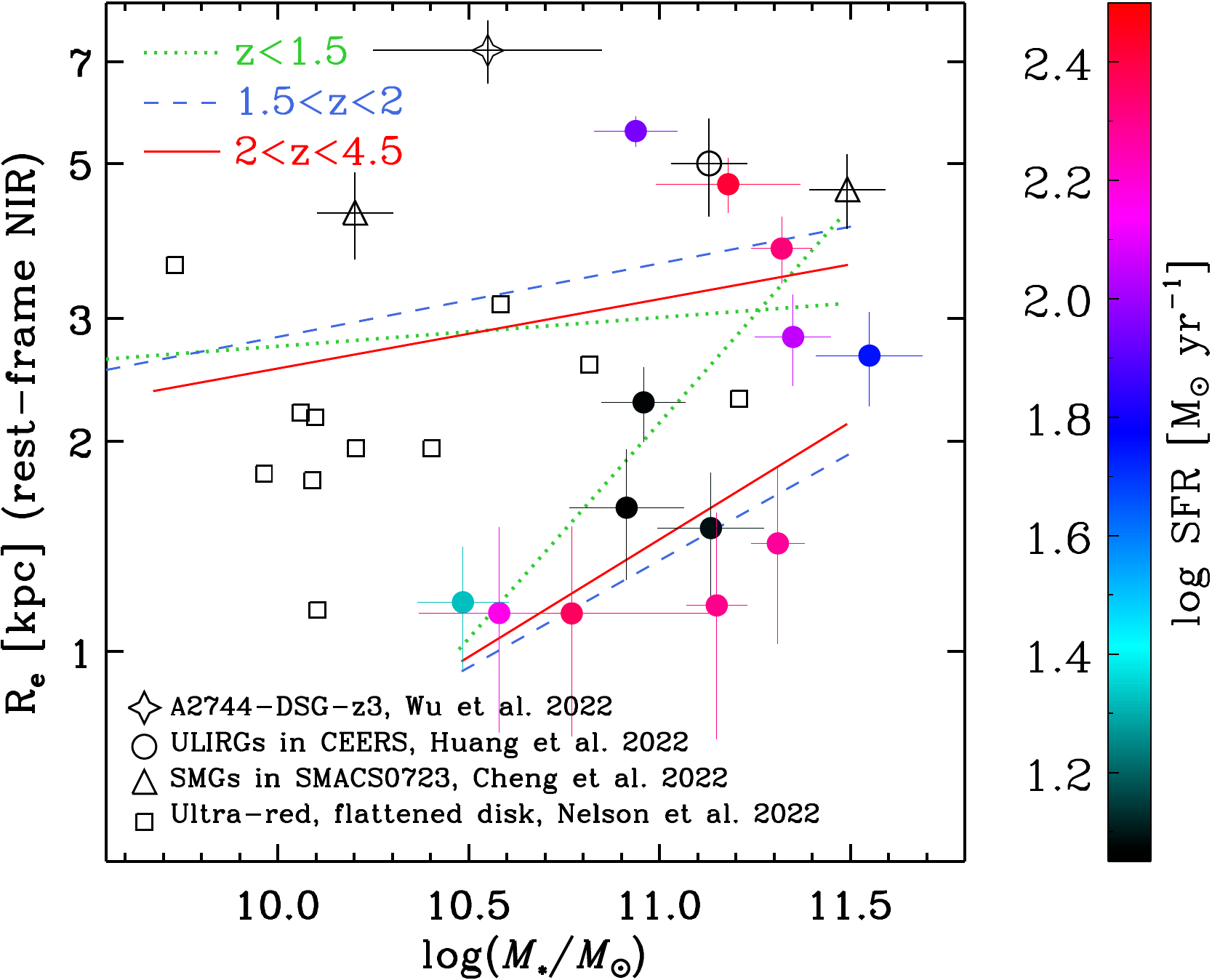}\hfill
    \caption{Stellar half-light radii versus stellar mass for ALMA source hosts
    in Figure \ref{galfit}.
    The filled symbols represent our objects, which are color-coded based on
    their SFRs.  The color bar on the right shows the coding. For comparison, the open symbols show the ultra-red, flattened disks \citep{2022arXiv220801630N}, SMGs in SMACS~0723 \citep{2022ApJ...936L..19C}, ULIRGs in CEERS 
    \citep{Huang2022}, and the SMG in A2744 \citep{Wu2022}, respectively. All 
    values are based on NIRCam data, and we re-measured the sources in SMACS~0723 
    using GALFIT for consistency. The color-coded straight lines, based on the CANDELS results 
    \citep{2019ApJ...877..103S}, show the relations for
    quiescent (lower group) and star-forming (upper group) galaxies in
    three redshift ranges as indicated. 
    Most of our sources, despite the
    fact that they are forming stars at high rates, have sizes in the range for quiescent galaxies.
    }
    \label{masssize}
\end{figure}

  To further study the morphologies, we ran \GALFIT\ 
\citep{2002AJ....124..266P} to fit S\'ersic profiles to the galaxies' F444W 
or F360M images (the latter only for the three TNJ1338 sources). The results
are shown in Figure~\ref{galfit} and detailed in Appendix~B\null. 
Briefly, the fitted S\'ersic indices $n$ are near~1 (median $\langle n\rangle = 1.1\pm 0.8$), 
consistent with disky galaxies. The \GALFIT\ run also computed half-light
radii ($R_e$) based on the best-fit profile and excluding a central 
point-source when one was present. At the source redshifts, F444W or F360M
samples rest-frame visible to near-IR, and therefore this $R_e$ reflects the
stellar mass distribution. In other words, $R_e$ is a proxy for the 
half-mass radius. 
The median $R_e$ of our sample is 1.6~kpc with bootstrap uncertainty 0.4~kpc.
The dispersion of the whole sample is 1.4~kpc.
For comparison, the median $R_e$ of the disk galaxies among the far-IR/SMG
sample ($0.5\lesssim z\lesssim 3$) of
\citet{LY2022} is 3.6~kpc.
Figure~\ref{masssize} shows the mass--size distributions of our ALMA sources
as well as those of the few recently published NIRCam results in other fields. The
mass--size relations of star-forming and quiescent galaxies from 
\citet[][their Fig.~7]{2019ApJ...877..103S} are also shown for comparison.
Qualitatively, most of our sources fall below the relation for star-forming 
galaxies and are more in line with that of quiescent galaxies, despite the fact 
that most of our sources are not quiescent. In contrast, the ultra-red, flattened, disky 
galaxies recently found by \citet{2022arXiv220801630N} have larger $R_e$ and follow 
more closely the relation for star-forming galaxies. A larger sample will be
needed to further investigate this problem.

\section{Discussion}

\begin{figure}
    \centering
    \includegraphics[width=0.7\textwidth]{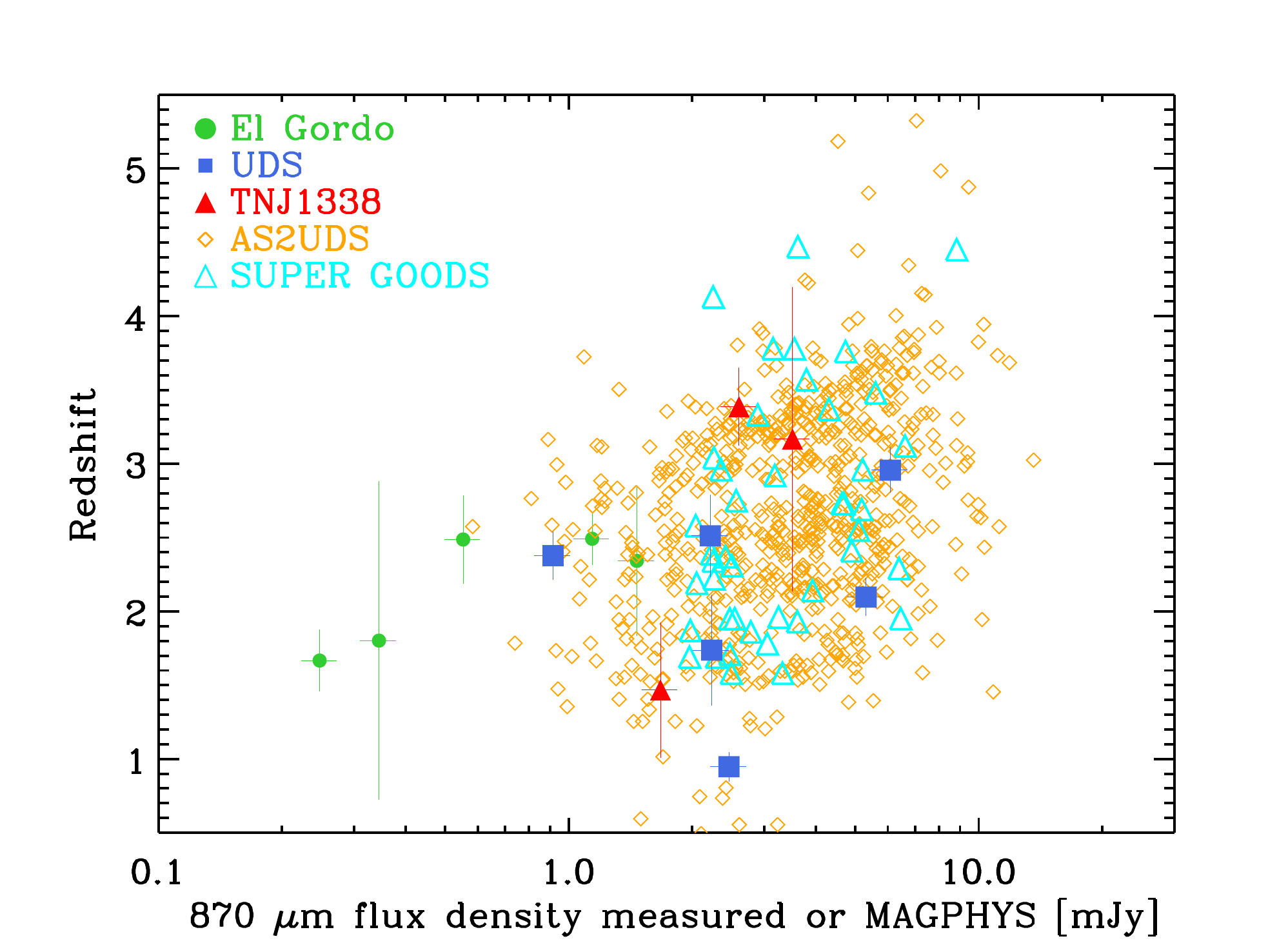}
    \caption{SMG redshifts vs 870~$\mu$m flux density. Green circles represent sources in El Gordo (with delensed flux densities), blue squares sources in UDS, and red triangles TNJ1338.
    For sources that do not have direct $S_{870}$ 
measurements (see Table~\ref{t:targets}), synthesized flux densities based on the best-fit templates from
{\sc magphys} are shown. Open orange diamonds show the AS2UDS sample \citep{2019MNRAS.487.4648S}, and open cyan triangles show the SUPER GOODS \citep{2018ApJ...865..106C} sample.
The two multiply lensed sytems, EG-ALMA 2a/2b and EG-ALMA 6a/6b/6c, are 
excluded, leaving 14 targets. 
}
    \label{870flux}
\end{figure}

\subsection{Sample selection effect}

   By virtue of their submm/mm emission, the ALMA sources discussed in this
work are SMGs. However, they include many sources fainter than 
the bulk of the SMG population previously studied in the literature. 
This is because of the combination of the gravitational lensing by a cluster
in one of our survey fields and that most of
our sources were not pre-selected by surveys using single-dish submm/mm
telescopes but were serendipitous discoveries in ALMA maps of various
depths. The ALMA maps are deeper than any single-dish survey, and therefore
most of our ALMA sources are fainter than those from the usual SMG samples.
This is demonstrated in Figure~\ref{870flux}. Specific to the El Gordo field,
the median $S_{870\micron}$ flux density of our sample is 1.1~mJy after 
correcting for the lensing magnification. If we exclude the two multiply 
imaged systems, the median $S_{870\micron}$ drops to 0.5~mJy in the El Gordo 
field, and the median $S_{870\micron}$ of the whole sample is 2~mJy. In 
contrast, the AS2UDS sample has median $S_{870\micron} = 3.6$~mJy and reaches
down to 0.6~mJy. Our sample therefore includes more SMGs with 
low-to-moderate SFRs, which explains why most of our sources exhibit 
properties (e.g., their non-disturbed disky morphologies) different from those
often described in the literature. On the other hand, it is not surprising that 
most $z\ga$ 2 galaxies should have some low-level, ongoing star formation. 
Due to the high sensitivity of ALMA, our sample is more sensitive than the 
single-dish-pre-selected SMG samples in probing such activity. 

\subsection{Disk galaxy evolution}

   After excluding the two multiply imaged systems (EG-ALMA~2a/2b and 
EG-ALMA~6a/6b/6c) and the quasar (UDS-ALMA~2), the remaining 13 ALMA source 
hosts in our sample are all disk galaxies, mostly showing little sign of 
disturbance suggestive of major mergers. The two ALMA sources of 
\cite{2022ApJ...936L..19C} that are large enough for morphological study are 
similar systems. This would be a surprising result in the pre-{\JWST} era. 
However,  recent {\JWST}/NIRCam morphological studies of the general galaxy 
population have shown that disk galaxies are already common at $z>1.5$ 
\citep{2022arXiv220709428F,2022arXiv221001110F, 2022arXiv220806516J}.
If star formation at $z\approx 1$--3 mostly happens in disk galaxies, it can
explain why our ALMA source hosts are mostly such systems. 

   All these 13 disk galaxies have already acquired large stellar masses, 
which range from 0.3 to $3.5\times 10^{11}$~\Msol. As they have no
indication of being major mergers, these galaxies most likely assembled the 
majority of their stars through secular growth \citep[e.g., ][]{2022arXiv221008658G}, although we cannot rule out
the role of minor mergers. Given their SFRs of 10 to 300~\Msol~yr$^{-1}$,
their sSFRs spread them over both the star-forming main sequence and the 
quiescent categories. Three of them are deemed  quiescent galaxies based on 
their sSFRs. These same three galaxies are also in the quiescent region in 
the rest-frame \textit{UVJ} diagram, which is to say that their UV-to-near-IR
emissions show no sign of ongoing star formation. Had there not been ALMA 
revealing their low-level star formation hidden by dust (SFRs of
10--15~\Msol~yr$^{-1}$), they would be viewed as ``red-and-dead'' and yet 
disky galaxies. These galaxies are similar to those recently reported by 
\cite{2022arXiv220801630N} in the sense that they are also red disky 
galaxies; the difference is that ours are still detected in the {\HST} bands.
Future morphological study within the same ALMA coverage will reveal whether 
there really are disk galaxies that have completely ceased star formation.

    Another interesting result is that our galaxies have a wide range of 
half-light (equivalently, half-mass) radii $R_e$. EG-ALMA~3 has the largest 
$R_e=5.35\pm0.27$~kpc, while EG-ALMA~13 has the smallest $R_e=1.08\pm0.21$~kpc. 
EG-ALMA~3
($z_{\rm phot}=2.34$, $M_*=8.7\times10^{10}$~\Msol, 
$\rm SFR =87$~\Msol~yr$^{-1}$) is
similar to the recently reported grand-design spiral galaxy A2744-DSG-z3 at 
$z=3.06$ \citep{Wu2022}, which has $R_e=7.3\pm0.8$~kpc and 
$M_*=4.0\times 10^{10}$~\Msol. The latter is also an ALMA source and has $\rm SFR = 85$~\Msol~yr$^{-1}$ based on its far-IR-to-mm luminosity \citep{Sun2022}.
Such large disks are rare in our sample, however, the median $R_e$ of our 
sample is 1.6~kpc, which means that our targets are predominantly small 
disks. The sizes of the general disk-galaxy population at 
$z>1.5$ is still awaiting investigation, and therefore it is unclear whether 
our sample being dominated by small disks is normal or is caused by some 
selection bias that is still unknown to us.  

\section{Summary}

    While our sample consists of only 16 unique objects, it is the
largest to date that has both ALMA and {\JWST}/NIRCam data. The ALMA 
positions enabled us to pinpoint unambiguous NIRCam counterparts. Due to the 
high sensitivities of the ALMA data, our sample is
more inclusive than the classic SMG samples in that our sources show a much
wider range of properties. We are able to probe dust-embedded star 
formation as low as ${\sim}10$~\Msol~yr$^{-1}$. The deep, high-resolution
NIRCam data detect the rest-frame near-IR emission from long-lived stars that
dominate the stellar masses, and most of our sources are high-mass
($M_*>10^{10.5}$~\Msol) galaxies with typically non-disturbed disks. 
Furthermore, most of them have small-to-medium half-mass radii 
(median of only 1.6~kpc), suggesting that they are small disks. We
postulate that secular growth can be a viable route to build high-mass disk 
galaxies and that we can now see snapshots of such processes in the submm/mm
regime in a similar way as in UV-to-IR wavelengths. Of course, we will need
a much larger sample to have sufficient statistics in different mass,
morphology, and redshift bins. This calls for the synergy of ALMA and {\JWST}.

\begin{acknowledgments}

We would like to thank the referee for the constructive comments. 
This work is based on observations made with the NASA/ESA/CSA {\it James Webb Space
Telescope}. The data were obtained from the Mikulski Archive for Space
Telescopes at the Space Telescope Science Institute, which is operated by the
Association of Universities for Research in Astronomy, Inc., under NASA
contract NAS 5-03127 for {\JWST}. These observations are associated with {\JWST}
programs 1176 and 2738 (PEARLS) and 1837 (PRIMER).
IRS acknowledges support from STFC (ST/T000244/1).
RAW, SHC, and RAJ acknowledge support from NASA {\JWST} Interdisciplinary
Scientist grants NAG5-12460, NNX14AN10G and 80NSSC18K0200 from GSFC. Work by
CJC acknowledges support from the European Research Council (ERC) Advanced
Investigator Grant EPOCHS (788113). BLF thanks the Berkeley Center for
Theoretical Physics for their hospitality during the writing of this paper.
MAM acknowledges the support of a National Research Council of Canada Plaskett
Fellowship, and the Australian Research Council Centre of Excellence for All
Sky Astrophysics in 3 Dimensions (ASTRO 3D), through project number CE17010001.
CNAW acknowledges funding from the {\JWST}/NIRCam contract NASS-0215 to the
University of Arizona.
We also acknowledge the indigenous peoples of Arizona, including the Akimel
O'odham (Pima) and Pee Posh (Maricopa) Indian Communities, whose care and
keeping of the land has enabled us to be at ASU's Tempe campus in the Salt
River Valley, where much of our work was conducted.

For the purpose of open access, the author has applied a Creative Commons Attribution (CC BY) licence to any Author Accepted Manuscript version arising from this submission.


Some of the data presented in this paper were obtained from the Mikulski Archive for Space Telescopes (MAST) at the Space Telescope Science Institute. The specific observations analyzed can be accessed via \dataset[https://doi.org/10.17909/c1x2-x453]{https://doi.org/10.17909/c1x2-x453}. STScI is operated by the Association of Universities for Research in Astronomy, Inc., under NASA contract NAS5–26555. Support to MAST for these data is provided by the NASA Office of Space Science via grant NAG5–7584 and by other grants and contracts.

This paper makes use of the following ALMA data: ADS/JAO.ALMA\#2013.1.01358.S, 2018.1.00035.L,  2013.1.01051.S, 2015.1.00530.S, 2012.1.00326.S, 2016.1.01184.S, 2013.1.00356.S, 2015.1.01105.S, 2013.1.00781.S, 2015.1.00442.S, 2015.1.01074.S,2015.1.01528.S, 2017.1.01027.S.
ALMA is a partnership of ESO (representing its member states), NSF (USA) and NINS (Japan), together with NRC (Canada), MOST and ASIAA (Taiwan), and KASI (Republic of Korea), in cooperation with the Republic of Chile. The Joint ALMA Observatory is operated by ESO, AUI/NRAO and NAOJ. The National Radio Astronomy Observatory is a facility of the National Science Foundation operated under cooperative agreement by Associated Universities, Inc. The ALMA data reduction and other data services of this work are fully or partially supported by China-Chile Astronomical Data Center (CCADC), which is affiliated to Chinese Academy of Sciences South America Center for Astronomy (CASSACA). 

\end{acknowledgments}

\software{astropy \citep{2013A&A...558A..33A,2018AJ....156..123A},  
          Source Extractor \citep{1996A&AS..117..393B},
          GALFIT \citep{2002AJ....124..266P},
          EAZY \citep{Bra08},
          CASA \citep{2007ASPC..376..127M},
          MAGPHYS\citep{Cun08}
          }

\facilities{ Hubble Space Telescope, James Webb Space Telescope, Mikulski Archive
\url{https://archive.stsci.edu}, Atacama Large Millimeter/Submillimeter Array }

\newpage

\appendix
\restartappendixnumbering 

\section{$z_{\rm phot}$ with different methods}
Figure~\ref{photzcheck}  compares the $z_{\rm phot}$ fitting results with {\sc eazy} and {\sc magphys+photo-z} \citep{2019ApJ...882...61B}, which take the ALMA flux into account. The results are consistent, implying that the main results of this work would not change if we used $z_{\rm phot}$, stellar mass, and SFR from {\sc magphys+photo-z} instead of taking $z_{\rm phot}$ from {\sc eazy}.

\setcounter{figure}{0}
\renewcommand{\thefigure}{A\arabic{figure}}
\begin{figure}
    \centering
    \includegraphics[width=0.6\textwidth]{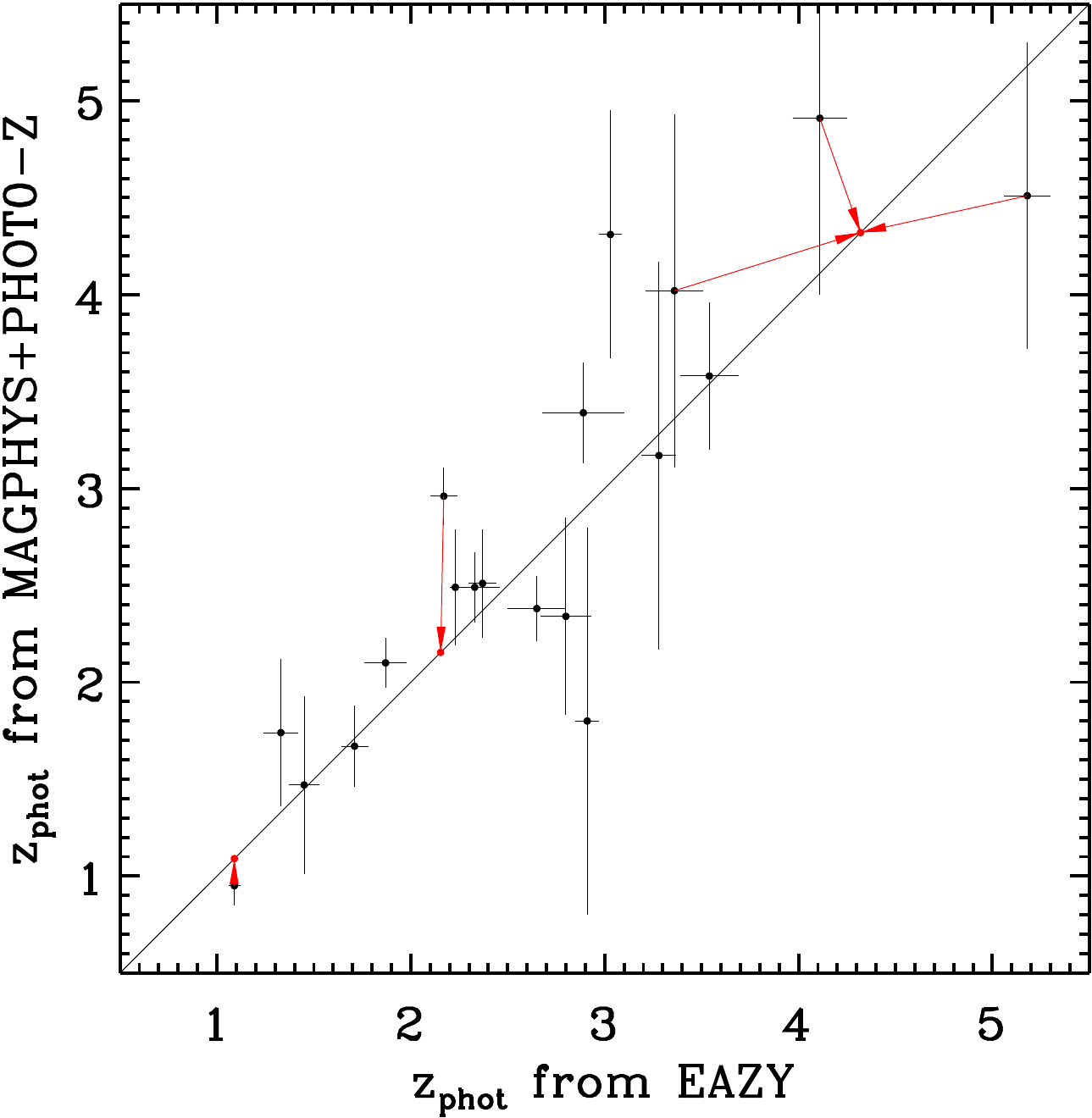}
    \caption{
    Comparison between  $z_{\rm phot}$ from {\sc eazy} and {\sc magphys+photo-z}. Diagonal line shows equality for the two forms of $z_{\rm phot}$. For three targets (UDS-ALMA 3, UDS-ALMA 5, EG-ALMA~6a/6b/6c) with known $z_{\rm spec}$, red arrows connect the $z_{\rm phot}$ to the measured $z_{\rm spec}$, indicated by red points.
    }
    \label{photzcheck}
\end{figure}

\section{S\'ersic profile fitting}

Figure~\ref{galfit} shows the fitted \GALFIT\ models, which are based on a
S\'ersic function plus a central point source. 
Figure~\ref{sersichist} shows the distribution of the derived S\'ersic 
indices with size.
The NIRCam images of TNJ1338 were obtained through medium-band filters
and are shallower. To check whether this biases the fitting, we combined
all three long-wavelength medium-band images and ran \GALFIT\ again on this
composite image. The $R_e$ values thus derived agree with the ones based on
the F360M image.

\setcounter{figure}{0}
\renewcommand{\thefigure}{B\arabic{figure}}

\begin{figure}
    \centering
    \includegraphics[width=0.6\textwidth]{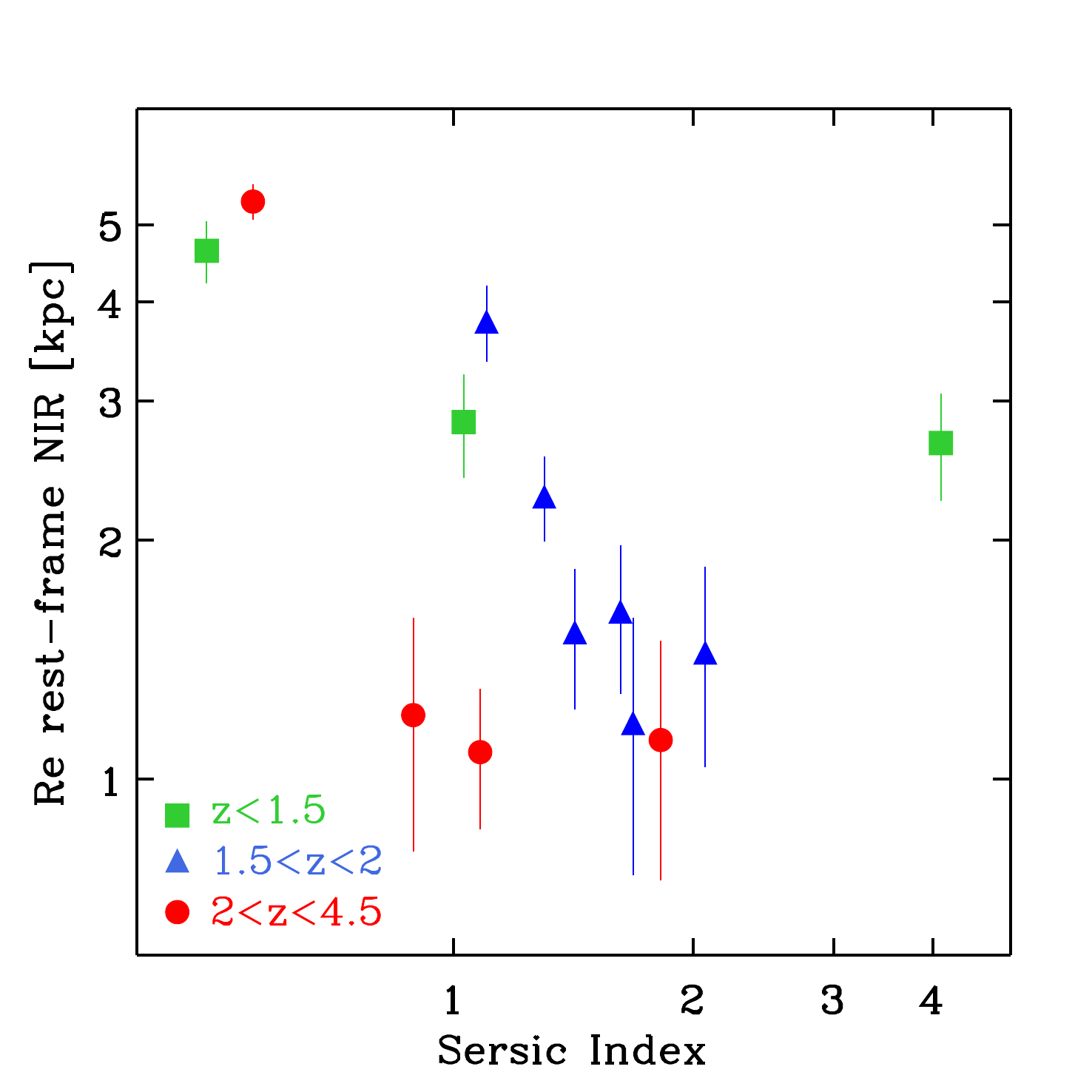}
    \caption{
Half-light radii vs S\'ersic indices for our SMG sample.
UDS-ALMA~1/3/4/5 have compact cores, which were subtracted when calculating the
half-light radii, as here we are interested only in their disk components. The 
SMG with $n\sim 4$ is UDS-ALMA~3, which has a point source, a bar, and spiral
components that skew the model fit.
}
    \label{sersichist}
\end{figure}

\setcounter{figure}{0}
\renewcommand{\thefigure}{B\arabic{figure}}

\section{EG-ALMA~6c: an F160W-faint SMG}

Previous studies have revealed a population of dusty near-IR-faint galaxies 
\citep{2004AJ....127..728F, 2011ApJ...742L..13H, 2019Natur.572..211W, 
2021MNRAS.502.3426S, 2021ApJ...922..114S} 
that are very faint or even invisible in {\HST}/WFC3 F160W (the reddest {\HST} 
band). This has led to the name  ``{\HST} dark,'' but they are prominent sources in 
{\it Spitzer}/IRAC 3.6 and 4.5~$\mu$m images. These could be either passive or 
dusty galaxies at $z\geq 3$. Our sample contains one such source, the triply imaged
system EG-ALMA~6a/6b/6c. EG-ALMA~6c is the most highly magnified of the 
three images. It is adjacent to the source ``ID~\#4a''
\footnote{In the naming scheme of
\citeauthor{2021ApJ...908..146C}, ID~\#4 refers to the lensed source, and 
suffixes a, b, and c refer to the three lensed images.} 
of \citet{2021ApJ...908..146C}, which is part of their galaxy group at 
$z\simeq4.32$.
The MUSE spectrum of ID~\#4a yields $z=4.3196$ based on absorption lines such as 
Ly$\alpha$, \ion{Si}{2}, and \ion{C}{2}\null. From our NIRCam images, ID~\#4a
and EG-ALMA~6c are separated by 1\farcs4, which corresponds to 9~kpc at $z=4.32$.
The host of this ALMA source is much redder than ID~\#4a. Fortunately, there are
archival ALMA Band-3 spectral scans in this region, which we reduced following 
the standard process. There is a strong, double-peaked emission line at 86.6~GHz
(Figure~\ref{ALMA2spec}). If EG-ALMA~6c is at a similar redshift to ID~\#4a, this 
line is CO $J=4$--3, and the redshift corresponds to $z=4.324$, showing that
ID~\#4a and EG-ALMA~6c are distinct sources.
The two peaks of the 86.6~GHz line are separated by 380~km~s$^{-1}$, which suggests
that the source is likely massive but similar to other submm galaxies 
\citep{2021MNRAS.501.3926B}
The integrated CO (4--3) flux is 0.79$\pm 0.08$~mJy~km s$^{-1}$, 
which corresponds to $M_{\rm H_2} = 10^{9.8\pm 1.0}$~\Msol\
(magnification-factor corrected), adopting $L'_{\rm CO (4-3)}/L'_{\rm CO (1-0)} = 0.46$, 
$\alpha_{\rm CO} = 0.8$~\Msol~$\rm (K~km~s^{-1}~pc^2)^{-1}$ 
for SMGs \citep{2013ARA&A..51..105C}. The depletion timescale $M_{\rm H_2}/\rm SFR$ is about 70~Myr, which is close to the typical depletion timescales for local ULIRGs.

The velocity difference between EG-ALMA~6c and ID~\#4a is $\sim$250~km~s$^{-1}$,
suggesting that the two might be interacting. As ID~\#4a is not seen in the ALMA
map, it must have very little cold dust. Its low gas content is consistent with 
the absence of emission lines in the ID~\#4a MUSE spectrum.

Figure~\ref{fig:eg9magphys} shows the \MAGPHYS\ SED-fitting results (de-magnified) for EG-ALMA~6c
at $z=4.324$. We highlight the F150W, F200W, and F277W SED to show that F150W-faint targets could be  dusty galaxies at $z\sim 4$.

\setcounter{figure}{0}
\renewcommand{\thefigure}{C\arabic{figure}}

\begin{figure}
    \centering
    \includegraphics[width=0.6\textwidth]{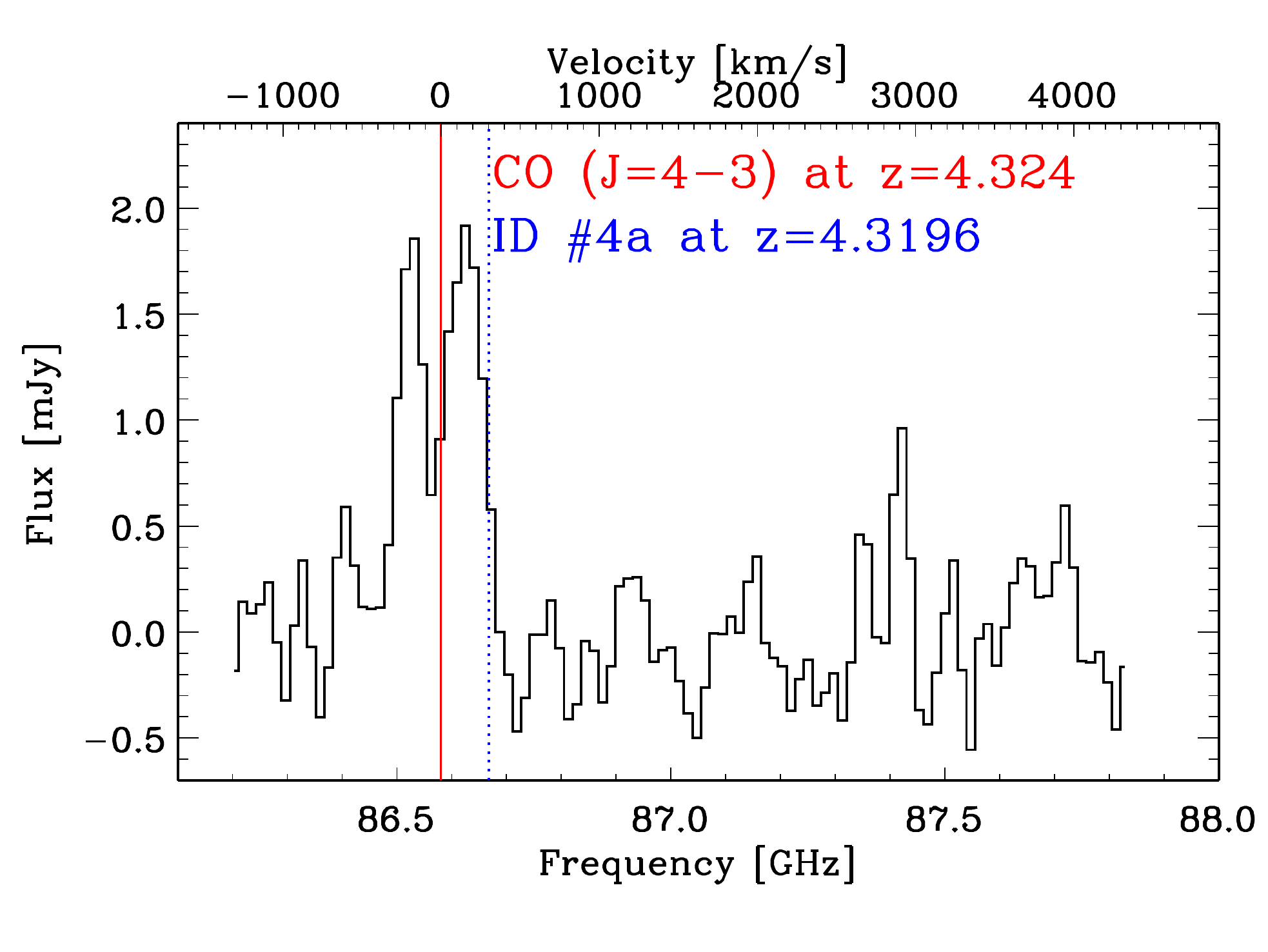}
    \caption{Spectrum of EG-ALMA~6c from the ALMA 92~GHz data cube. The solid red
    line shows the flux-weighted central frequency of the observed emission
    line.  If this line is CO $J=4$--3, its redshift is 4.324. This is
    consistent with the lensing model of \cite{2021ApJ...908..146C}. The dashed blue line shows where the CO $J=4$--3 line would be centered at the redshift of source ID~\#4a \citeauthor{2021ApJ...908..146C}.
    }
    \label{ALMA2spec}
\end{figure}

\begin{figure}
    \centering
    \includegraphics[width=0.6\textwidth]{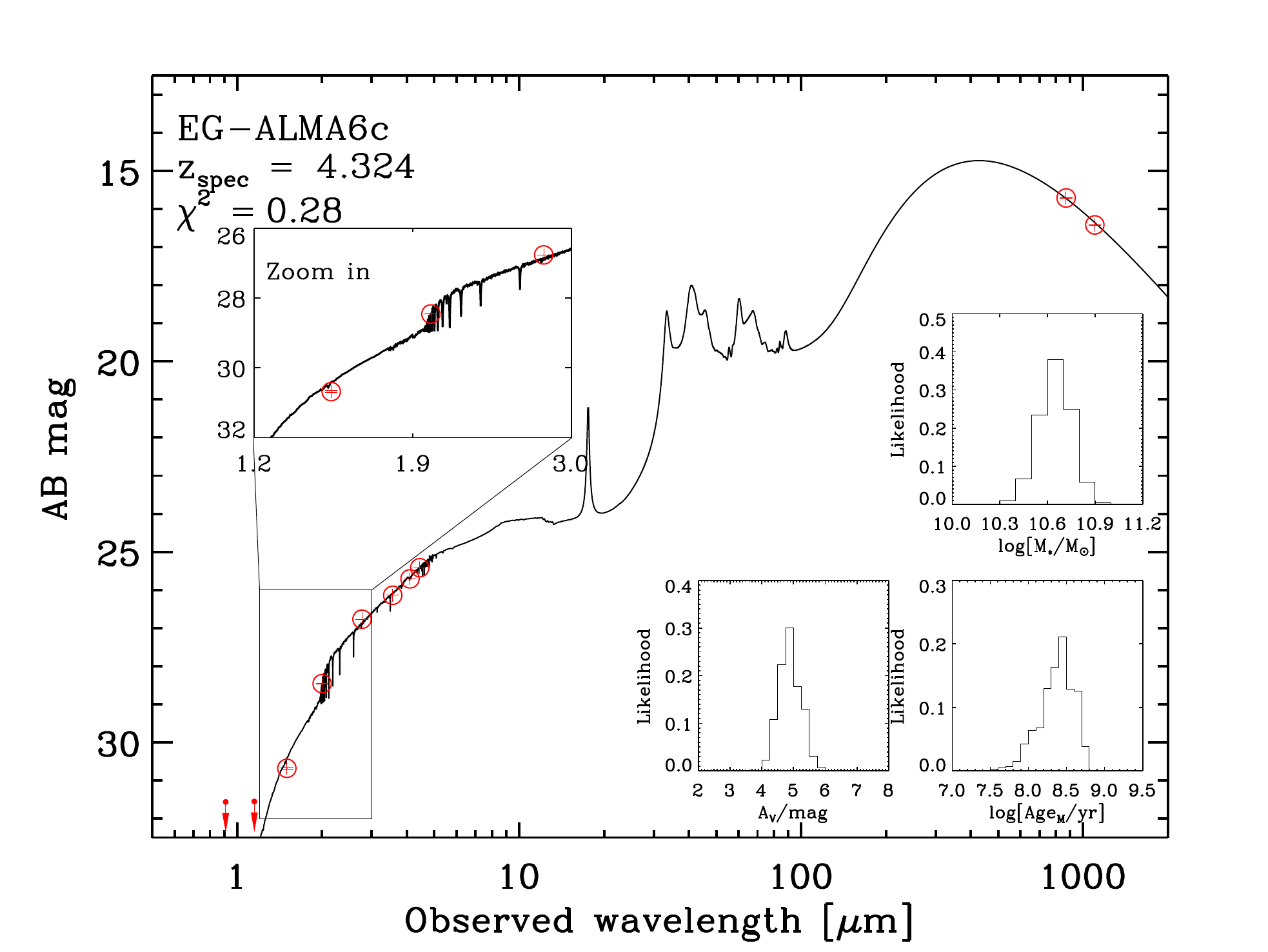}
    \caption{\MAGPHYS\ fitting results for EG-ALMA~6c. Red circles show the NIRCam photometry (using a 0\farcs6 circular aperture) and the ALMA photometry. The two red arrows indicate the 3$\sigma$ upper limits in the F090W and F115W bands. All values are de-magnified. The black curve is the best-fitting model at the spectroscopic redshift $z=4.324$. The three insets on the right are the probability density functions of the stellar mass $M_*$, extinction $A_V$, and mass-weighted age. The inset on the left shows the SED near the Balmer break.}
    \label{fig:eg9magphys}
\end{figure}

\end{document}